\documentclass[a4paper,fleqn,usenatbib]{mnras}

\usepackage[T1]{fontenc}
\usepackage{ae,aecompl}
\usepackage{upgreek}
\usepackage{cancel}

\usepackage{graphicx}	
\usepackage{caption}
\usepackage{subcaption}
\usepackage{amsmath}	
\usepackage{physics}
\usepackage{amssymb,bm}	
\usepackage{xspace}	
\def\app#1#2{%
    \mathrel{%
    \setbox0=\hbox{$#1\sim$}%
    \setbox2=\hbox{%
    \rlap{\hbox{$#1\propto$}}%
    \lower1.1\ht0\box0%
    }%
    \raise0.25\ht2\box2%
  }%
}

\def\Sref#1{Sect.~\ref{#1}\xspace}
\def\Fref#1{Fig.~\ref{#1}\xspace}
\def\Tref#1{Table~\ref{#1}\xspace}
\def\Eref#1{Eq.~\ref{#1}\xspace}

\graphicspath{{./figures/}}

\usepackage[usenames, dvipsnames]{color}
\usepackage[table,xcdraw,svgnames]{xcolor}

\usepackage{newtxtext,newtxmath}

\title[GL of GWs: ML effects on inferred parameters]
{Exploring the Impact of Microlensing on Gravitational Wave Signals: Biases, Population Characteristics, and Prospects for Detection}

\author[Anuj Mishra et al.]{Anuj Mishra,$^{1}$\thanks{E-mail: anuj@iucaa.in}
Ashish Kumar Meena,$^{2}$%
Anupreeta More,$^{1,3}$\thanks{E-mail: anupreeta@iucaa.in}
Sukanta Bose$^{1,4}$ 
\\
\\
$^{1}$The Inter-University Centre for Astronomy and Astrophysics (IUCAA), Post Bag 4, Ganeshkhind, Pune 411007, India
\\
$^{2}$Physics department, Ben Gurion University of the Negev, P.O. Box 653, Be'er-Sheva 84105, Israel
\\
$^{3}$Kavli Institute for the Physics and Mathematics of the Universe (IPMU), 5-1-5 Kashiwanoha, Kashiwa-shi, Chiba 277-8583, Japan
\\
$^{4}$Department of Physics \& Astronomy, Washington State University, Pullman, WA 99164, USA
}

\date{Accepted XXX. Received YYY; in original form ZZZ}

\pubyear{2021}

\begin{document}
\label{firstpage}
\pagerange{\pageref{firstpage}--\pageref{lastpage}}
\maketitle

\begin{abstract}
In this study, we investigate the impact of microlensing on gravitational wave (GW) signals in the LIGO$-$Virgo sensitivity band.
Microlensing caused by an isolated point lens, with (redshifted) mass ranging from $M_\mathrm{Lz}\in(1,~10^5)~{\rm M}_\odot$ and impact parameter $y\in (0.01,~5)$, can result in a maximum mismatch of $\sim 30\%$ with their unlensed counterparts.
When $y<1$, it strongly anti-correlates with the luminosity distance enhancing the detection horizon and signal-to-noise ratio (SNR).
Biases in inferred source parameters are assessed, with in-plane spin components being the most affected intrinsic parameters. 
The luminosity distance is often underestimated, while sky-localisation and trigger times are mostly well-recovered. Study of a population of microlensed signals due to an isolated point lens primarily reveals: (i) using unlensed templates during the search causes fractional loss~($20\%$ to $30\%$) of potentially identifiable microlensed signals; 
(ii) the observed distribution of $y$ challenges the notion of its high improbability at low values ($y\lesssim 1$), especially for $y\lesssim 0.1$; (iii) Bayes factor analysis of the population indicates that certain region in $M_\mathrm{Lz}-y$ parameter space have a higher probability of being detected and accurately identified as microlensed. Notably, the microlens parameters for the most compelling candidate identified in previous microlensing searches, GW200208\_130117, fall within a 1-sigma range of the aforementioned higher probability region. 
Identifying microlensing signatures from $M_\mathrm{Lz}<100~$M$_\odot$ remains challenging due to small microlensing effects at typical SNR values.
Additionally, we also examined how microlensing from a population of microlenses influences the detection of strong lensing signatures in pairs of GW events, particularly in the \textit{posterior-overlap} analysis.
\end{abstract} 

\begin{keywords}
gravitational lensing: strong -- gravitational lensing: micro -- gravitational waves
\end{keywords}

\section{Introduction} 
\label{sec:intro}

Direct detection of gravitational wave~(GW) signals by Laser Interferometer Gravitational-wave Observatory~\citep[LIGO;][]{LIGOScientific:2014pky}, Virgo ~\citep{VIRGO:2014yos}, and Kamioka Gravitational Wave Detector~\citep[KAGRA;][]{Somiya:2011np} detectors which are coming from the binary black hole~(BBH), binary neutron star~(BNS), and NS-BH mergers opened a new window to probe the Universe. At a fundamental level, direct detection of GW signals allows us to test various theories of gravity~\citep[e.g,][]{LIGOScientific:2016lio, LIGOScientific:2017zic, LIGOScientific:2019fpa, LIGOScientific:2021sio}. Observation of GW signals from BBH mergers let us probe the properties of BHs in the Universe~\citep[e.g.,][]{LIGOScientific:2020kqk} and their possible contribution to the dark matter~\citep[e.g.,][]{Bird:2016dcv}. BNS or BH-NS mergers, in addition to GW signal, also emit electromagnetic~(EM) signal, which becomes an excellent tool for multi-messenger astrophysics~\citep[e.g.,][]{Poggiani:2019enk, Margutti:2020xbo}. So far, a total of 90 GW signals have been detected by LVK detector network coming from BBH, BNS and BH-NS mergers in the first three observing runs~\citep{LIGOScientific:2018mvr, LIGOScientific:2020ibl, LIGOScientific:2021usb, KAGRA:2021vkt}. 
Many more such events are expected to be detected in the future observing run~\citep[e.g.,][]{KAGRA:2013rdx} and with new detectors like LIGO-India~\citep{Saleem:2021iwi}, Cosmic Explorer~\citep[CE;][]{Evans:2021gyd}, Deci-hertz Interferometer GW Observatory~\citep[DECIGO;][]{Kawamura:2020pcg}, Einstein Telescope~\citep[ET;][]{Maggiore:2019uih}, and Laser Interferometer Space Antenna~\citep[LISA;][]{Barausse:2020rsu}.

Since GWs couple very weakly with matter, there is no absorption and scattering as they move in space. However, since GWs move along the geodesics, their path can still be altered if they encounter a matter distribution along their path, a phenomenon known as gravitational lensing~\citep[e.g.,][]{1971NCimB...6..225L, Ohanian:1974ys}. For GW signals in the LIGO frequency band,~$f\in[10,10^4]$~Hz, gravitational lensing by galaxy or galaxy cluster scale lenses can lead to the formation of multiple copies of the GW signal (de-)magnified by different factors and arriving with a certain time delay~($\tau_\mathrm{d}$) between them. 
These lensed signals can have time delays ranging from a few hours to several months~\citep[e.g.,][]{Oguri:2018muv, More:2021kpb}. 
In such cases,~$f \tau_\mathrm{d} \gg 1$ and we can study the gravitational lensing using the \textit{geometric} optics approximation~\citep[e.g.,][]{Bernardeau:1999mh}. The extra (de-)magnification introduced by gravitational lensing can introduce bias in the estimation of source distance and binary component masses~\citep[e.g.,][]{Broadhurst:2018saj, Oguri:2018muv, Smith:2017jdz, Hannuksela:2019kle, Broadhurst:2020moy, Diego:2021fyd}. 
In addition, strong lensing also introduces a constant phase shift in the lensed GW signal ($e^{-i n \pi}$ with~$n=0, 1/2, 1$ for type-I, type-II, type-III lensed images; \citealt{Dai:2017huk}). 
Interestingly, this phase shift can be a useful aid in the search of type-II lensed GWs~\citep[][]{Dai:2020tpj, Ezquiaga:2020gdt, Vijaykumar:2022dlp}.
In the context of GWs, strong lensing has been investigated in several works recently~\citep[e.g.,][]{Liao:2017ioi, Takahashi:2016jom, Dai:2017huk, Haris:2018vmn, Li:2018prc, Smith:2017jdz, Broadhurst:2018saj, Broadhurst:2019ijv, 2020arXiv200613219B, Shan:2020esq, Ezquiaga:2020gdt, Cremonese:2021puh, Cremonese:2021ahz, Caliskan:2022wbh, Shan:2023ngi}. 
Various searches have also been carried out for signatures of strong lensing in the existing LVK data~\citep[e.g.,][]{Hannuksela:2019kle, Smith:2018kbc, Dai:2020tpj, Liu:2020par, LIGOScientific:2021izm, LIGOScientific:2023bwz, Janquart:2023mvf}.
However, none of these searches has provided any indications of the existence of a strongly lensed GW signal.

If the GW signal encounters an isolated microlens with mass in the range~$\sim[1,~10^5]{\rm M_\odot}$, the typical time delay~($\tau_\mathrm{d}$) between different lensed signals is such that,~$f\tau_\mathrm{d}\sim1$, where~$f$ is the frequency of GW signal~\citep[see fig. 4 in][]{Meena:2019ate}. In such cases, these multiple lensed signals interfere with each other giving rise to non-negligible frequency-dependent effects in the observed signal~\citep[e.g.,][]{1986ApJ...307...30D, Nakamura:1997sw, Baraldo:1999ny, Nakamura:1999uwi, Jung:2017flg, Seo:2021psp, Bulashenko:2021fes, Caliskan:2022hbu} and we study the gravitational lensing under the \textit{wave} optics approximation. Owing to this frequency dependence, the effect of microlensing is not just limited to the luminosity distance or chirp mass but can extend to other GW signal parameters~\citep[e.g.,][]{Meena:2019ate, Diego:2019lcd, Kim:2023scq}. In galaxy or galaxy cluster scale lenses, instead of isolated microlenses, a whole population (made of stars and stellar remnants like NS and BHs) of microlenses resides. This can lead to complex frequency-dependent microlensing effects in the already strongly lensed GW signal. As shown in~\citet{Diego:2019lcd} and~\citet{Mishra:2021xzz}, strong lensing magnification is an important parameter in determining the strength of these microlensing effects. 
However, microlensing effects are expected to be negligible due to stellar-mass microlenses in strongly lensed GW signals lensed by galaxy scale lenses with magnification below ten as shown in~\citet{Meena:2022unp}. Later~\citet{Meena:2023qdq}, by studying the mismatch between un-lensed and microlensed GW signals, showed that microlensing effects due to stellar mass microlenses are expected negligible in nearly~90\% of global minima in a sample of strongly lensed GW systems with magnification~$<50$.

Since microlensing can introduce complex frequency-dependent features in the observed GW signal, it is important to understand and model these lensing features so that we can properly construct the unlensed GW signal and deduce the properties of the microlens itself. Previous studies~\citep[e.g.,][]{Cao:2014oaa, Lai:2018rto, Christian:2018vsi, Urrutia:2021qak, Basak:2021ten, Bondarescu:2022srx} have made valuable contributions in studying the microlensing effects caused by isolated point-lenses.
However, most of the aforementioned studies examined only a restricted microlens or GW parameter space. For example, the region $y<0.1$ has not been studied well due to its low improbability. 
However, as we show below, our findings demonstrate that selection bias during detection amplifies the probability density in this region due to the extended detection horizon for such signals. 
These results are consistent with previous studies conducted by \cite{takahashi2003wave} and more recently by \cite{Bondarescu:2022srx}.
Moreover, due to the computational expense involved in performing a full parameter estimation run, only a few studies have been conducted in this direction \citep[e.g.,][]{Christian:2018vsi, LIGOScientific:2021izm}. 
Furthermore, some studies only worked primarily in geometrical optics for simplicity. 
Additionally, most studies lack a comprehensive population-wide study that could provide a broader understanding of the phenomenon and make scientific predictions. 
These limitations pose challenges to gaining a thorough understanding of the phenomenon.
Hence, more detailed studies are required to further improve our understanding of microlensing effects in GW signals.

In our current work, we aim to address these gaps by studying the effect of microlensing in a more exhaustive manner: utilizing tools and techniques such as fitting factor, Bayesian analysis and Fisher-information matrix. 
We begin by conducting a fitting factor-based study to investigate the detectability of microlensed signals and demonstrate how the non-inclusion of microlensing effects during the search can affect the observed SNR. Furthermore, we examine how the presence of isolated microlenses can enhance the true source SNR.
Next, we explore the bias in the parameter estimation of GW source parameters when the true signal is microlensed due to an isolated point lens, but the recovery model assumes the usual unlensed signal without incorporating any microlensing effects.
To provide a broader perspective, we perform a population study of microlensed signals, inferring the properties of the population and making predictions about the most likely microlensing parameter space that will be detected and correctly identified as a microlensed signal.
Additionally, we investigate the identification of microlensed signals using a Bayes factor study, considering various scenarios such as varying SNR values, lens masses, and impact parameter values.
Finally, we discuss the crucial aspect of how microlenses in lensing galaxies can affect strongly lensed signals, thereby influencing the searches for strongly lensed GW signals.
Through these comprehensive analyses, we aim to shed light on the multifaceted nature of microlensing effects and their implications for GW signals.

This manuscript is organised as follows. In \Sref{sec:basic_theory}, we review the relevant basics of gravitational lensing and GW data analysis. 
Inspired by recent searches of microlensing, we focus on the point lens model in Sections \ref{sec:ml_effect_on_detection}$-$\ref{sec:ul_vs_ml_study}. 
In \Sref{sec:ml_effect_on_detection}, we study the effect of individual microlenses on the detection of GW signals. In \Sref{sec:ml_effect_on_PE}, we study the bias in the estimation of source parameters of observed microlensed GW signals when recovered using the usual unlensed waveform model. In \Sref{sec:ml_pop_study}, we study the properties of a mock microlensed population. In \Sref{sec:ul_vs_ml_study}, we study the challenges in identifying microlensing signatures in real data.  Lastly, in \Sref{sec:ml_effect_on_PO_analysis}, we investigate a more complex scenario of microlensing of strongly lensed GW signals intervened by a population of microlenses and assess its impact on strong lensing searches. In \Sref{sec:conclusion}, we conclude our work and discuss its implications. 
Throughout this work, we use $H_0=70\:{\rm km\:s^{-1}\:Mpc^{-1}}$, $\Omega_{m}=0.3$, and $\Omega_{\Lambda}=0.7$ to estimate various cosmological quantities. We focus exclusively on transient GW signals originating from compact binary coalescence (CBCs). All mass-related quantities, including $M_\mathrm{Lz}$, are consistently reported in solar mass units (M$_\odot$).

\section{Basic Theory}
\label{sec:basic_theory}

\subsection{Gravitational lensing}
\label{subsec:basics_of_lensing}
Since both gravitational waves and electromagnetic (EM) waves follow null geodesics, the gravitational lensing~(GL) theory for the two is the same.
However, some remarkable differences arise between the GL of GWs and GL of EM waves because of the following:
(i) The GWs are usually coherent as opposed to the EM waves that we observe from astrophysical sources;
(ii) The difference in the frequency range of interest: current ground-based detectors operate at $10-10^4$ Hz, whereas observations in the EM domain are conducted in a much higher frequency range  $\sim 10^{6}-10^{20}$ Hz;
(iii) We measure the amplitude of GWs, as compared to the flux for EM waves, thereby preserving the phase information in the case of the former.
As a result, wave effects can arise in the lensing of GWs due to intervening compact objects in the mass range~$[1,~10^5]~{\rm M_\odot}$ owing to the formation of extra images (microimages) of the signal with time-delay values such that~$f \tau_\mathrm{d} \sim 1$~\citep[e.g.,][]{Nakamura:1999uwi, 2003ApJ...595.1039T}. In such scenario, the amplification factor~(i.e., ratio of lensed and unlensed signal) is given as~\citep[e.g.,][]{Bernardeau:1999mh, goodman2005introduction},
\begin{equation}
F(f, \pmb{y}) = \frac{f}{i}\int d^2\pmb{x} ~ e^{i2\pi f \tau_{\rm d}(\pmb{x}, \pmb{y})} ,
\label{eq:amp_fac}
\end{equation}
where $f$ is the frequency of the signal. $\pmb{x}\equiv(x_1, x_2)$ and $\pmb{y}\equiv(y_1, y_2)$ denote the dimensionless image plane and source plane coordinates, respectively, and the (double) integral is over the whole lens plane. $\tau_{\rm d}(\pmb{x}, \pmb{y})$ denotes the time delay function (measured with respect to its unlensed counterpart) given as
\begin{equation}
    \tau_{\rm d}(\pmb{x}, \pmb{y}) = \frac{(1+z_{\rm d})\xi_0^2}{c}\frac{D_{\rm s}}{D_{\rm d} D_{\rm ds}} 
    \left[\frac{1}{2}(\pmb{x} -\pmb{y})^2 - \psi\left(\pmb{x}\right)
    + \phi_{\rm m}\left(y\right)\right]\,,
\end{equation}
where $D_{\rm d}$, $D_{\rm s}$, and $D_{\rm ds}$ denote the angular diameter distances between the observer and the lens, the observer and the source and the lens and the source, respectively. $z_{\rm d}$ is the lens redshift, $c$ is the speed of light, and $\xi_0$ is an arbitrary reference unit of length on the lens plane responsible for making the coordinates dimensionless. 
Unless otherwise noted, $\xi_0$ is chosen to be the Einstein radius of the lens corresponding to the total lens mass.
$\psi(\pmb{x})$ is the lens potential, $\phi_m\left(y\right)$ is a constant independent of lens properties, and $y \equiv |\pmb{y}|$. If~$f \tau_d \gg 1$ in Eq.~\eqref{eq:amp_fac}, known as geometric optics limit, the integral becomes highly oscillatory and only the stationary points of the integrand have a non-zero contribution to the integral. In such cases, Eq.~\eqref{eq:amp_fac} can be written as
\begin{equation}
    F\left(f\right)\big|_{\rm geo} = \sum_j \sqrt{|\mu_j|} \exp\left(2\pi i f \tau_{\rm d, j} - i \pi n_j \right),
    \label{eq:amp fac point geo}
\end{equation}
where $\mu_j$ and $\tau_{d,j}$ are, respectively, the magnification factor and the time delay for the $j$-th image. Also, $n_j$ is the Morse index, with values 0, 1/2, and 1 for stationary points corresponding to minima, saddle
points and maxima of the time-delay surface, respectively.

\begin{figure}
    \centering
    \includegraphics[width=0.48\textwidth]{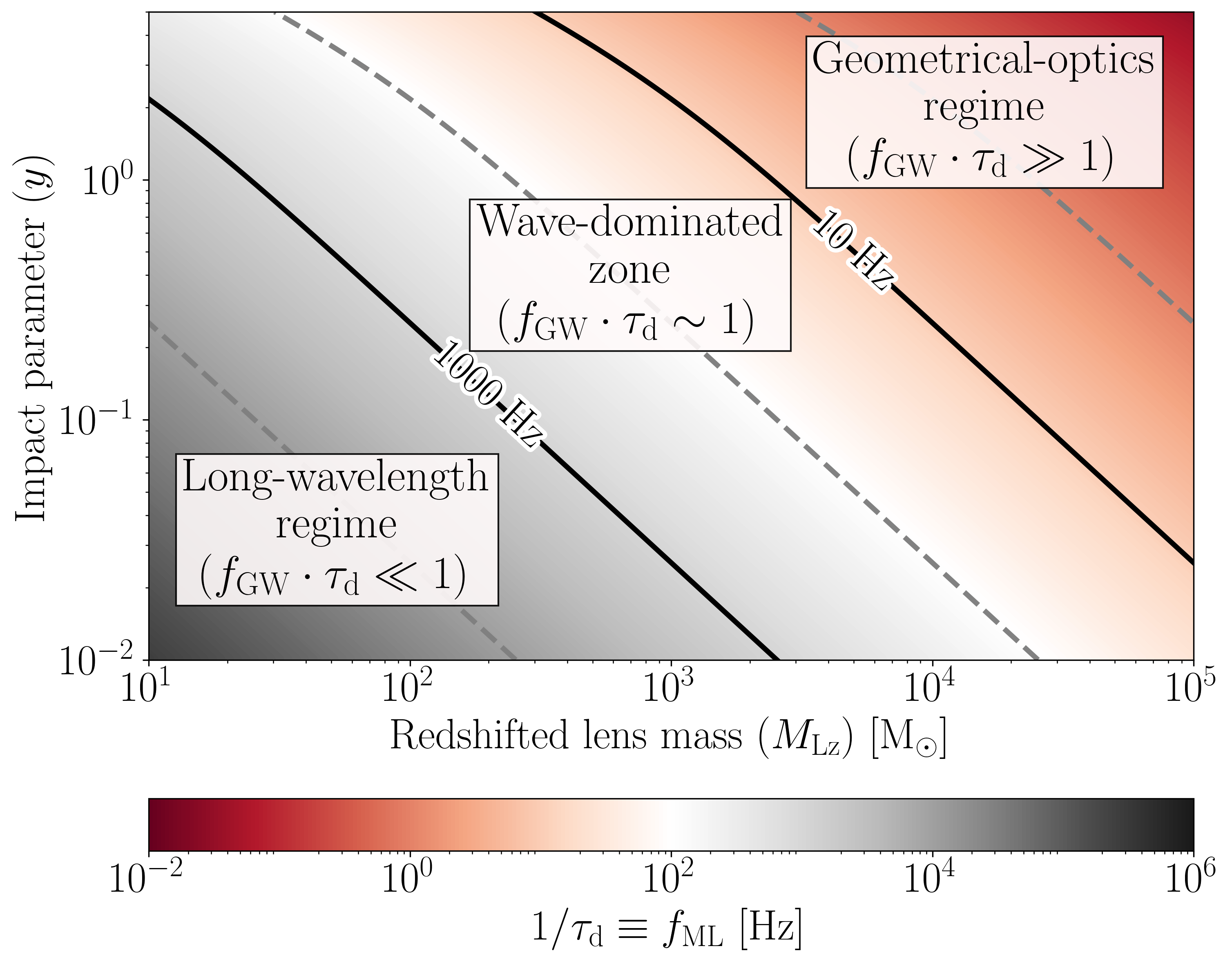}
    \caption{Contour plot of the characteristic frequency, $f_{\rm ML}$, indicating the onset of significant microlensing effects for varying point lens parameters $M_{\rm Lz}$ and $y$ within the LIGO$-$Virgo sensitivity band ($10$-$10^3$ Hz). Contours at $10$ and $10^3$ Hz denote the rough transition regions, dividing the parameter space into three zones: (i) \textit{Long-wavelength regime (left-panel)}, where GW frequency $f_\mathrm{GW}$ is significantly lesser than $f_\mathrm{ML}$, i.e., $f_\mathrm{GW} \ll f_\mathrm{ML}$, resulting in minimal interaction; (ii) \textit{Wave dominated zone (middle-panel):} region where $f_\mathrm{GW} \sim f_\mathrm{ML}$, leading to significant interferenfce effects on GWs. (iii) \textit{Geometrical-optics regime (right panel):} region where $f_\mathrm{GW} \gg f_\mathrm{ML}$. This region is inclusive of milli-lensing and strong-lensing scenarios.}
    \label{fig:f_ML_cplot_for_Mlz_vs_y}
\end{figure}

The diffraction integral, \Eref{eq:amp_fac}, can be solved analytically only for some trivial lens models. For example, the solution for a point-mass lens of mass $M_{\rm L}$ is given by
\begin{equation}
    \begin{split}
        F\left(\omega,y\right) = \exp\bigg\{\frac{\pi \omega}{4} + 
        \frac{i\omega}{2}\left[\ln\left(\frac{\omega}{2}\right) - 
        2\phi_{\rm m}\left(y\right)\right]\bigg\} \\
        \times\ \Gamma\left(1-\frac{i\omega}{2}\right)
        {}_{1}{F}_{1}\left(\frac{i\omega}{2},1;\frac{i\omega y^2}{2}\right),
    \end{split}
    \label{eq:amp fac point}
\end{equation}
where $\omega = 8\pi G(1+z_{\rm d})M_{\rm L} f/c^3$, $\phi_{\rm m}(y) = (x_{\rm m} - y)^2/2 - \ln(x_{\rm m})$ and $x_{\rm m} = \left(y+\sqrt{y^2+4}\right)/2$. The scale factor, $\xi_0$, has been chosen as equivalent to the Einstein radius of the point mass lens. On the other hand, if we have a population of point mass lenses embedded in a galaxy or galaxy cluster, then we can only solve Eq.~\eqref{eq:amp_fac} numerically. The corresponding lensing potential is given as\citep[e.g.,][]{Suyu:2023jue, Saha:2010gb}
\begin{equation}
    \begin{split}
        \psi(\pmb{x}) = \sum_{\rm k}\frac{m_{\rm k}}{M_0}\ln|\pmb{x}-\pmb{x}_{\rm k}| + 
        \frac{\kappa}{2}\left(x_1^2+x_2^2 \right) + 
        \frac{\gamma_{1}}{2}\left(x_1^2-x_2^2 \right) + 
        \gamma_{2}x_1 x_2,
    \end{split}
    \label{eq:potential pop}
\end{equation}
where $m_{\rm k}$ and $\pmb{x}_{\rm k}$ denote the mass and position of the $k$-th point-mass lens in the population, respectively. $M_0$ is an arbitrary mass corresponding to the Einstein radius of $\xi_0$. $(\kappa, \gamma_1, \gamma_2)$ represents the external effects introduced by the galaxy or galaxy cluster.

In simple words, wave effects typically arise when the wavelength of the signal, $\lambda$, becomes comparable to the time delay between microimages, $c\tau_\mathrm{d}$, i.e., when $\lambda \gtrsim c\tau_\mathrm{d}$ (or $f \tau_\mathrm{d}\lesssim 1$). 
For a lens mass $M$, this condition translates to, roughly, $f/{\rm Hz}\lesssim 10^5 (M_\odot/M)$. 
A more careful representation has been plotted in \Fref{fig:f_ML_cplot_for_Mlz_vs_y} in case of an isolated point lens, where we explicitly plot contours of the characteristic frequency $f_{\rm ML}\equiv \tau_{\rm d}^{-1}$ at which wave effects become dominant as a function of different lens parameters, namely, the redshifted lens mass $M_{\rm Lz}$ and the impact parameter $y$. 
The range of $M_{\rm Lz} \in (10, 10^5)M_\odot$ has been chosen based on the condition stated above. The range of $y\in (0.01, 5)$ has been chosen so as to avoid regions which are highly improbable ($y<0.01$) as $p(y)\propto y$, and regions which are not interesting ($y\gg 1$), where the magnification of the micro-images is not significant to cause any appreciable interference as $\mu_{\rm I}(y>5)\sim 1$ and $\mu_{\rm II}(y>5)\sim 0$. 

The detected GW signals coming from chirping solar mass binaries have most of their power in the inspiral and pre-merger phase, i.e., in the frequency range $f_{\rm low}< f\lesssim \mathcal{O}(10^2)\,$Hz. Here, $f_{\rm low}$ denotes the initial frequency from which the analyses commence, typically around $\mathcal{O}(10)~$Hz for ground-based gravitational wave detectors.
\footnote{Where the symbol $\mathcal{O}(n)$ represents `of the order of n'.} 
The contours corresponding to $10$ and $10^3$ Hz in \Fref{fig:f_ML_cplot_for_Mlz_vs_y} are then representative of the transition regions and demarcate the parameter space into roughly three regimes: 
(i) \textit{Long wavelength regime} $(\lambda_\mathrm{GW} \gg  R_s)$, (ii) \textit{Wave dominated zone} $(\lambda_\mathrm{GW} \lesssim  R_s)$, and (iii) \textit{Geometric optics regime} $(\lambda_\mathrm{GW} \ll  R_s)$. 
On account of the above discussion, it is expected that microlensing will be significant in the wave zone, especially the lower right region between these two contours. Consequently, a microlensed GW signal in wave zone would undergo frequency modulations.
The region indicated by the ``Long wavelength regime" basically means the microlenses in this parameter space will only weakly interact with the incoming GWs, owing to $\lambda_\mathrm{GW} \gg R_s$. In this limit, the GW signal will only encounter the initial low-frequency part of the amplification factor where it is only gradually amplifying.
Whereas the region marked with ``geometrical optics regime" is indicative of the parameter space where we would usually find $\lambda_\mathrm{GW} \ll  R_s$. The signals in this region would undergo modulations corresponding to the geometrical optics limit. This region is inclusive of milli-lensing and strong-lensing scenarios \citep[e.g.,][]{Liu:2023ikc, Janquart:2023mvf}.
The contour lines in the plot exhibit a negative gradient, indicating an increase in the time delay as we progress diagonally in an upward-right direction, from lower values of $M_\mathrm{Lz}$ and $y$ to higher values.

\begin{figure*}
    \centering
    \includegraphics[width=\textwidth]{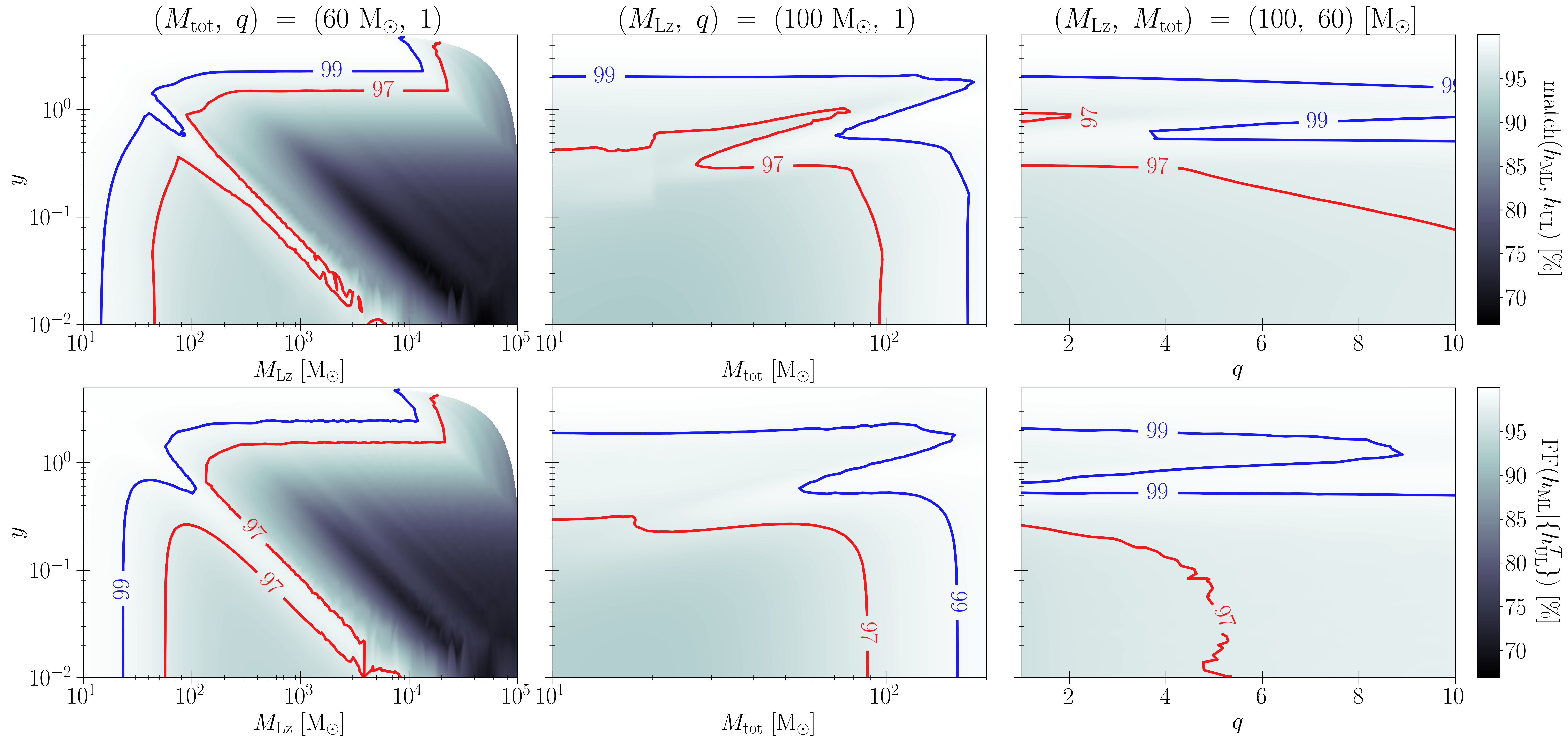}
    \caption{Effect of microlensing on GW WFs for different microlensing and CBC parameters. 
    The top panel show match values between the unlensed and the corresponding microlensed WFs, whereas the bottom panel show fitting factor values (or, the maximum match) for the microlensed WFs when recovering with the unlensed WFs corresponding to the 4D aligned-spin template WFs modelled by parameters  $\{\mathcal{M}_c,~q,~\chi_{1z},~\chi_{2z}\}$. 
    The analysis has been done for: (i) varying redshifted lens mass $(M_{\rm Lz})$ and impact parameter $(y)$ for a fixed binary mass of $M_{\rm tot} = 60~{\rm M}_{\odot}$ and mass ratio $q=1$ (\textit{left panel}), 
    (ii) $M_{\rm tot}$ vs. $y$ parameter space for fixed $(M_{\rm Lz},~q) = (\rm 100~M_{\odot},~1)$ (\textit{middle panel}), (iii)  $q$ {\it vs.} $y$ parameter space for fixed $(M_{\rm Lz},~M_{\rm tot})= (100~{\rm M}_{\odot},~60~{\rm M}_{\odot})$ (\textit{right panel}). 
    The injected spins are kept zero.
    }
    \label{fig:match_FF_analysis}
\end{figure*}

\subsection{GW data analysis and parameter estimation}
\label{subsec:basics_of_GW_data_analysis}
In this work, we restrict ourselves to GW signals arriving from binary black holes (BBHs) and modelled in accordance with General Relativity (GR). 
The corresponding GW waveforms (WFs) are 15 dimensional, modelled by a set of parameters $\pmb{\lambda} \equiv \{\pmb{\lambda}_{int}, \pmb{\lambda}_{ext}\}$, with 8 intrinsic parameters $(\pmb{\lambda}_{int})$ that depend only on the properties of the two BHs, and 7 extrinsic parameters $(\pmb{\lambda}_{ext})$ that are related to how the source is located and oriented relative to the GW detector \citep[e.g.,][]{Husa:2009zz}. 
The 8 intrinsic parameters comprise the two masses, $m_1$ and $m_2$, and the 6 spin components of the two spin angular momenta, $\vec{s_1}$ and $\vec{s_2}$
of the heavier and lighter binary components, respectively. The spin parameters are usually defined in the frame aligned with the total angular momentum $\vec{J}$, as it remains approximately constant for simple precession cases \citep{Fairhurst:2019vut}. These are - (dimensionless) spin magnitudes, $|\vec{s_1}|=a_1$ and $|\vec{s_2}|=a_2$, the tilt angles between the spin vectors and the orbital angular momentum vector ($\vec{L}$), $\theta_1 = \arccos\left(\hat{L}\vdot \hat{s_1}\right)$ and $\theta_2 = \arccos\left(\hat{L}\vdot \hat{s_2}\right)$ (where $\hat{s_{\rm i}}\equiv\vec{s_{\rm i}}/|\vec{s_{\rm i}}|$), the difference between the azimuthal angles of the individual spin vector projections onto the orbital plane, $\phi_{12}$, and the azimuthal angle of $\vec{L}$ on its cone about $\vec{J}$, $\phi_{JL}$). The fact that $\pmb{\lambda}_{int}$ is only 8 dimensional in this case is a consequence of the no-hair theorem, while ignoring electric charge. 
The rest 7 extrinsic parameters are: 4 spacetime coordinates for the coalescence event (arrival time of the signal as it enters the sensitivity band of the detectors, $t_\mathrm{c}$,
luminosity distance to the source, $d_\mathrm{L}$, and two coordinates for specifying its sky location, i.e., right ascension, $\alpha$,  and declination, $\delta$), and 3 Euler angles for the binary’s orientation relative to the Earth (inclination, $\iota$, polarisation, $\psi$, and coalescence phase, $\phi$).

The strain amplitude of GW signals is often much smaller than the random noise present in the detectors. 
However, with the knowledge of physical models describing the dynamics of compact binary mergers and their associated GW waveforms, one can employ an optimal filter to devise a very sensitive search methodology to detect such signals, called the \textit{matched-filtering} technique \citep{Sathyaprakash:1991mt}. 
This involves computing correlations between the detector data and millions of waveform \textit{templates}, covering the parameter space of possible masses and spins of the binary components, and identifying instances of signal-to-noise ratio (SNR, or $\rho$) above a certain threshold\footnote{Current (network) SNR threshold used in the search of GW signals in the LIGO$-$Virgo collaboration is 8.} \citep[e.g.,][]{Usman:2015kfa, Allen:2005fk}.

For detector time-series data $d(t) = h(t) + n(t)$, where $h(t)$ is a GW signal 
and $n(t)$ is the detector noise, the matched-filter SNR of a waveform template  $h_{\rm T}(t, \pmb{\lambda})$ with $d(t)$ 
is given by
\begin{equation}
    \rho =  \max_{\{\phi, t\}}\bra{d}\ket{\hat{h}_{\rm T}}  
    \equiv \max_{\{\phi, t\}}\frac{\bra{d}\ket{h_{\rm T}}}{\sqrt{\bra{h_{\rm T}}\ket{h_{\rm T}}}},
\end{equation}
where the maximisation is done over phase $\phi$ and time $t$
, and $\bra{.}\ket{.}$ is the noise-weighted inner product, called \textit{overlap},
defined as \citep{Usman:2015kfa}\footnote{In writing this, we assume that noise is stationary, i.e.,
$\expval{\tilde{n}(f)^*\tilde{n}(f')}=(1/2)S_{\rm n}(|f|)\delta(f-f')$.}
\begin{equation}
    \bra{h_1}\ket{h_2} \equiv 4~\Re \left[\int^{f_{\rm high}}_{f_{\rm low}} {\rm d}f~\frac{\tilde{h}_1^*(f)\tilde{h}_2(f)}{S_{\rm n}(f)}\right],
    \label{eq:overlap}
\end{equation}
where $\tilde{h}(f) = \mathcal{F}\{h(t)\}(f) = \int dt~h(t)e^{i 2\pi f t}$ 
and $S_{\rm n}(f)$ is the single-sided power spectral density (PSD) of the detector noise. 
Note that since extrinsic parameters primarily affect the signal by introducing a constant phase shift and amplitude, they need not be incorporated explicitly; they are implicitly accounted for during the phase maximization process\footnote{However, this is not always true. For example, in the presence of higher-order modes, the effect of coalescence phase cannot be absorbed into a constant phase shift.}.
In this work, we mostly use the target PSDs for the fourth observing run (O4) of the advanced LIGO and Virgo detectors \citep{KAGRA:2013rdx}\footnote{For LIGO detectors, we used the PSD given in \url{https://dcc.ligo.org/public/0165/T2000012/002/aligo_O4high.txt}. While for Virgo, we used the PSD available at \url{https://dcc.ligo.org/public/0165/T2000012/002/avirgo_O4high_NEW.txt}.}.

Without loss of generality, one can also assume the expected value of noise is zero, i.e., $\overline{n(t)}=0$, in which case the expected value of the SNR of a signal $h(t)$ using a template $h_{\rm T}(t)$ is given by
\begin{equation}
    \expval{\rho} =  \max_{\{\phi, t\}} \bra{h}\ket{\hat{h}_{\rm T}} = 
    \norm{h}~ \mathcal{M}(\hat{h}, \hat{h}_{\rm T}) = \norm{h}\cos{\theta},
    \label{eq:expval_snr}
\end{equation}
where we denote the norm as $\norm{h}\equiv\sqrt{\bra{h}\ket{h}}$, and $\theta$ is the angle between $h$ and $h_{\rm T}$ in the Hilbert space of GW signals; the term
\begin{equation}
   \mathcal{M}(\hat{h}, \hat{h}_{\rm T}) \equiv \max_{\{\phi, t\}} \bra{\hat{h}}\ket{\hat{h}_{\rm T}} 
   \label{eq:match_definition}
\end{equation}
is referred to as \textit{match}\footnote{Unless otherwise noted, the overlap is always maximised over time $t$, phase $\phi$. A natural measure of deviation between any two waveforms can also be defined using \textit{mismatch}, $\mathcal{MM} \equiv 1 - \mathcal{M}$.}, defined as the overlap maximised over time and phase.
In the context of GW searches, the function $\mathcal{M}(\pmb{\lambda_{T}}) \equiv \bra{\hat{h}}\ket{\hat{h}_{\rm T}(\pmb{\lambda_{T}})}$ is called the \textit{ambiguity function},
where the vector $\pmb{\lambda_{T}}\subseteq \pmb{\lambda}$ represents parameters of the template vector and $\pmb{\lambda_{T}}\in \mathcal{T}$, where $\mathcal{T}$ is the discrete set of parameter grid employed for searching \citep[e.g.,][]{Creighton:2011zz, Droz:1998ge}.
From \Eref{eq:expval_snr}, one can see that the optimal value of SNR, $\rho_{\rm opt}$, is simply $\norm{h}$. However, in realistic scenarios, the expected value of the observed matched-filter SNR will be some fraction of the optimal SNR. This fraction is called the (effective) \textit{fitting factor} \citep{Ajith:2012mn, canton2017designing},
\begin{equation}
   {\rm FF} = \max_{\pmb{\lambda_{T}}, t, \phi}{\mathcal{M}(\pmb{\lambda_{T}})} = \frac{\expval{\rho}}{\rho_{\rm opt}},
   \label{eq:FF_definition}
\end{equation}
which is the maximum match obtained among all the templates. The FF value then corresponds to the match with the nearest template to the actual signal (one that subtends the minimum angle to it). The reason why ${\rm FF} < 1$ is primarily three folds - (i) parameter grid of the templates are discretely spaced. 
(ii) limited dimensionality of the template WFs: implying the template signals usually live on a sub-manifold of the actual signal. Hence, the signal can only have a fraction of the projection along that subspace. 
(iii) incomplete model of the template WFs (or some missing physics): in addition to the previous point, if the true waveform contains some physics not incorporated in our template WFs, such as microlensing, eccentricity, non-GR effects, etc., the non-inclusion of these physical effects in the template WFs can further decrease the FF value.

Throughout this work, we perform parameter estimation using nested sampling~\citep{Wu:2009yr}. Specifically, we utilize the \texttt{Dynesty} sampler~\citep{Speagle:2019ivv} as implemented in the \texttt{Bilby} package~\citep{Ashton:2018jfp, ashton2020bilby}. Additionally, for computing microlensing effects, whether for generating microlensed injections (simulated observations) or inferring microlens parameters, we employ a custom frequency domain source model. This model incorporates the two microlensing parameters $M_\mathrm{Lz}$ and $y$, in addition to the standard 15 BBH parameters, and is made publicly available through the \texttt{Python/Cython} package \texttt{GWMAT} (Mishra, A., in prep.).

\section{Effect of microlensing on the detection of GWs: Matched-filtering Analysis}
\label{sec:ml_effect_on_detection}
In this section, we examine the potential impacts of microlensing on GW detection. 
We adopt an isolated point lens model for our microlens analysis.
The lensed GW signal $h_{\rm L}(t)$ is obtained from the unlensed signal $h_{\rm U}(t)$ by using the net amplification factor $F(f)$ caused by the intervening lens system(s), given by the expression,
\begin{equation}
    \tilde{h}_{\rm L}(f)= F(f)\vdot \tilde{h}_{\rm U}(f),
    \label{Eq:F(f)_effect_fdwf}
\end{equation}
where $\tilde{h}_{\rm L}$ and $\tilde{h}_{\rm U}$ are the Fourier transforms of the timeseries $h_{\rm L}$ and $h_{\rm U}$, respectively.
Since $F(f)$ maps real numbers to complex numbers, it causes modulations in both the amplitude and the phase of the signal, thereby affecting the morphology of the WF. As intrinsic parameters are mainly determined from GW phasing, it is highly likely that the intrinsic parameters of a detected microlensed signal will be biased. Similarly, the modulations in  the amplitude are likely to affect the extrinsic parameters. We shall now study this in more detail.

As mentioned earlier, there are several reasons why we anticipate $\rm FF<1$ (or $\expval{\rho}< \expval{\rho}_{\rm opt}$) in practical scenarios. In the context of microlensing, our objective is to determine the reduction in SNR resulting from the exclusion of this physical effect in the search process.
Motivated by real GW searches, we use 4D aligned-spin template WFs to recover the microlensed WFs, which are modelled by the parameters: chirp mass $(\mathcal{M}_c)$, mass ratio $(q)$, and aligned spin components of the two component masses $(\chi_{1z},~\chi_{2z})$. 
To estimate minimum loss of SNR during the search, we compute the maximum match (\Eref{eq:FF_definition}), or the fitting factor (FF), between the microlensed and the unlensed waveforms in the 4D parameters listed above. 
We use the \texttt{PyCBC} package (\citealt{alex_nitz_2020_4355793}, 
\citealt{ Usman:2015kfa}) for computing match values (\Eref{eq:match_definition}), and work with the approximant \texttt{IMRPhenomPv3} \citep{Khan:2018fmp} with an $f_\text{\scriptsize low}$ value of $20$ Hz, where $f_\text{\scriptsize low}$ is the lower frequency cutoff in the evaluation of the overlap (see \Eref{eq:overlap}). It is also the starting frequency for the generation of $(l=2,~m=2)$ mode of the GW waveform. 
The power spectral density (PSD) used is \texttt{aLIGOZeroDetHighPower}\footnote{\url{https://dcc.ligo.org/LIGO-T070247/public} ;\\ \url{https://dcc.ligo.org/T1800044-v5}. }, which is analogous to O4 targeted PSDs of LIGO detectors \citep{KAGRA:2013rdx}\footnote{\url{https://dcc.ligo.org/LIGO-T2000012/public};\\ \url{https://dcc.ligo.org/LIGO-T1500293/public}.}. 
The FF values have been computed using the Nelder-Mead algorithm as implemented in the `optimization' module of the \texttt{Scipy} library \citep{Wu:2009yr}.
It's important to note that since we employ a maximization algorithm to compute FF, we do not account for any additional reduction in FF due to the discrete placement of templates (point (i) below \Eref{eq:FF_definition} does not apply in our case).

A fitting factor value of $x$ ensures that the maximum fractional loss of possible astrophysical signals is not more than a factor of $(1-x^3)$.\footnote{Assuming the rate of mergers $R\propto d^3$ and that $\rho \propto d^{-1}$, where d is the luminosity distance to the source.}
The template banks typically used for searching GW signals from compact binary coalescences (CBCs) have a minimum fitting factor threshold of around 97$\%$ for WFs within the parameter space, which implies no more than $\sim 10\%$ of possible astrophysical signals are lost due to the discrete nature of the bank.

\begin{figure*}
    \centering
    \includegraphics[width=\textwidth]{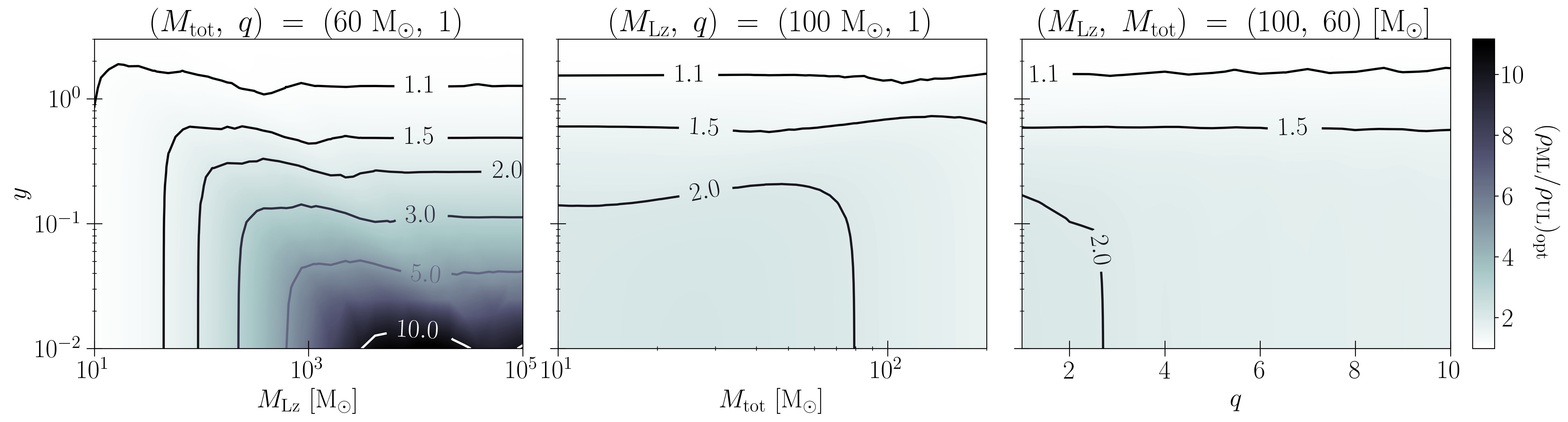}
    \caption{The figure depicts the variation in the optimal SNR in the no-lens versus microlens cases. The variation is shown as a function of point-lens mass ($M_\mathrm{Lz}$; left panel), binary mass ($M_\mathrm{tot}$; middle panel), and mass-ratio ($q$; right panel). The presence of an isolated microlens always increases the SNR relative to the no lensing case.}
    \label{fig:ul_vs_ml_SNR}
\end{figure*}

The results are shown in \Fref{fig:match_FF_analysis}, where we study the effect of microlensing on GW signals for different microlensing and CBC parameters. 
The top panel shows match values between the unlensed and the corresponding microlensed WFs, quantifying the amount by which a WF changes due to microlensing. The bottom panel shows fitting factor values (or, the maximum match) for the microlensed WFs when recovering with the unlensed WFs corresponding to the 4D aligned-spin template WFs modelled by parameters  $\{\mathcal{M}_c,~q,~\chi_{1z},~\chi_{2z}\}$. As discussed above,
this loss can lead to a drop in the detection rate owing to the influence of microlensing on the signals. 
The analysis has been done for: (i)~varying redshifted lens mass $(M_{\rm Lz})$ and impact parameter $(y)$ for a fixed binary mass of $M_{\rm tot} = 60~{\rm M}_{\odot}$ and mass ratio $q=1$, which represents a "golden"\footnote{Binaries with total mass $\sim 50 - 200~{\rm M}_\odot$.} black-hole binary to which our detectors are sensitive from inspiral to ringdown phase of the coalescence \citep{hughes2005golden, Nakano:2015uja, Ghosh:2016qgn} (\textit{left panel}), 
(ii)~$M_{\rm tot}$ {\it vs.} $y$ parameter space for fixed $(M_{\rm Lz},~q)$ = $(\rm 100~M_{\odot},~1)$ (\textit{middle panel}), 
(iii)~$q$ {\it vs.} $y$ parameter space for fixed $(M_{\rm Lz},~M_{\rm tot})$ = $(100~{\rm M}_\odot,~60~{\rm M}_{\odot})$ (\textit{right panel}). 
The differences between the match and FF values in the top and bottom panels suggest that microlensing of GW signals can lead us to infer biased or inaccurate source parameters.

In the leftmost panel, the ranges of $M_{\rm Lz}$ and $y$ have been kept the same as in \Fref{fig:f_ML_cplot_for_Mlz_vs_y}. 
The match and FF values are close to $1$ for low $M_{\rm Lz}$ and high $y$, consistent with the unlensed scenario, and decrease almost diagonally as $M_{\rm Lz}$ and $1/y$ are increased.  
However, we notice both match and FF plots have oscillations across equal time delay contours as shown in \Fref{fig:f_ML_cplot_for_Mlz_vs_y}. Also, the worst match values (or the highest mismatch) come from a region where wave effects are large, i.e., the region depicted between the two contours $f_{\rm ML}=\{10,~100\}$ shown in \Fref{fig:f_ML_cplot_for_Mlz_vs_y}. 
In the middle panel, the match and FF values are close to $1$ for high $M_{\rm tot}$ and $y$ values, and decrease almost diagonally down for low values of $M_{\rm tot}$ and $y$. 
The reason we see a high match for higher BBH masses is because the signal length becomes comparable to the time delay between the microimages associated with those lensing parameters. This leads to fewer modulations that can affect the GW signal.
It is also worth noting that the match and the FF contours corresponding to a value of $97\%$ (red curves) differ significantly in a region where $y\in\sim (0.1,~1)$ and $M_{\rm tot}\in\sim(20,~60)~{\rm M}_{\odot}$.
Similarly, in the rightmost panel, the FF contours corresponding to a value of $97\%$ (red curves) change drastically between the top and the bottom panel. 
This variation between match and FF values hints towards a strong degeneracy between microlensing the CBC intrinsic parameters and the microlensing parameters.
 
Although \Fref{fig:match_FF_analysis} is important to determine the effect of microlensing on the detection of GWs, it is not sufficient. 
Since microlensing will also affect the inferred (effective) \textit{luminosity distance}, the horizon distance to a microlensed GW signal will also shift accordingly relative to the unlensed case, and so will the inferred rate of mergers. 
Therefore, in \Fref{fig:ul_vs_ml_SNR}, we show how optimal SNR can vary in the presence of microlens. The leftmost panel shows variation in the microlensing parameter space of $M_\mathrm{Lz}$ and $y$, while keeping the binary parameters fixed to $(M_{\rm tot},~q)=(60~{\rm  M}_\odot,~1)$. Same with the middle and the right panels, except we vary $M_{\rm tot}$ and $q$, respectively, while fixing $M_\mathrm{Lz} = 100{\rm M}_\odot$.
In all the panels, the contours represent the ratio of the optimal SNRs between the case when the microlens is present {\it vs.} when it is absent, i.e., $\rho_\mathrm{ML}/\rho_\mathrm{UL}$. This value should tend to unity at higher impact parameter values, consistent with the darker regions at high values of $y$ where microlensing effects are insignificant.
In the leftmost panel, we observe a drastic change in the optimal SNR. The SNR in the presence of microlens increases almost monotonically as we increase $M_\mathrm{Lz}$ and $1/y$, reaching a value of more than 10 times the unlensed SNR in the bottom-right corner of the plot. Even for modest values of microlensing such as $(M_\mathrm{Lz}, y)=(10, 1)$, we observe a $10\%$ increase in the SNR.
In the middle and the right panels, since contour corresponding to the value $1.5$ is almost flat, we observe that the change in SNR is not correlated with varying $M_{\rm tot}$ and $q$ for higher values of $y\gtrsim 0.5$. However, for lower values of $y\lesssim 0.5$, we do see a correlation in both panels. SNR tends to increase for lower mass binaries, which is a consequence of longer signal duration, i.e., the integrated effect of microlensing over the signal. 
Although not visible explicitly, when we examine the variation of the SNR with $q$, we find oscillatory behaviour of SNR as $q$ increases. This oscillatory behavior is a consequence of amplitude oscillations in $F(f)$. As we fix the microlensing parameters, $M_\mathrm{Lz}$ and $y$, and only vary $q$, the GW frequency at ISCO would decrease monotonically, and so does the strain at maximum strain amplitude. The oscillations in $|F(f)|$ would then translate to oscillations in the optimal SNR as the ISCO frequency varies.  

In \Fref{fig:match_FF_analysis}, we observed that although microlensing would further decrease the SNR due to a decrease in fitting factor values, \Fref{fig:ul_vs_ml_SNR} suggests that the SNR itself increases due to the presence of an isolated microlens. Thus, the effect of microlensing on the detection of GWs is non-trivial.
These figures suggest that the observed distribution of the microlens population can differ significantly from the expected prior distribution. 
We study this statistically in more detail in \Sref{sec:ml_pop_study}.

\section{Effect of microlensing on parameter estimation of GWs: Bayesian Analysis}
\label{sec:ml_effect_on_PE}
In this section, we study the biases caused by the microlensing effects on inferred source parameters. We also investigate the correlations between the recovered parameters  when microlensed signals are either recovered with a microlensed waveform model or the usual 15D unlensed waveform model. We again employ the isolated point lens model for microlensing. 

A frequency domain GW waveform, $\tilde{h}(f)$, for a chirping BBH system can be written in the form \citep{Allen:2005fk}:
\begin{equation}
    \tilde{h}(f) = A(f; \mathcal{M}, D_\mathrm{eff})e^{-i\Psi(f;\mathcal{M}, \eta, \vec{s_1}, \vec{s_2})},
    \label{eq:GW_WF_general_form}
\end{equation}
Here, $\mathcal{M}$ and $\eta$ represent the chirp mass and the symmetric mass ratio, while $\vec{s_{1,2}}$ denote the binary constituent spin vectors, and $D_\mathrm{eff}$ is the effective luminosity distance of the source which is related to the true luminosity distance, $D_o$, as
\begin{equation}
    D_\mathrm{eff} = D_\mathrm{o}\left[F_{+}^2\left(\frac{1+\cos^2\iota}{2}\right)^2 + F^2_\mathrm{x}\cos^2\iota \right]^{-1/2} 
\end{equation}
where $\iota$ is the inclination angle defined in the orbital angular momentum frame between the direction to the observer and the orbital angular momentum axis of the binary system; $F_{{+}/\mathrm{x}} \equiv F_{{+}/\mathrm{x}}(\alpha, \delta, \psi, t_c)$ are the antenna pattern response functions that relate the source orientation to the detector orientation, described by the right ascension and declination of the source, $(\alpha,~\delta)$, the trigger time at the detector, $t_\mathrm{c}$, and on the polarisation angle $\psi$.
The induced strain on the detector is then related to the pure polarised components as
\begin{equation}
    h = F_{+}(\alpha,~\delta,~\psi) h_{+} + F_\mathrm{x}(\alpha,~\delta,~\psi) h_\mathrm{x}
\end{equation}
Therefore, from \Eref{eq:GW_WF_general_form}, the inference of GW phasing is extremely important to study the intrinsic source properties, like their masses and spins, while the extrinsic parameters mainly affect the amplitude of the signal and result in an effective luminosity distance. As already pointed out earlier, since microlensing induces modulations in both the amplitude and phase, \Eref{Eq:F(f)_effect_fdwf}, it is expected it will affect most of the GW parameters.

To understand the effect of microlensing on the inferred source parameters, we do a Bayesian analysis by performing a set of parameter estimation runs \citep{cutler1994gravitational, Husa:2009zz, Thrane:2018qnx, Christensen:2022bxb}. 
We inject zero-noised microlensed BBH signals into the three detectors (LIGO Livingston, LIGO Hanford, and Virgo) having PSDs corresponding to the target sensitivities of the upcoming O4 runs \citep{KAGRA:2013rdx}. The injected signals are non-spinning with extrinsic parameters corresponding to $\rm GW150914$, except for the luminosity distance, $d_\mathrm{L}$, which is scaled to obtain a desired SNR. 
Firstly, we inject and recover a microlensed signal using a microlensed WF model to see the correlation between the 17D parameters, especially between the 15D BBH parameters and the two lensing parameters.
Then, to understand the biases in the inferred BBH parameters and their degeneracies with microlensing, we inject a set of microlensed signals and recover using the usual unlensed templates, i.e., assuming no microlensing is present in the signal. 
Parameter estimation (PE) runs are performed using the publicly available package \texttt{\textsc{Bilby-Pipe}} \citep{ashton2020bilby, Ashton:2018jfp, Romero-Shaw:2020owr}, keeping all parameters free while recovery.
For both the injection and the recovery templates, we use \texttt{IMRPhenomXPHM} \citep{London:2017bcn} waveform approximant with $f_\mathrm{low}=20$ Hz as the lower frequency cutoff for the likelihood evaluation.
As mentioned previously, we use the \texttt{Dynesty} sampler with the following settings: \{nlive=2048, nact=50\}, and use n-parallel=4 to combine four independent parallel chains to get the final posterior sample.

\begin{figure*}
    \centering
    \includegraphics[width=\linewidth]{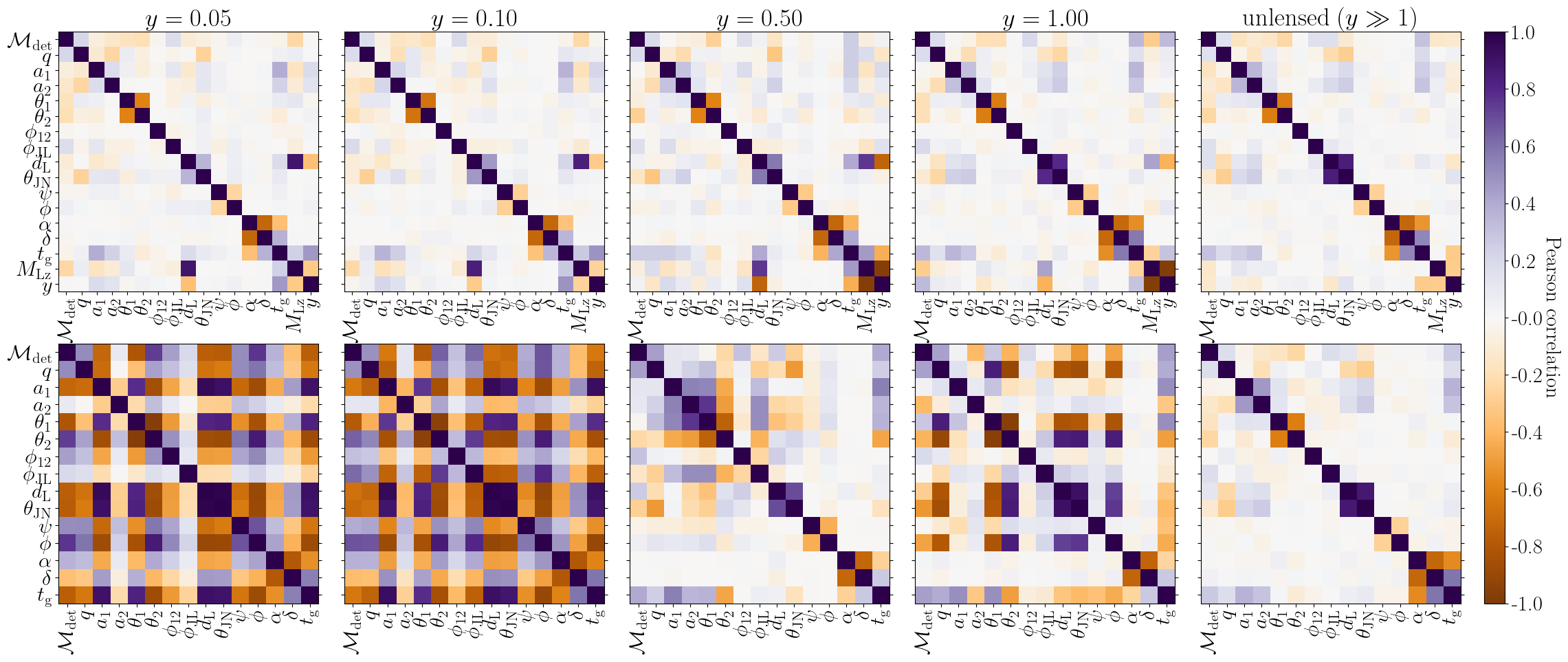}
    \caption{Correlations between the parameters of microlensed waveforms. The injected waveforms correspond to a BBH with a total binary mass of $M_\mathrm{tot} = 60~$M$_\odot$, mass-ratio $q=1$, and having observed network SNR of $\sim 50$. Among the microlens parameters, the redshifted lens mass is fixed to $M_\mathrm{Lz}=100~$M$_\odot$ and the impact parameter increases from left to right as $y\in \{0.05,~0.1,~0.5,~1.0,\gg 1\}$ (also indicated at the top of each column). 
    \textit{Top:} correlations when the recovery model corresponds to 17D microlensed waveforms. \textit{Bottom:} correlations when the recovery model is the usual 15D unlensed BBH waveforms. }
    \label{fig:corr_1}
\end{figure*}

\begin{figure*}
    \centering
    \includegraphics[width=\linewidth]{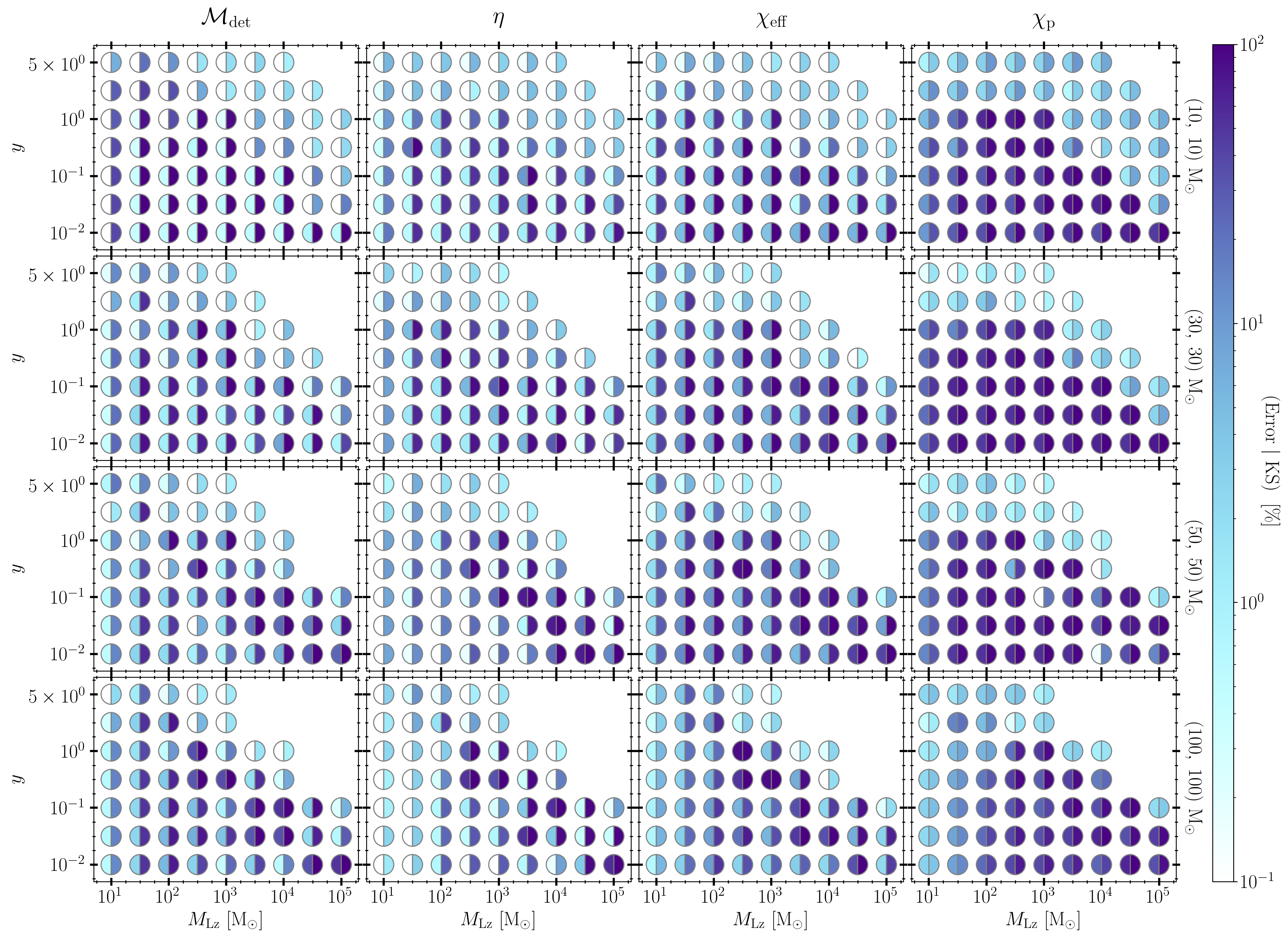}
    \caption{  
    Bias in the inferred intrinsic source parameters of a microlensed signal when recovered using the usual unlensed waveform model, characterised by 15 parameters.
    Each column represents the bias in a specific parameter, indicated at the top, while different rows correspond to different source binaries, indicated on the right side of each row.
    The intrinsic parameters are represented by the detected chirp mass ($\mathcal{M}_\mathrm{det}$), symmetric mass-ratio ($\eta$), projected effective spin ($\chi_\mathrm{eff}$), and precession effective spin ($\chi_\mathrm{p}$).
    The source parameters correspond to equal mass binaries with a mass ratio of $q=1$ and a total mass of $M_\mathrm{tot}/~{\rm M}_\odot\in\{20,~60,~100,~200\}$.
    Each subplot shows the bias in the microlensing parameter space of the redshifted lens mass, denoted as $M_\mathrm{Lz}/~{\rm M}_\odot\in\{\rm 1e1,~5e1,~1e2,~5e2,~1e3,~5e3,~1e4,~5e4,~1e5\}$, and the impact parameter values denoted as $y\in\{0.01,~0.05,~0.1,~0.5,~1.0,~3.0,~5.0\}$.
    Each circular marker in the plot has two halves: the left half represents the relative percentage error (absolute error for $\chi_\mathrm{eff}$ and $\chi_\mathrm{p}$) between the median values of the microlensed and unlensed recoveries, while the right half represents the two-sample Kolmegorov-Smironov (KS) Statistic value between the 1D marginalised posteriors of the microlensed and unlensed recoveries (in percentage).
    We further ignore points in the parameter space corresponding to a time-delay between microimages greater than the signal duration, indicating cases where strong lensing is observed (empty region in the top-right corner of each subplot).  
}
    \label{fig:PE_bias_1}
\end{figure*}

\begin{figure*}
    \centering
    \includegraphics[width=\linewidth]{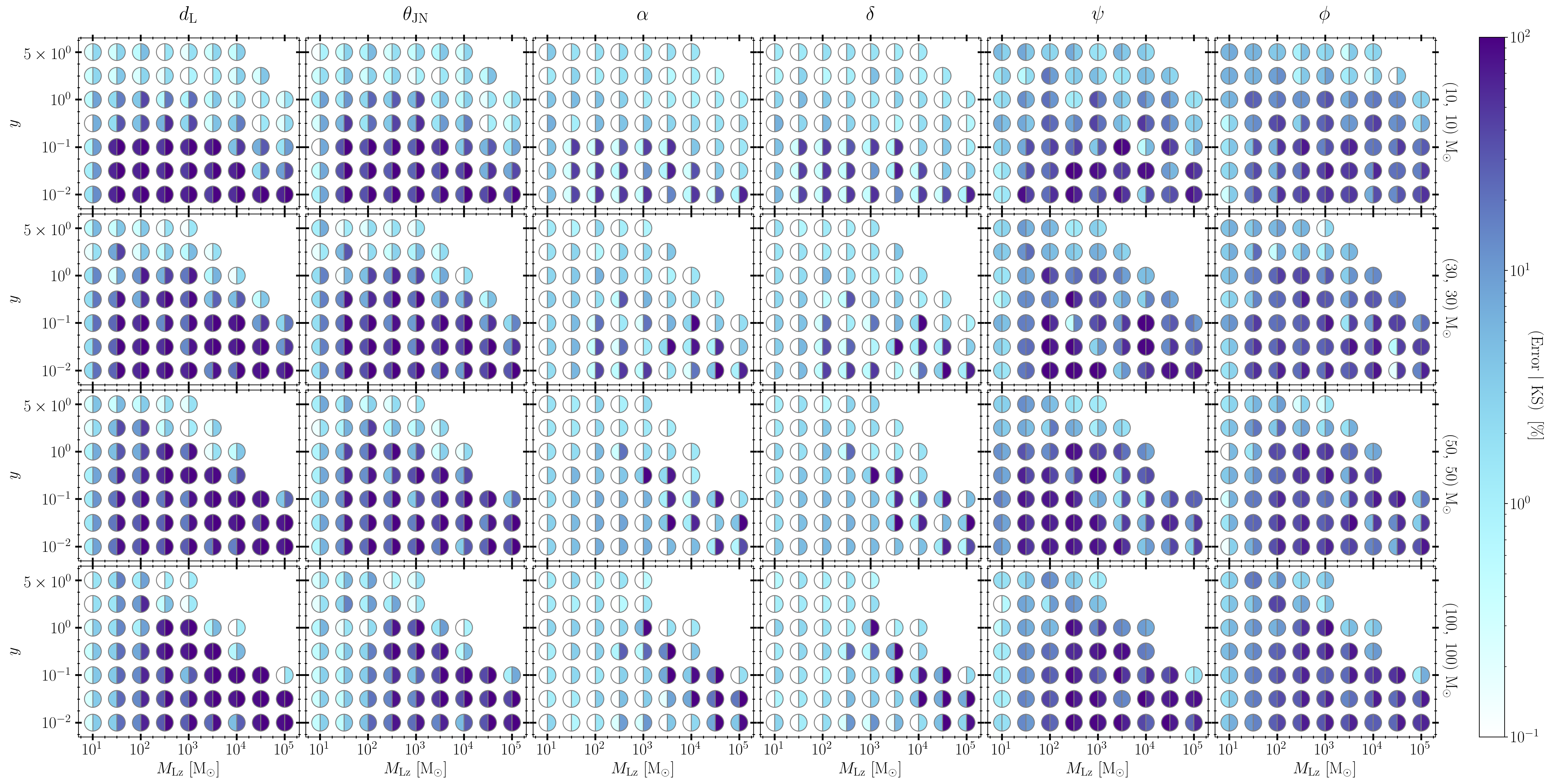}
    \caption{ Same as \Fref{fig:PE_bias_1}, but for extrinsic parameters.
    The extrinsic parameters are represented by the luminosity distance ($d_\mathrm{L}$), inclination ($\theta_\mathrm{JN}$), sky location (RA and Dec; $\alpha$ and $\delta$), polarization angle ($\Psi$), and coalescence phase ($\phi$).
    The trigger time $t_\mathrm{c}$ is not shown as it is well recovered throughout the parameter space.
    }
    \label{fig:PE_bias_2}
\end{figure*}

\begin{figure*}
    \centering
     \includegraphics[width=\textwidth]{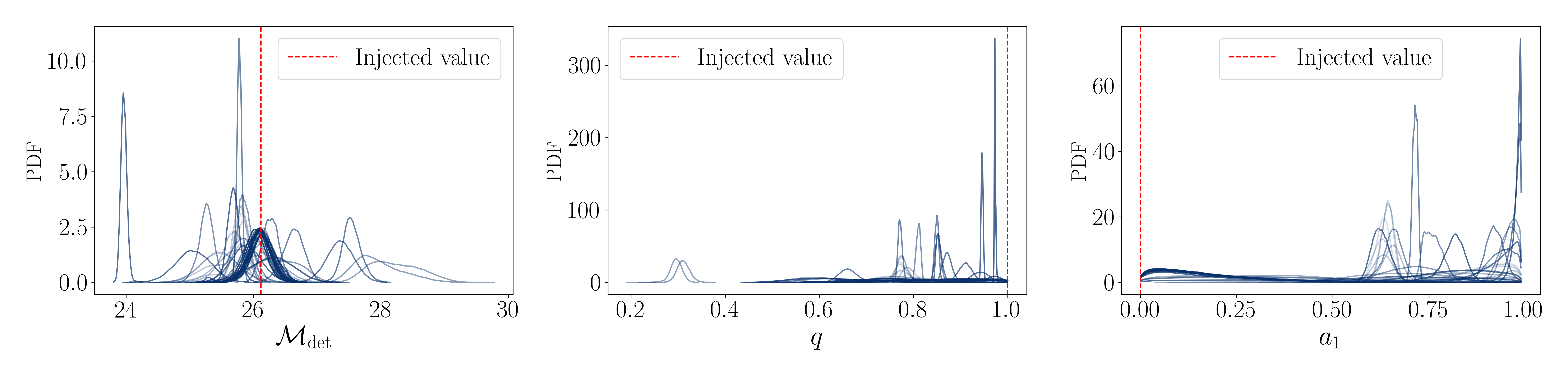}
    \caption{1D marginalised posterior distributions for the recoveries of $\{\mathcal{M}_\mathrm{det},~q,~a_1\}$ of the $60{\rm M}_\odot$ binary system shown in \Fref{fig:PE_bias_1} and \ref{fig:PE_bias_2}. The dashed red line represents the injected value} 
     \label{fig:mtot60_PE_1D_study}
\end{figure*}

\begin{figure*}
     \centering
     \begin{subfigure}[t]{0.49\textwidth}
         \centering
         \includegraphics[width=\textwidth]{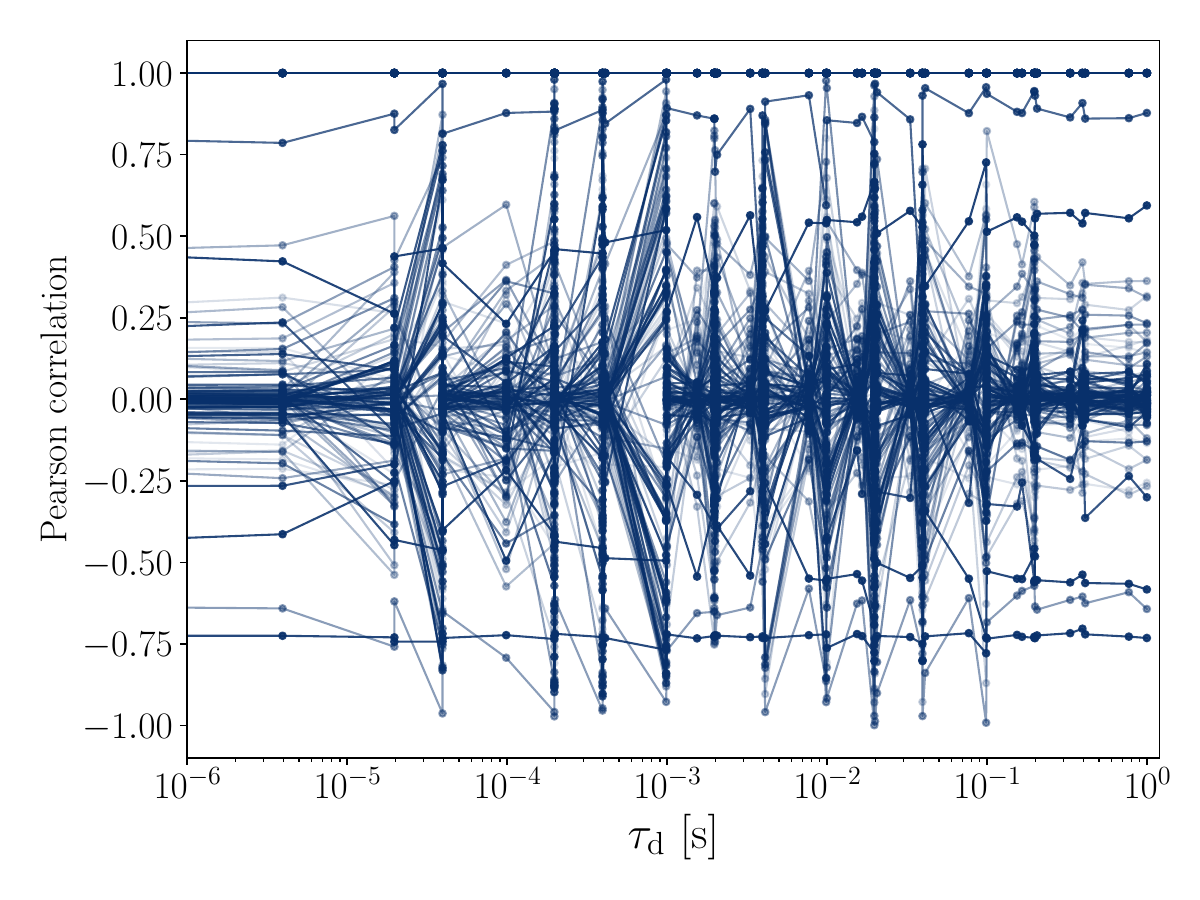}
         \caption*{}
     \end{subfigure}
     \hfill
     \begin{subfigure}[t]{0.49\textwidth}
         \centering
         \includegraphics[width=\textwidth]{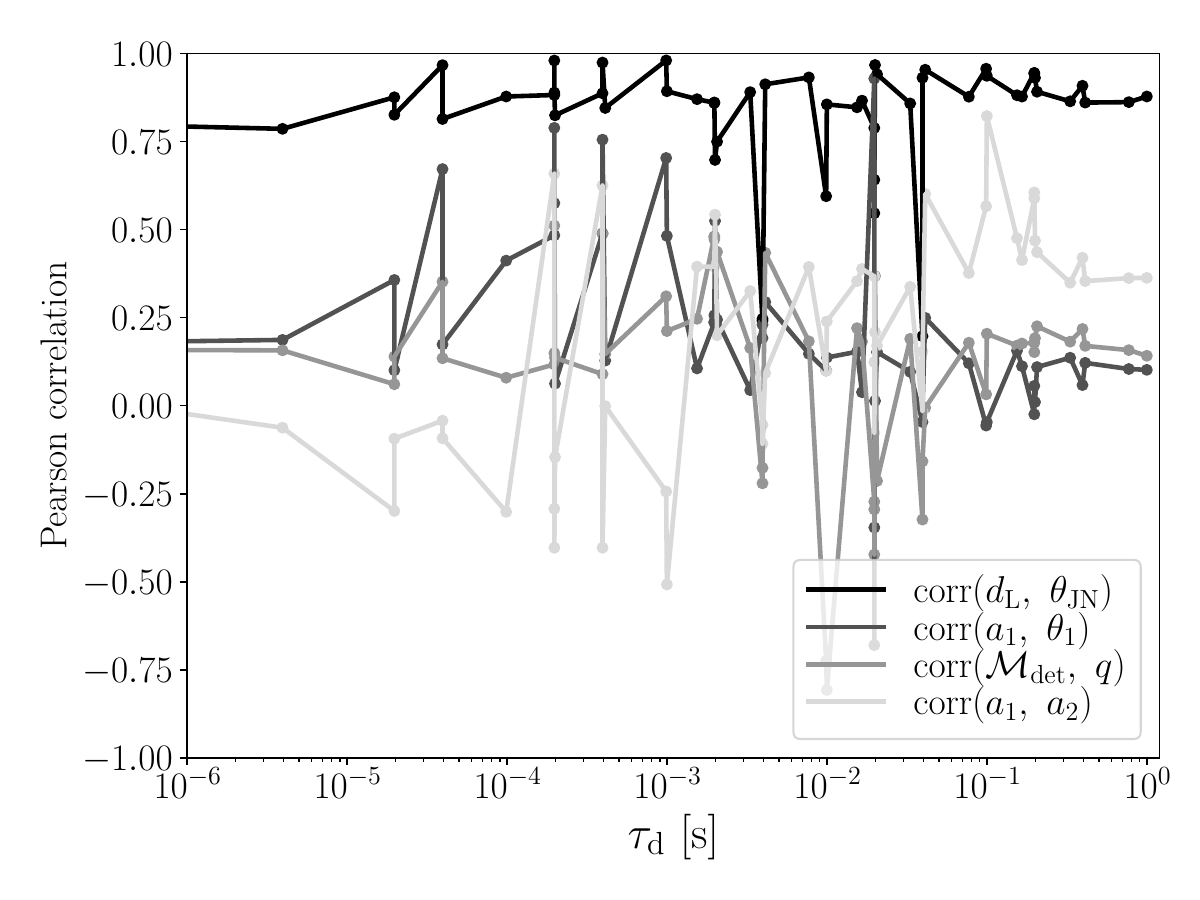}
         \caption*{}
     \end{subfigure}    
     \caption{ 
    Variation in the correlation values between the 15 parameters of a BBH system due to microlensing effects. The studied BBH and lens system corresponds to the $60~{\rm M}_\odot$ binary shown in Figures \ref{fig:PE_bias_1} and \ref{fig:PE_bias_2}. \textit{Left:} The variation is displayed for each parameter pair, resulting in a total of 120 possible pairs represented as individual lines. These lines illustrate the correlation values as a function of the time-delay between micro-images, determined by the microlens parameters in the grid shown in \Fref{fig:PE_bias_1} and \ref{fig:PE_bias_2}. Each line is associated with 63 data points, represented as dots, indicating the specific time-delay values used to construct the line.
    \textit{Right:} Same as the left panel but explicitly showing the correlations for four pairs as written in the legend.
    Moreover, the lines originate from $\tau_\mathrm{d}=0$, showcasing the unlensed case.
     }
     \label{fig:comb_PE_correlation_study}
\end{figure*}

In \Fref{fig:corr_1}, we study correlations between the 17 parameters of a microlensed waveform $\pmb{\lambda}_\mathrm{ML} \in \{\pmb{\lambda}_\mathrm{UL},~M_\mathrm{Lz},~y\}$, where $\pmb{\lambda}_\mathrm{UL}$ represents the 15 parameters corresponding to an unlensed BBH waveform as described in \Sref{subsec:basics_of_GW_data_analysis}.     
The injected signals are non-spinning equal mass binaries having a total mass of $M_\mathrm{tot}=60~$M$_\odot$ and a network SNR of roughly $50$. 
Among the microlens parameters, the redshifted lens mass is fixed to $M_\mathrm{Lz}=100~$M$_\odot$ and the impact parameter increases from left to right as $y\in \{0.05,~0.1,~0.5,~1.0,\gg 1\}$ (also indicated at the top of each column). 
The top row shows correlations when the recovery model corresponds to 17D microlensed waveforms, while the bottom row depicts correlations when the recovery model is the usual 15D unlensed BBH waveforms. 
The colours depict Pearson correlation coefficients ranging from $-1$ to $1$, where blue cells represent positive correlations $(>0)$ and orange cells represent negative correlations $(<0)$. The matrix is symmetric by construction. 
Studying the top row of the figure, we first notice that the correlations of $M_\mathrm{Lz}$ and $y$ with other parameters are mostly of opposite sign (see, e.g., opposite tonalities in the bottom two rows of each subpanel for almost all the parameters where correlations are significant.). This is expected as the effect of increasing $M_\mathrm{Lz}$ and decreasing $y$ favours microlensing effects, and vice-versa.  
\textit{It is worth noting that the luminosity distance $d_\mathrm{L}$ shows maximum correlation with the microlensing parameters, which can even exceed $\sim\pm90\%$}. 
For lower values of $y<1$, the anti-correlation can increase significantly. This is because as $y$ decreases, the magnification of micro-images increases even more rapidly, which is compensated by an increase in the effective distance of the binary, thereby showing strong negative correlations.
This justifies our earlier result of how optimal SNRs (or distance estimates) get significantly affected due to microlensing, as shown in \Fref{fig:ul_vs_ml_SNR}. 
Next, the (detected) chirp mass $\mathcal{M}_\mathrm{det}$\footnote{We explicitly write ``detector frame" here to avoid confusion with the source frame chirp mass, which will be highly biased due to a biased inference of the luminosity distance (or, the redshift).} also shows interesting correlations and can be as high as $\sim\pm 30\%$, while the spin components show only weak correlations with the microlensing parameters. However, even small correlations can have severe effects in the parameter estimation considering the sensitivity of WFs to these parameters.  
To give an idea, the mismatch between a non-spinning waveform, as considered in this exercise, with a waveform (i) having a small effective spin $\chi_\mathrm{eff}=0.05$ is $>5\%$, (ii) having only a slight variation of $1\%$ in the chirp mass is $\sim 3\%$\footnote{A mismatch value above $1\%$ is large enough to bias the inferred parameters for an event with SNR $\gtrsim 15$. }.
Apart from the luminosity distance $d_\mathrm{L}$ and the trigger time\footnote{Similar behaviour is observed for the `jitter time' when the time-marginalisation is used.}, $t_\mathrm{c}$, other extrinsic parameters such as the sky location parameters, right ascension and declination $(\alpha, \delta)$, the polarisation angle $(\psi)$, the phase of coalescence $(\phi)$, and the inclination $\theta_\mathrm{JN}$ show only negligible correlation with microlensing.
Another important thing to note is that the two microlensing parameters $M_\mathrm{Lz}$ and $y$ show a strong negative correlation among themselves, which increases with increasing $y$ from left to right up to $y=1$, reaching a value $< -95\%$ for $y=1$. This suggests that the correlation between the two parameters increases in the geometric optics limit. Therefore, while doing the 17D microlens parameter estimation, sampling in these two parameters directly will not be the most efficient choice, and one may resort to different combinations of these two parameters like the relative magnification and the time delay between the micro-images, as used in \cite{Liu:2023ikc}.

Now focusing on the bottom row of \Fref{fig:corr_1}, if only the unlensed WF model is used in recovering a microlensed signal, there will be several indirect (yet significant) correlations between the BBH parameters and the microlensing parameters, as parameters that are directly correlated with microlensing parameters will further affect other parameters that are strongly correlated with them, and so on. 
When this cascade effect is in place, it will drastically affect parameter estimation.
Such an effect can be seen for the two leftmost cases $y=\{0.05, 0.10\}$ in the bottom row of \Fref{fig:corr_1}, where almost all the BBH parameters have become strongly correlated. 
It is worth noting that these two cases fall in the long-wavelength regime. This behaviour could be a \textit{general characteristic of the bottom-right corner of the long wavelength regime, where microlensing effects are strong but slowly varying.}

As a reference, the rightmost panel show the correlations when the injected waveform is unlensed, while the recovery model is either microlensed (top-right panel) or unlensed (bottom-right panel), highlighting the usual correlations present among the BBH parameters. 
The variation in the correlation coefficient values compared to the unlensed case is clearly visible as we vary the impact parameter.
With changing $y$, the correlations become intertwined in different ways in an attempt to absorb microlensing effects. It is interesting to note that the sign of correlations can also change, i.e., \textit{the correlation between parameters can rotate due to microlensing effects}. For example, the variation in the correlation between $\mathcal{M}_\mathrm{det}$ and $t_\mathrm{c}$ changes from a slightly positive correlation to a strong negative correlation as we decrease the impact parameter $y$.
A similar effect can be observed for spin components which show strong correlations with other parameters and among themselves as we compare the leftmost panel with the other three panels.  

In \Fref{fig:PE_bias_1} and \ref{fig:PE_bias_2}, we show biases in the inferred parameters when the injected microlensed signals are recovered under the assumption of unlensed hypotheses. The injected signals are such that their observed SNR is roughly $50$, which is achieved by tweaking the luminosity distance accordingly. \Fref{fig:ul_vs_ml_SNR} then implies that we keep the signals at higher distances as the lens mass increase and the impact parameter decrease.
We choose specific grid points to cover the microlensing and source parameter space. 
For microlensing, we choose the redshifted lens mass values as $M_\mathrm{Lz}\in\{\rm 1e1,~5e1,~1e2,~5e2,~1e3,~5e3,~1e4,~5e4,~1e5\}$, and the (possible) impact parameter values as $y\in\{0.01,~0.05,~0.1,~0.5,~1.0,~3.0,~5.0\}$. We further ignore those points in the parameter space that correspond to a time delay between microimages greater than the signal duration, i.e., cases where strong lensing is observed (see the empty region in the top-right corner of each subplot).  
The source parameters correspond to the equal mass binaries having mass ratio $q=1$ and total mass $M_\mathrm{tot}\in\{20,~60,~100,~200\}$. The $x$ and $y$ axes in each subplot represent varying $M_\mathrm{Lz}$ and $y$ values, respectively. Each column represents the bias in a specific parameter as indicated at its top, while different rows correspond to different source binaries as indicated on the right side of each row. Each circular marker in the plot has two halves, with the left half representing the relative percentage error between the median values of the microlensed vs unlensed recoveries, while the right half represents the two-sample Kolmegorov-Smironov (KS) statistic value between the 1D marginalised posteriors of the microlensed and the unlensed recoveries (in percentage). 
For parameters $\chi_\mathrm{eff}$ and $\chi_\mathrm{p}$, the left half represents the percentage absolute error rather than the relative error as their injected values were zero.
The KS statistic value for two cumulative distribution functions, $C_1(x)$ and $C_2(x)$, is defined as
\begin{equation}
    KS = \max_{x}|C_1(x) - C_2(x)|,
\end{equation}
which is more sensitive to the change in the distribution itself compared to the change in the median values.
So in simple words, the left half indicates the bias in the recovery, while the right half indicates the change in the 1D marginalised posterior distribution, both converted to percentages.

In \Fref{fig:PE_bias_1}, we show bias in the intrinsic parameters, i.e., in masses and spins of the binaries. 
The detected chirp mass $\mathcal{M}_\mathrm{det}$ and symmetric mass-ratio $\eta$ represent the masses, while the projected effective spin $\chi_\mathrm{eff}$ \citep{Racine:2008qv, Ng:2018neg} and precession effective spin $\chi_\mathrm{p}$ \citep{gerosa2021generalized, Schmidt:2014iyl} represent the spins of the binary black holes.
These two-dimensional effective spin quantities, $\chi_\mathrm{eff}$ and $\chi_\mathrm{p}$, offer a simplified interpretation of the six-dimensional spin parameters. 
For all the parameters and binary masses (i.e., all the subplots in the figure), we see negligible biases in the recoveries for low $M_\mathrm{Lz}$ and $1/y$ values as the leftmost and topmost array of markers suggest errors to be $\lesssim 1 \%$ in most of the subplots (see the left halves of the markers for the data points having either $M_\mathrm{Lz}=10$ or $y=5$). 
This is expected as it corresponds to negligible microlensing effects that are difficult to be detected with the current sensitivities of the ground-based detectors. However, even in this negligibly-lensed regime, the biases in the spin parameters reach $\sim 10\%$ in a few cases, which is a result of bad recoveries of the spin parameters in general, as they appear higher in the post-Newtonian orders. 
Additionally, the two mass-related parameters, $\mathcal{M}_\mathrm{det}$ and $\eta$, are usually well recovered for $M_\mathrm{Lz}\lesssim 100~$M$_\odot$, where we expect a larger number of microlenses. 

As suspected in \Fref{fig:f_ML_cplot_for_Mlz_vs_y} of \Sref{subsec:basics_of_lensing}, we indeed observe the biases to increase in the wave zone for all the parameters. This is especially clear if we look at the bottom-most row corresponding to the $(100,~100)~$M$_\odot$ binary, where biases seem to be more streamlined and increasing along the diagonal from lower $M_\mathrm{Lz}$ and $1/y$ values to higher values in each column. As we move up the rows to lower binary masses, this pattern along the diagonal broadens and eventually covers up a large parameter space, even spanning regions in the long-wavelength regime, as we see in the case of $(10,~10)~$M$_\odot$ binary. One of the reasons for such broadening of the biases along the diagonal from the wave zone toward the long-wavelength regime is due to the fact that lower mass binaries tend to cover a broader frequency spectrum, i.e., they have a higher power in high frequencies as compared to heavier binaries. 
From \Fref{fig:f_ML_cplot_for_Mlz_vs_y}, it is then expected that such a signal with significant contribution from high frequencies will also show bias in the long-wavelength regime, i.e., for lower $M_\mathrm{Lz}$ and $y$ values.

Of the four parameters, the spin parameters seem to be biased the most, especially the precession effective spin $\chi_\mathrm{p}$, as compared to the chirp mass $\mathcal{M}_\mathrm{det}$ and the symmetric mass-ratio $\eta$. This is also true in the usual parameter estimations of the unlensed signals owing to their appearance in the different post-Newtonian orders. Chirp mass and symmetric mass-ratio give the most dominant effect at 0th PN order of GW phasing while $\chi_\mathrm{eff}$ and $\chi_\mathrm{p}$ appear at 1.5 PN and 2.5 PN orders, respectively \citep{arun2005parameter, Schmidt:2014iyl, Isoyama:2020lls}. 
The recoveries for $\chi_\mathrm{p}$ exhibit biases across most of the parameter space in each row, particularly in the wave zone, and these biases appear to decrease as we move down the row toward heavier mass binaries. This suggests that the biases in $\chi_\mathrm{p}$ are correlated with the length of the microlensed signal.
It is also worth noting that even for the modest values in our microlensing parameter space, such as $(M_\mathrm{Lz},~y)\sim (10^2,~1)$, the recovery of $\chi_\mathrm{p}$ show significant biases.  
Thus, \textit{the bad recoveries for longer signals indicate that microlensing and spin-precession are degenerate with each other. Therefore, any signal showing signs of precession must also be analysed for the presence of microlensing signatures to break the degeneracy. However, vice-versa may not be true, i.e., it is unlikely that the presence of precession can bias microlensing searches. This is because the unlensed parameter space is always a subset of the microlensed parameter space. }


Similarly, in \ref{fig:PE_bias_2}, we show the bias in the recoveries of the extrinsic parameters, i.e., luminosity distance ($d_\mathrm{L}$), inclination ($\theta_\mathrm{JN}$), RA ($\alpha$), Dec. ($\delta$), polarisation angle ($\psi$) and coalescence phase $\phi$. 
We do not show the recoveries of trigger time $t_\mathrm{c}$ as we do not see any appreciable bias in its recovery. Among all the cases studied here, the absolute errors for $t_\mathrm{c}$ never exceeded $5\%$.
We notice that the most affected parameter is the luminosity distance, as it gets directly affected due to the modulations in the amplitude induced by microlensing. On the other hand, the sky position parameters Ra and Dec are among the best-recovered parameters not affected by microlensing. This is expected since the localization of GW sources is mainly based on the observed time delays between each pair of interferometers. Since microlensing does not affect the observed trigger times, the localisation is not affected except when microlensing effects are extreme.  

The KS values (right halves of the markers) show a similar trend as that of errors. It is interesting to note that in most cases, the right half of the circle is darker than the one on the left, indicating that KS values are more sensitive to microlensing effects than the bias in the inferred parameters. This can be seen in \Fref{fig:mtot60_PE_1D_study}, where we explicitly show the marginalised 1D posterior distributions for three parameters $\{\mathcal{M}_\mathrm{det},~q,~a_1\}$ for the case having binary mass $60~$M$_\odot$. In the leftmost panel, one can notice several cases where the distribution shifts because of microlensing effects even when the posterior mode itself hasn't changed much. These cases are examples that result in a high KS-value but a low relative error. In contrast, there are several cases in the middle and the rightmost panels where the recovered distribution is significantly biased and is also well converged (e.g., see well-converged distributions in the rightmost panel for $a_1$ away from the injected value of $0$). Such cases result in a high relative error as well as a high KS-value (markers with both left and right halves coloured as dark blue in \Fref{fig:PE_bias_1} and \ref{fig:PE_bias_2}).

Lastly, in \Fref{fig:comb_PE_correlation_study}, we examine the variation in Pearson correlation values between GW parameters for the $(30,~30)~M_\odot$ binary discussed earlier (second row in Figures \ref{fig:PE_bias_1} and \ref{fig:PE_bias_2}). The left panel illustrates the correlation values for all possible $120$ pair combinations of the $15$ parameters (represented by different shades of blue). The $x$-axis corresponds to the time delay between microimages associated with the microlens parameters considered for that binary. The data points, indicated by circular dots, consist of $54$ data points for each of the 120 lines. Additionally, the lines originate from $\tau_\mathrm{d}=0$, representing the unlensed case.
The line with a Pearson correlation value of unity indicates the diagonal elements of the correlation matrix, which represents the correlation of a parameter with itself. The purpose of this plot is to demonstrate how correlations can vary based on the microlens parameters. It is evident that correlations can significantly fluctuate depending on the specific microlens parameters. For instance, there is a notable concentration of lines around zero at low time delay values ($\tau_\mathrm{d} < 10^{-5}$ s), which becomes sparser at $\tau_\mathrm{d} = 10^{-3}$ s.
Since it is not possible to follow which line corresponds to which correlation pair, we specifically show the correlations for four pairs in the right panel, as written in the legend.


\begin{figure*}
    \centering
    \includegraphics[width=1.0\linewidth]
    {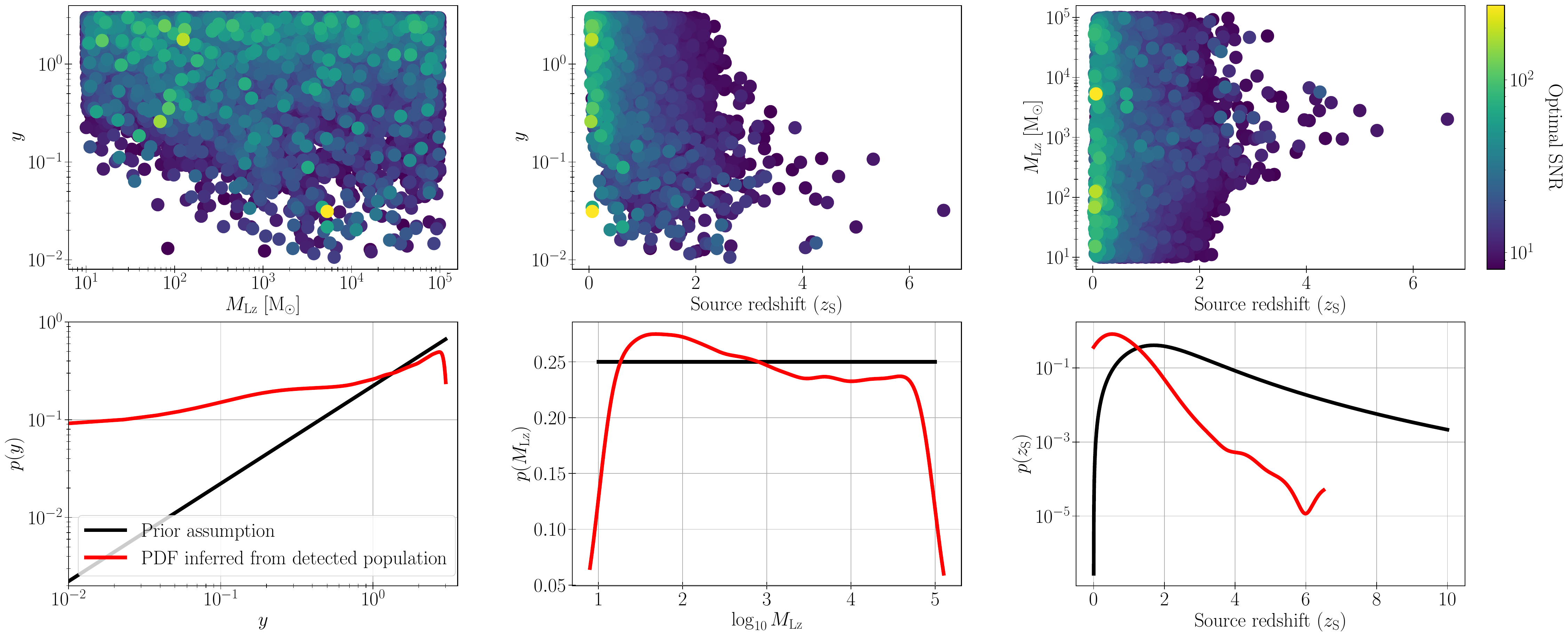}
     \caption{Population study of microlensed BBH signals for the joint network of LIGO$-$Virgo detectors assuming the targeted O4 sensitivities. \textit{Top:} Distribution of population is shown in the microlensing parameter space of $M_\mathrm{Lz}$ and $y$ (left), and for $y$ vs. $z_\mathrm{S}$ (right), where $M_\mathrm{Lz}$, $y$ and $z_\mathrm{S}$ denote the redshifted lens mass, impact parameter and source redshift, respectively. The colour bar represents the observed network optimal SNR.
     \textit{Bottom:} Comparison of probability density functions (PDFs) between the ones that were used while generating the population (black coloured curves), and the ones inferred from the observed population themselves (red coloured curves). The PDF comparison is shown for three parameters, $y$ (left), $M_\mathrm{Lz}$ (middle) and $z_\mathrm{S}$ (right).}
     \label{fig:ml_population_study}
\end{figure*}


\section{Study of Microlensed population} 
\label{sec:ml_pop_study}
In this section and the subsequent section, we investigate a population of microlensed signals.
In contrast to the previous sections, where we either fixed certain parameters while varying others or chose a grid to cover the parameter space, in this section, we sample the sources realistically to infer the population-wide distributions of parameters. We will pay particular attention to the microlens parameters, which is the focus of this study. Furthermore, analyzing the population statistics allows us to explore various aspects, such as the effectualness of unlensed waveforms in detecting microlensed waveforms and the potential parameter space for microlensing detection.

We generate mock GW data of around $2.5\times10^4$ microlensed BBH signals, where BBH parameters are derived from the population model constructed using the GWTC-3 catalogue \citep{KAGRA:2021vkt,KAGRA:2021duu}. We put an observed network SNR threshold of 8 when using the unlensed templates for recovery, and the detector noise PSDs used correspond to the target O4 sensitivities \citep{KAGRA:2013rdx}.
The population model basically provides a fit to the distribution of observed parameters, particularly masses, spin magnitudes, spin tilts, and the redshift distribution of the BBH mergers. All other BBH parameters are sampled uniformly from their respective domains. For microlens parameters, we assume a log-uniform prior in $M_\mathrm{Lz}$ 
(in units of $M_\odot$) 
and a power-law prior for $y$ with an index of unity (a linear prior):
\begin{equation}
\begin{aligned}
    p(M_\mathrm{Lz}) &\propto \text{LogUniform}(10^1,~10^5), \\
    p(y) &\propto y,~~y\in(0.01,~3.00),\
    \label{eq:mlz_y_priors}
\end{aligned}
\end{equation}
where the motivation to use $p(y) \propto y$ comes from geometry and isotropy \citep{Lai:2018rto}.
To wit, the probability of a source having an impact parameter $y$ relative to a microlens will be proportional to the area of a ring of infinitesimal width having radius $y$, i.e., $p(y)dy = 2\pi y dy$.
We assume Madau-Dickison profile for the merger rate density in the universe, giving source-redshift density model as \citep{Madau:1996hu,Fishbach:2018edt}:
\begin{equation}
\begin{aligned}
    p(z_\mathrm{S}) &\propto \frac{dV_c}{dz_\mathrm{S}}\frac{1}{1+z_\mathrm{S}}\psi(z_\mathrm{S}), \\
    \text{where}~~~ \psi(z_\mathrm{S}) &= 0.015 \frac{(1+z_\mathrm{S})^{2.7}}{1+[(1+z_\mathrm{S})/2.9]^{5.6}}~.
\end{aligned}    
\end{equation}
The source redshift range was set to be $z_\mathrm{S} \in (0.001,~10)$, with the lower limit of $z_\mathrm{S}\equiv z_\mathrm{min}=0.001\sim \mathcal{O}(1)$ Mpc corresponding to a value below which merger rate is negligible due to low cosmological volume and star formation rate. 
The upper limit of $z_\mathrm{S} \equiv z_\mathrm{max} = 10$ serves as an approximate representation of the maximum distance from which a microlensed signal can be detected using current ground-based detectors. This limit assumes ideal conditions such as low impact parameters, a high lens mass, and a massive binary system as the source. For instance, a system characterized by parameters $(M_\mathrm{Lz},~y,~M_\mathrm{tot},~\iota) = (10^4,~10^{-2},~200,~0)$ exemplifies these ideal conditions.

In \Fref{fig:ml_population_study}, we show a mock sample of the detectable microlensed population - its distribution and the inferred properties. The top row shows the network optimal SNR as a function of  $M_\mathrm{Lz}$ and $y$ (left panel),  $y$ vs. $z_\mathrm{S}$ (middle panel), and  $M_\mathrm{Lz}$ vs. $z_\mathrm{S}$ (right panel). 
Firstly, we note that using only unlensed templates during the search of these microlensed signals, we detected around $91.6\%$ of total signals (using \Eref{eq:FF_definition}) in the parameter space considered here. 
We observe that most of the detected signals tend to have higher impact parameters, which is expected based on our initial assumption given in \Eref{eq:mlz_y_priors}. However, as predicted in \Fref{fig:ul_vs_ml_SNR} of \Sref{sec:ml_effect_on_detection}, it is worth noting that we do detect a significant number of events in the range $y\in(0.01,~0.1)$ as well, which is usually considered to be a probabilistically insignificant region. 
The top-middle panel confirms the hypothesis that these signals with low-impact parameters can indeed arrive from far away regions ($z\gtrsim 2$) as opposed to the current detection horizons for unlensed BBH signals ($z\lesssim 1$). This is a consequence of an increase in their SNR values because of microlensing, and hence an increase in their detection horizon (see the left-most panel in \Fref{fig:ul_vs_ml_SNR}). 
We also notice that even in the case of population, the behaviour of FF in $M_\mathrm{Lz}$ and $y$ plane (top-left panel) is similar to the behaviour of FF shown in \Fref{fig:match_FF_analysis}, where we had kept the binary mass fixed.

The bottom row of \Fref{fig:ml_population_study} illustrates the comparison of probability density functions (PDFs). The black curves represent the PDFs used for sampling during the population generation (referred to as the "prior"; see \Eref{eq:mlz_y_priors}), while the red curves represent the PDFs inferred from the detected population itself.
The PDF comparison is shown for three parameters, $y$ (left panel) and $M_\mathrm{Lz}$ (middle panel) and $z_\mathrm{S}$ (right panel). 
We use kernel density estimation (KDE) to obtain the PDFs from the observed data.
The comparison of PDFs for the impact parameter, $p(y)$, shows very interesting behaviour. 
At low values of $y$ ($\lesssim 1$), the observed signals have roughly a flat density profile in $y$, instead of the linear profile 
used as the prior (\Eref{eq:mlz_y_priors}; see black curve). 
The reason for this behaviour can be attributed to the behaviour of magnification due to a point-lens, which is only a function of $y$, given by
\begin{equation}
    \mu_{\pm}(y) = \frac{1}{2} \pm \frac{y^2 + 2}{2y\sqrt{y^2+4}}. 
\end{equation}
It can be seen that in the limit $y \ll 1$, $\mu_{\pm}(y)$ becomes proportional to $1/y$, which in turn increases the detection horizon, thereby increasing the relative probability density in that region. 

On the other hand, the probability density profile for $M_\mathrm{Lz}$ (bottom-middle panel) is roughly similar to our initial assumption of a Log-uniform distribution with only a slight preference for lower masses compared to heavier masses. This slight preference for lower masses $(\log_{10}M_\mathrm{Lz}\lesssim 2.5)$ is a result of better FF recovery values in that region as shown in the bottom left panel of \Fref{fig:match_FF_analysis}, owing to smaller microlensing effects. If we instead recover with the microlensed templates instead of the unlensed ones, we find $p(M_\mathrm{Lz})$ to be even more consistent with the Log-Uniform distribution showing no special preference for any mass values. This indicates that the behaviour of FF is indeed the reason behind the slight preference for lower mass values in case of unlensed recoveries.
The probability density of the source redshift $p(z_\mathrm{S})$ (bottom-right panel) shows a similar trend as if there were no microlensing \citep[e.g., see Fig. 2 in][]{Fishbach:2018edt} but with a longer tail reaching much higher values up to $z\sim 5$ as opposed to the current detection horizons for BBH signals ($z\lesssim 1$) \citep[also see][]{KAGRA:2021duu}. 

We showed that the selection bias incurred during detection would significantly affect the properties of the observed population compared to the true microlensed population. However, not all the observed microlensed events will be correctly identified as being microlensed, i.e., having significant evidence for the microlensing hypothesis over the unlensed hypothesis. 
For example, a low SNR event that has low $M_\mathrm{Lz}$ and high $y$ values, such as $(M_\mathrm{Lz},~y)=(10~M_\odot, 3)$, would not be correctly identified as being a microlensed event with the current sensitivities of the detectors.
Therefore, in order to predict the parameter space which has a higher potential of being detected and also identified as a microlensed event, we should anticipate a further selection bias on the detected events. This involves weighing events according to their microlensing effects, i.e., the events with higher microlensing effects are more probable to be correctly identified as microlensed. In the next section, we investigate this bias and present a combined PDF that incorporates both the detection and identification aspects.

\begin{figure*}
    \centering
    \includegraphics[width=\linewidth]{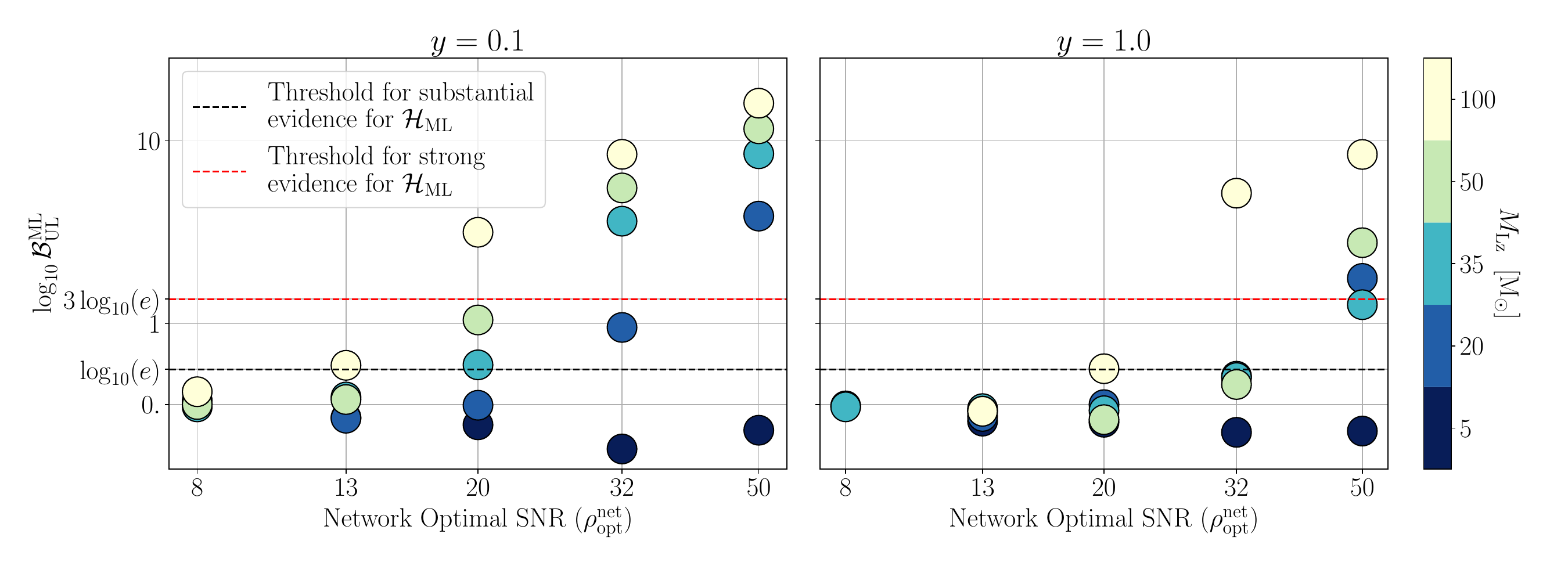}
    \caption{The figure shows log Bayes factor values, $\log_{10}\mathcal{B}^{\rm ML}_{\rm UL}$, for the evidence for microlensed hypothesis ($\mathcal{H}_\mathrm{ML}$) over unlensed hypothesis for varying network SNR, $\rho^\mathrm{net}_\mathrm{opt} \in \{8,13,20,32,50\}$, redshifted lens mass, $M_\mathrm{Lz}\in \{5,20,35,50,100\}M_\odot$, and impact parameter values, $y\in \{0.1,1.0\}$.}
    \label{fig:ul_vs_ml_hypothesis_comb}
\end{figure*}

\section{Model Comparison: Unlensed vs. Microlensed hypothesis}
\label{sec:ul_vs_ml_study}

In order to determine if a GW event has been microlensed or not, one can propose two models for data analysis - (i) UL: the unlensed hypothesis, described by the usual 15 parameters, and (ii) ML: microlensed hypothesis, described using two additional (microlensing) parameters, $M_\mathrm{Lz}$ and $y$, to the unlensed hypothesis, i.e., a total of 17 parameters. \footnote{ It is important to note that this approach inherently assumes $\mathcal{M}-$closedness \citep{Kass:1995loi, schad2022workflow}, which means that one of the proposed models is indeed the true model. This assumption is not true in the case of real GW triggers, as there could be various systematics or incomplete physical models which we do not know {\em a priori}. The possible consequences of this assumption in model selection should be explored in future studies.}
To determine which model is more preferred by the GW data $\mathcal{D}$, we can use the Bayes' theorem to calculate the \textit{odds ratio}, defined as \citep{Deutsch:1999gs}:
\begin{equation}
    \mathcal{O}^\mathrm{ML}_\mathrm{UL}\equiv\frac{p(\mathrm{ML}|~\mathcal{D})}{p(\mathrm{UL}|~\mathcal{D})} = \frac{p(\mathcal{D}|~\mathrm{ML})}{p(\mathcal{D}|~\mathrm{UL})} \vdot \frac{p(\mathrm{ML})}{p(\mathrm{UL})}\,.
    \label{eq:odds_ratio}
\end{equation}
Under the assumption that all models are equally likely {\rm a priori}~\footnote{This assumption makes sense for the initial set of searches. However, based on such results, an informed prior on models can be used, which should incorporate our belief that the number of microlensed signals is much smaller than the unlensed signals, i.e., $N_\mathrm{ML}\ll N_\mathrm{UL}$. }, we set $p(\mathrm{UL})=p(\mathrm{ML})$ in \Eref{eq:odds_ratio}.  
The odds ratio then reduces to the Bayes factor, which is simply the ratio of the evidences $(\mathcal{Z})$, or marginalised likelihoods, of the two models:
\begin{equation}
\begin{aligned}
    \mathcal{O}^\mathrm{ML}_\mathrm{UL}=\mathcal{B}^\mathrm{ML}_\mathrm{UL}
    &\equiv\frac{p(\mathcal{D}|\mathrm{ML})}{p(\mathcal{D}|\mathrm{UL})} \\
    &=\frac{ \int_{\mathcal{V}_{\lambda_\mathrm{ML}}}~d\pmb{\lambda}_\mathrm{ML}~p(\mathcal{D}|~\pmb{\lambda}_\mathrm{ML})~ p(\pmb{\lambda}_\mathrm{ML}|~\mathrm{ML}) }{ \int_{\mathcal{V}_{\lambda_\mathrm{UL}}}~d\pmb{\lambda}_\mathrm{UL}~p(\mathcal{D}|~\pmb{\lambda}_\mathrm{UL})~p(\pmb{\lambda}_\mathrm{UL}|~\mathrm{UL}) } \equiv \frac{\mathcal{Z}_\mathrm{ML}}{\mathcal{Z}_\mathrm{UL}}, 
    \label{eq:bayes_factor}
\end{aligned}    
\end{equation}
where $\pmb{\lambda}_\mathrm{UL}\in \mathbb{R}^{15}$ and $\pmb{\lambda}_\mathrm{ML}\equiv \{ \pmb{\lambda}_\mathrm{UL},~M_\mathrm{Lz},~y \} \in \mathbb{R}^{17}$ are the model parameters of $\mathcal{H}_\mathcal{UL}$ and $\mathcal{H}_\mathcal{ML}$ hypotheses; 
$\mathcal{V}_{\pmb{\lambda}}$ represents the parameter space volume; 
$p(\mathcal{D}|~\pmb{\lambda})$ is the likelihood of observing 
$\mathcal{D}$ for a certain 
$\pmb{\lambda}$; $p(\pmb{\lambda}|~\mathrm{Model})$ is our prior belief which we aim to update after a parameter estimation run.

It is rather convenient to work with the logarithm of the \textit{Bayes factors}, which can be expressed as:
\begin{equation}
    \log_{10} \mathcal{B}^{\rm ML}_{\rm UL} = \log_{10}\mathcal{Z}_\mathrm{ML} -  \log_{10}\mathcal{Z}_\mathrm{UL} = \log_{10} \mathcal{B}^{\rm ML}_{\rm noise} - \log_{10} \mathcal{B}^{\rm UL}_{\rm noise},
\end{equation}
where $\log\mathcal{B}^{\rm UL}_{\rm noise}$ represents the (logarithmic) Bayes factor indicating the likelihood of an unlensed GW signal being present in the data, as opposed to the null hypothesis of pure noise (similarly, $\log \mathcal{B}^{\rm ML}_{\rm noise}$ represents evidence for the presence of a microlensed signal over pure noise). This second equality stems from the fact that the evidence computation for the ``noise model", $\mathcal{Z}_\mathrm{noise}$, does not depend upon the waveform models being compared but only on the data segment and the assumed estimate of the noise PSD profile of the detectors. Therefore, $\mathcal{Z}_\mathrm{noise}$ will result in equivalent values for both $\mathrm{ML}$ and $\mathrm{UL}$ models given the same settings for the likelihood evaluation and other sampler settings.
When the (absolute) logarithmic value of Bayes factor is large ($\log_{10} \mathcal{B}^{\rm ML}_{\rm UL}\gtrsim 1.3$), we say the microlensed model is preferred over the unlensed model. When the value is negative or only slightly positive ($\log \mathcal{B}^{\rm ML}_{\rm UL}\lesssim 0.4$), the ML model is discarded in favour of the UL model.\footnote{This interpretation is based on Kass-Raftery's scale \citep{Kass:1995loi}.} 
This follows from Occam's razor argument, which favours less complicated models, i.e., models with fewer parameters are more preferred among similarly performing models. 

In the high-SNR limit, when most of the posterior volume is confined around a particular value, i.e., when the posterior density is highly peaked around the posterior mode, the evidence can be approximated using the Laplace approximation and the resulting Bayes factor can be written as \citep{Cornish:2011ys, Vallisneri:2012qq}:
\begin{equation}
    \ln \mathcal{B}^\mathrm{ML}_\mathrm{UL} \approx \frac{1}{2}\rho_\mathrm{res}^2  +   \ln \mathscr{O}^\mathrm{ML}_\mathrm{UL} 
    \label{eq:bayes_factor_estimate}
\end{equation}
where $\rho_\mathrm{res}^2 = (1 - FF^2)\rho_\mathrm{ML}^2$ is the SNR of the \textit{residual} waveform ($h_\mathrm{ML} - h^{'}_\mathrm{UL}$) after the best fit UL waveform has been subtracted out from the true ML waveform, $FF$ is the fitting factor as defined in \Eref{eq:FF_definition}, and $\mathscr{O}$ represents the Occam's factor defined as the ratio of the posterior volume $\Delta V$ to the prior volume $V$, i.e, $\mathscr{O}\equiv \Delta V/V \propto \sqrt{|F^{-1}|}$, where $F$ is the \textit{Fisher information matrix}. 
Thus, up to a difference of (logarithmic value of) Occam's factors, the log Bayes factor should scale as $\rho_\mathrm{res}^2$, which grows as a square of the SNR. Thus, SNR becomes crucial in determining the subtle effects buried in the signal. 

In \Fref{fig:ul_vs_ml_hypothesis_comb}, we show Bayes factor values, $\log_{10}\mathcal{B}^{\rm ML}_{\rm UL}$, for the evidence of microlensed hypothesis over unlensed hypothesis for varying network SNR, $\rho^\mathrm{net}_\mathrm{opt} \in \{8,~13,~20,~32,~50\}$, redshifted lens mass, $M_\mathrm{Lz}\in \{5,~20,~35,~50,~100\}$, and impact parameter values, $y\in \{0.1,~1.0\}$. A low impact parameter value of $0.1$ is chosen to foresee results in the case of best-case scenarios while $y=1$ represents the characteristic value of $y$. 
For microlens parameter recoveries, we set priors as $p(M_\mathrm{Lz}) \propto \text{LogUniform}(10^{-1},~10^5)$ and $p(y) \propto y,~~y\in(0.01,~3.00)$.
We limit ourselves to only lower microlens masses ($M_\mathrm{Lz}\le 100~$M$_\odot$) because, from an astrophysical standpoint, more massive BH lenses are less probable. The black and red lines correspond to a Bayes factor value of $e$ and $e^3$, respectively, and mark the threshold for the positive and strong evidence for microlensing. This threshold has been set following the interpretation of Bayes factors as given in \cite{Kass:1995loi}, which sets a higher cutoff for the strong evidence as compared to the Jeffrey's scale \citep{Deutsch:1999gs}\footnote{However, we note that a better approach to interpreting the Bayes factor values would be to do a background injection study.}. 
In this and subsequent sections, we will use the terms "positive" and "strong" to characterise the strength of evidence in accordance with the terminology used in the aforementioned references.
We can see how SNR exponentially increases the Bayes factor values, especially in the left panel for $y=0.1$, where microlensing effects are higher than in the right panel. We find that up to an SNR of $13$, microlensing effects due to $M_\mathrm{Lz}<100~$M$_\odot$ do not show any interesting Bayes Factor recoveries in favour of microlensing. It is important to note that an SNR of $13$ is above the expected average SNR of the detected events ($\approx 12$; see \cite{Schutz:2011tw}), as PDF for the SNR goes as $p(\rho)\propto \rho^{-4}$.
Considering the fact that in the real GW data noise will bring in additional complexities, it seems highly unlikely that we will detect microlensing for $M_\mathrm{Lz}<100~$M$_\odot$ with current sensitivities of the detectors. However, for high SNR events, microlensing effects from even small mass microlenses become detectable, such as $M_\mathrm{Lz}\gtrsim 20~$M$_\odot$ for SNR $50$. 

\begin{figure*}
    \centering
    \includegraphics[width=\linewidth]{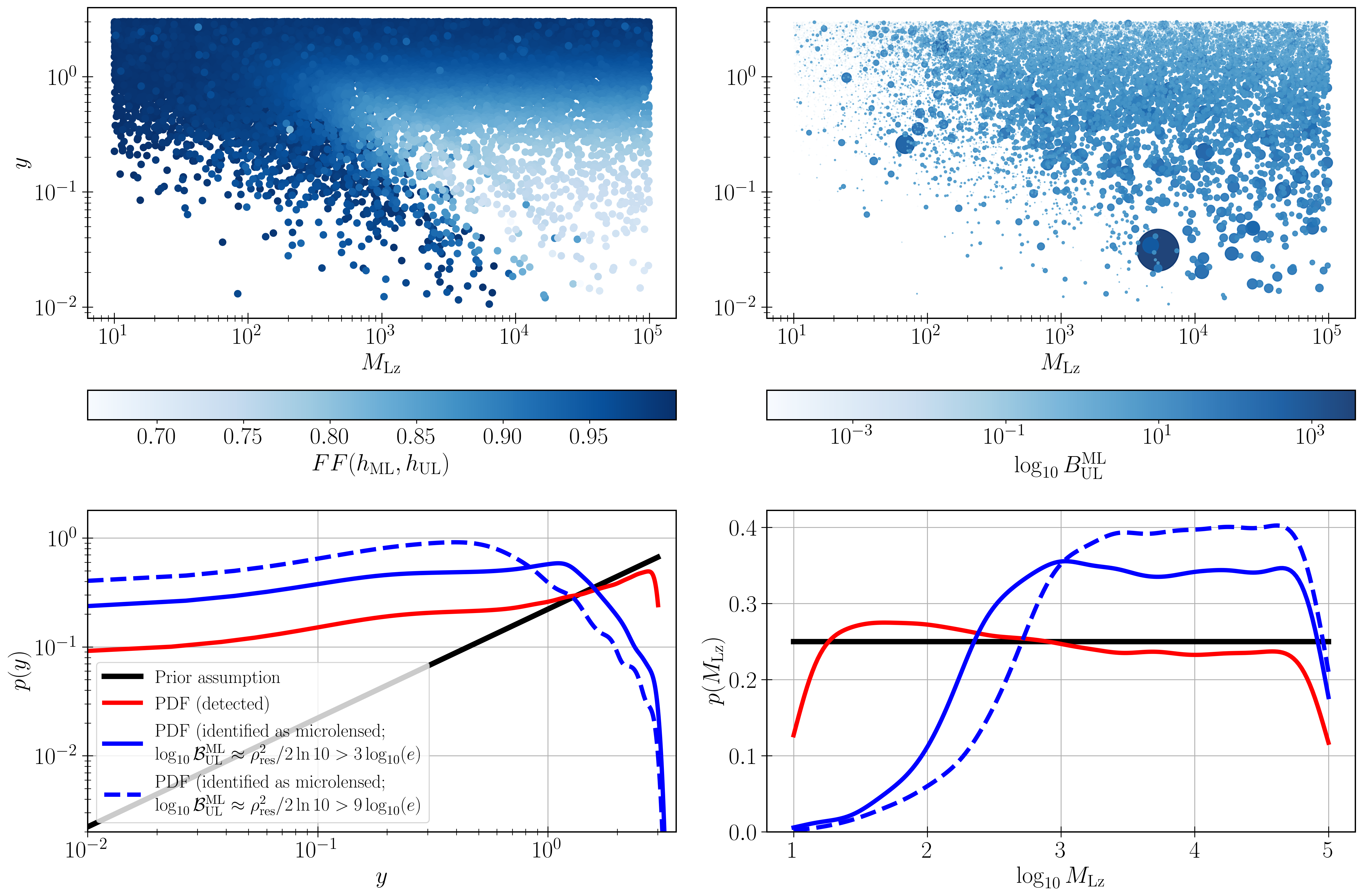}
    \caption{Bayes factor study for the microlensed population estimated using the fitting factor (FF) and the SNR $(\rho)$, i.e., $\ln \mathcal{B}^{\rm ML}_{\rm UL} \approx (1 - FF^2)\rho^2/2$. \textit{Top row:} FF values and the Bayes factor values for the evidence of microlensing over the unlensed hypothesis $(\log_{10}\mathcal{B}^\mathrm{ML}_\mathrm{UL})$ are shown in the microlens parameter space of the observed population. \textit{{Bottom row:}} The probability density functions (PDFs) for the microlens parameters are shown for four cases: (i) our prior assumption, (ii) the detected population (selection bias), (iii) the population that is detected and also correctly identified as being microlensed assuming a threshold of $\log_{10}\mathcal{B}^\mathrm{ML}_\mathrm{UL}>3\log_{10}(e)$, and (iv) same as (iii) but with a higher threshold of $\log_{10}\mathcal{B}^\mathrm{ML}_\mathrm{UL}>9\log_{10}(e)$. These threshold values are chosen to investigate various confidence levels in a microlensed event detection while also taking into account the uncertainties related to using the approximation for $\mathcal{B}^{\rm ML}_{\rm UL}$.}
    \label{fig:pop_FF_BLU_analysis}
\end{figure*}
\begin{figure*}
     \centering
     \begin{subfigure}[t]{0.49\textwidth}
         \centering
         \includegraphics[width=\textwidth]{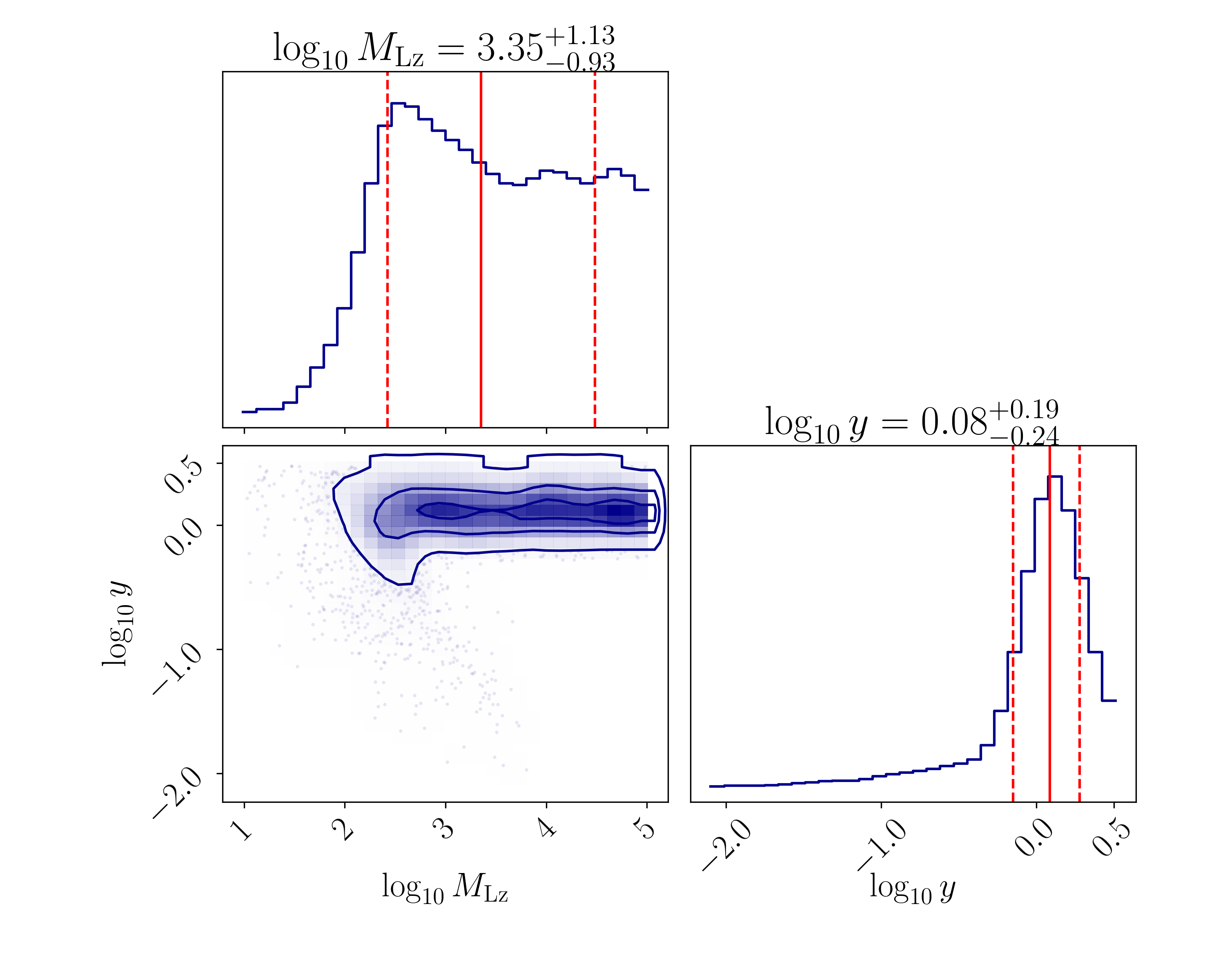}
         \caption{$\log_{10}\mathcal{B}^\mathrm{ML}_\mathrm{UL}\in(3\log_{10}(e),~9\log_{10}(e))$; Expected microlensed population detectable with low confidence. }
         \label{fig:y equals x}
     \end{subfigure}
     \hfill
     \begin{subfigure}[t]{0.49\textwidth}
         \centering
         \includegraphics[width=\textwidth]{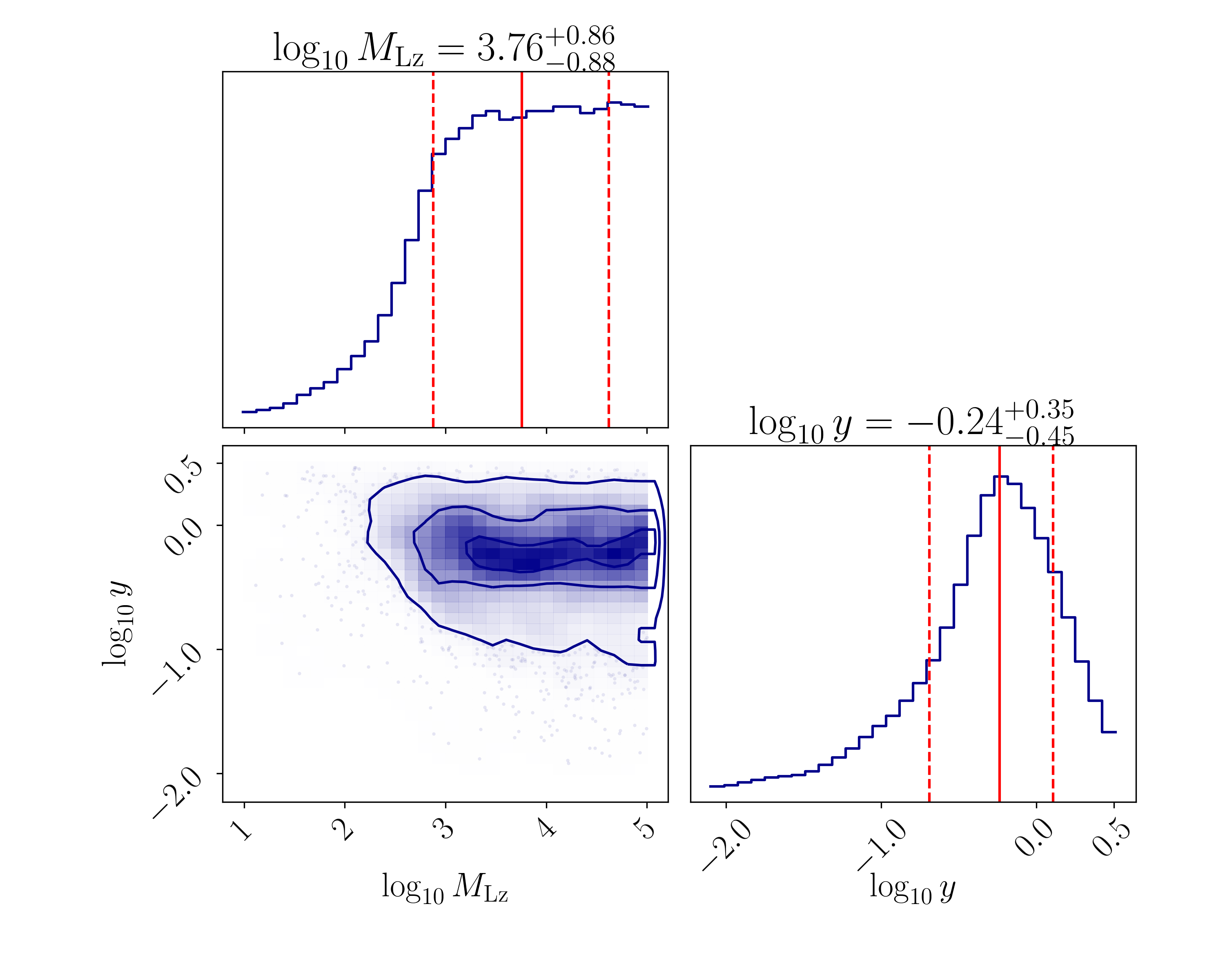}
         \caption{$\log_{10}\mathcal{B}^\mathrm{ML}_\mathrm{UL}>9\log_{10}(e)$. Expected microlensed population detectable with high confidence. }
         \label{fig:three sin x}
     \end{subfigure}
     \caption{The corner plots depict the probable region in the microlensing parameter space for a point lens that can be detected and identified as a microlensed event in the joint network of LIGO$-$Virgo detectors, assuming the targeted O4 sensitivities. The Bayes factor values of the population and the probability density functions (PDFs) of the microlens parameters are explicitly shown in \Fref{fig:pop_FF_BLU_analysis}. The panels on the \textit{left} and \textit{right} use different threshold values for $\log_{10}\mathcal{B}^\mathrm{ML}_\mathrm{UL}$, which is estimated using the expression $\ln \mathcal{B}^{\rm ML}_{\rm UL} \approx (1 - FF^2)\rho^2/2$. These threshold values are chosen to explore different confidence levels for claiming a microlensed event detection while also considering the uncertainties associated with this approximation. 
     The contour plot in the 2D space and the red lines in the 1D distributions represent credible regions with quantile values of $16\%$, $50\%$, and $84\%$.
     }
     \label{fig:corner_plots}
\end{figure*}

In the right panel of \Fref{fig:ul_vs_ml_hypothesis_comb}, we notice that for the characteristic value of the impact parameter $y=1$, the microlensing model is not favoured for $M_\mathrm{Lz}<50~$M$_\odot$ upto an SNR of 32, which further showcases the difficulty of correctly identifying a microlensed event. An important consequence of this is the fact that in dark matter constraint studies using microlensing, one should, in principle, incorporate SNR dependence. That is, the microlens parameter space that can be correctly identified to be a microlensed event is SNR dependent. Hence, a non-detection of the microlensed event can only put a constraint on the fraction of dark matter in the parameter space where it is sensitive to detecting those microlensing effects. If such an SNR dependence is not included, it will result in an over-constraint on the dark matter fractions. 

The aforementioned observation inspires us to inquire about the microlens parameters that are most likely to be detected and also correctly identified as microlensed. In \Fref{fig:ml_population_study}, we presented the distribution of the detected microlensed population. Now in \Fref{fig:pop_FF_BLU_analysis}, we conduct a more thorough analysis of this population to determine such parameter space where microlensing is most likely to be detected. Although a rigorous approach would require computing Bayes factors for the population using nested sampling algorithms, it will be highly expensive computationally. Therefore, we exploit the expression given in \Eref{eq:bayes_factor_estimate} to estimate the Bayes factors. We further neglect the Occam's factor term and approximate Bayes factors as simply
\begin{equation}
\ln \mathcal{B}^{\rm ML}_{\rm UL} \approx\rho_\mathrm{res}^2 / 2=(1 - FF^2)\rho^2/2. 
\label{eq:Bayes_fac_crude_estimate}
\end{equation}
However, a more rigorous study would require estimating Occam's factor as well.  As mentioned before, since the Occam's factor term is just the ratio of the posterior to the prior volume, one can estimate it using the (inverse of) Fisher matrix by computing the ratio of the uncertainty in the recovered value of an extra parameter to the prior volume for that parameter. These uncertainties in the parameter roughly scale inversely with the SNR. 

Assuming the prior volume to be a unit hypercube of $d$ dimensions, in the case of a true microlensed signal with a sufficient SNR value, the information content in the posteriors would be higher for the microlensing (ML) hypothesis compared to the unlensed (UL) hypothesis. Consequently, the ratio of the posterior volume to the prior volume would be smaller for the $\mathcal{H}_\mathcal{ML}$ hypothesis. Thus, the second term in \Eref{eq:bayes_factor_estimate} becomes 
\begin{equation}
   \ln \mathscr{O}^\mathrm{ML}_\mathrm{UL} \approx  \ln \frac{\Delta V_\mathrm{ML}}{\Delta V_\mathrm{UL}} < 0.
\end{equation}
Since we are neglecting this term, \textit{we conclude that we are mostly over-estimating the Bayes factors when we use} \Eref{eq:Bayes_fac_crude_estimate}.  
This reasoning is supported by the observations in \Fref{fig:ul_vs_ml_hypothesis_comb_BF_err_study}. In this figure, we compare the numerically computed Bayes factors (using nested sampling) displayed in \Fref{fig:ul_vs_ml_hypothesis_comb} with the theoretically estimated $\mathcal{B}^\mathrm{ML}_\mathrm{UL}$ values (indicated by cross marks) obtained using \Eref{eq:Bayes_fac_crude_estimate}. We observe that, in almost all cases, our approximation tends to overestimate the true value of $\mathcal{B}^\mathrm{ML}_\mathrm{UL}$, and its performance improves as the SNR increases.
For example, one can see that for SNR values of $32$ and $50$ in \Fref{fig:ul_vs_ml_hypothesis_comb_BF_err_study}, the difference between the numerically computed Bayes factors with that of theoretically estimated ones is quite small compared to what we notice  for lower SNR values ($\lesssim 20$). 

Our investigation demands that when estimating $\log_{10}\mathcal{B}^\mathrm{ML}_\mathrm{UL}$ using \Eref{eq:Bayes_fac_crude_estimate}, we employ a higher threshold for positive/strong evidence for microlensing compared to established scales like Kass-Raftery's scale \citep{Kass:1995loi}. 
For our purpose, we choose this threshold heuristically based on our observation in \Fref{fig:ul_vs_ml_hypothesis_comb_BF_err_study}.
Specifically, we set the threshold values to be three times that of the Kass-Raftery's scale. For positive evidence for microlensing, we consider $\log_{10}\mathcal{B}^\mathrm{ML}_\mathrm{UL} \in (3\log_{10}(e),~9\log_{10}(e))$. For strong evidence, we require $\log_{10}\mathcal{B}^\mathrm{ML}_\mathrm{UL} > 9\log_{10}(e)$. By setting these higher thresholds, we aim to ensure that our assessment of positive or strong evidence for microlensing is conservative and accounts for the potential overestimation indicated by our analysis.

We now apply the method described above to analyse the distribution of Bayes factors using fitting factor values, as given by \Eref{eq:Bayes_fac_crude_estimate}, for the population generated in \Sref{sec:ml_pop_study}.
In the top row of \Fref{fig:pop_FF_BLU_analysis}, we present the fitting factor values and the corresponding Bayes factor values $(\log_{10}\mathcal{B}^\mathrm{ML}_\mathrm{UL})$ in the microlens parameter space of the observed population. 
While in the bottom row, the probability density functions (PDF) for the microlensing parameters, $M_\mathrm{Lz}$ and $y$, are shown for four cases: (i) our prior assumption (black line), (ii) the detected population (red line; as also shown in the bottom row of \Fref{fig:ml_population_study}), (iii) population that is detected and also correctly identified as being microlensed assuming thresholds as discussed above, i.e., $\log_{10}\mathcal{B}^\mathrm{ML}_\mathrm{UL}>3\log_{10}(e)$ (solid blue line; representing threshold for positive evidence for ML), and (iv) a higher threshold of $\log_{10}\mathcal{B}^\mathrm{ML}_\mathrm{UL}>9\log_{10}(e)$ (dashed blue line; representing threshold for strong evidence for ML). 
Here we employ two distinct threshold values to study how different levels of evidence for microlensing influence the probability density in the lensing parameter space.
We observe that the fitting factor values exhibit similar behaviour and range of values as shown in \Fref{fig:match_FF_analysis}, where the source binary was kept fixed. 
The Bayes factor values also show interesting values in the region where FF values are low (darker and bigger circles), especially the lower end of the \textit{wave zone} as discussed in \ref{fig:f_ML_cplot_for_Mlz_vs_y}. 
On the other hand, we do not observe any appreciable $\mathcal{B}^\mathrm{ML}_\mathrm{UL}$ values for lower $M_\mathrm{Lz}$ and $y$ values, where the number of detections is also relatively low. 
When we set a threshold of $3\log_{10}(e)$ for the recovered $\log_{10}\mathcal{B}^\mathrm{ML}_\mathrm{UL}$, we find that the distribution of $p(y)$ peaks around $y=1$, while $p(M_\mathrm{Lz})$ peaks at $\log_{10}M_\mathrm{Lz} = 3$. Additionally, there is a slight bimodality in the distribution of $p(y)$, with another peak observed at a lower $y$ value around $0.2$. This bimodality arises from the fact that although lower $y$ values lead to stronger microlensing effects, their detection probability is lower compared to those with higher $y$ values.
When we increase the threshold to $\mathcal{B}^\mathrm{ML}_\mathrm{UL}>10$, we find that the bimodal distribution in $y$ converges to a value close to the lower peak at $y\sim 0.3$. Moreover, for $p(M_\mathrm{Lz})$, a higher threshold  causes the peak to shift towards higher $M_\mathrm{Lz}$ values, around $M_\mathrm{Lz} \sim 4.5$.

\begingroup
  \renewcommand*{\thefootnote}{\alph{footnote}}
\begin{table}
\centering
\begin{center}
\caption{Effect of using unlensed templates during the search for microlensed signals. Below, ``Total" refers to all the events; ``ML" (``UL") refers to the case when microlensed (unlensed) templates are employed for search; $N_1$ depicts the total number of events with $\log_{10}\mathcal{B}^\mathrm{ML}_\mathrm{UL} > 3\log_{10}(e)$, indicating events with mostly positive evidence in favour of microlensing; $N_2$ depicts the total number of events with $\log_{10}\mathcal{B}^\mathrm{ML}_\mathrm{UL} > 9\log_{10}(e)$, indicating events with strong evidence in favour of microlensing. Here we estimate $\log_{10}\mathcal{B}^\mathrm{ML}_\mathrm{UL}$ using \Eref{eq:Bayes_fac_crude_estimate}. $\epsilon^\mathrm{ML}_\mathrm{UL}$ denotes the fractional loss of microlensed signals when unlensed templates are used during the search, as defined in \Eref{eq:fractional_loss}.}
\label{table:ML_pop_statistics}
\begin{tabular}{c c c c c c}
\hline
& Total & $N_1$ & $N_2$ &  $N_1/$Total & $N_2/$Total\\
&  &&  &  [\%] & [\%]\\
\hline
ML & 25458 & 8137 & 3734 & 32.0  &  14.7\\
UL & 23318 & 6481 & 2653 & 27.8 &  11.4 \\
\vspace{0.1cm}
$\epsilon^\mathrm{ML}_\mathrm{UL}$ [\%] & $8.4$ & $20.4$ & $29.0$ & $13.1$ & $22.4$ \\
\hline
\end{tabular}
\end{center}
\end{table}
\endgroup

\begin{figure}
    \includegraphics[width=0.49\textwidth]{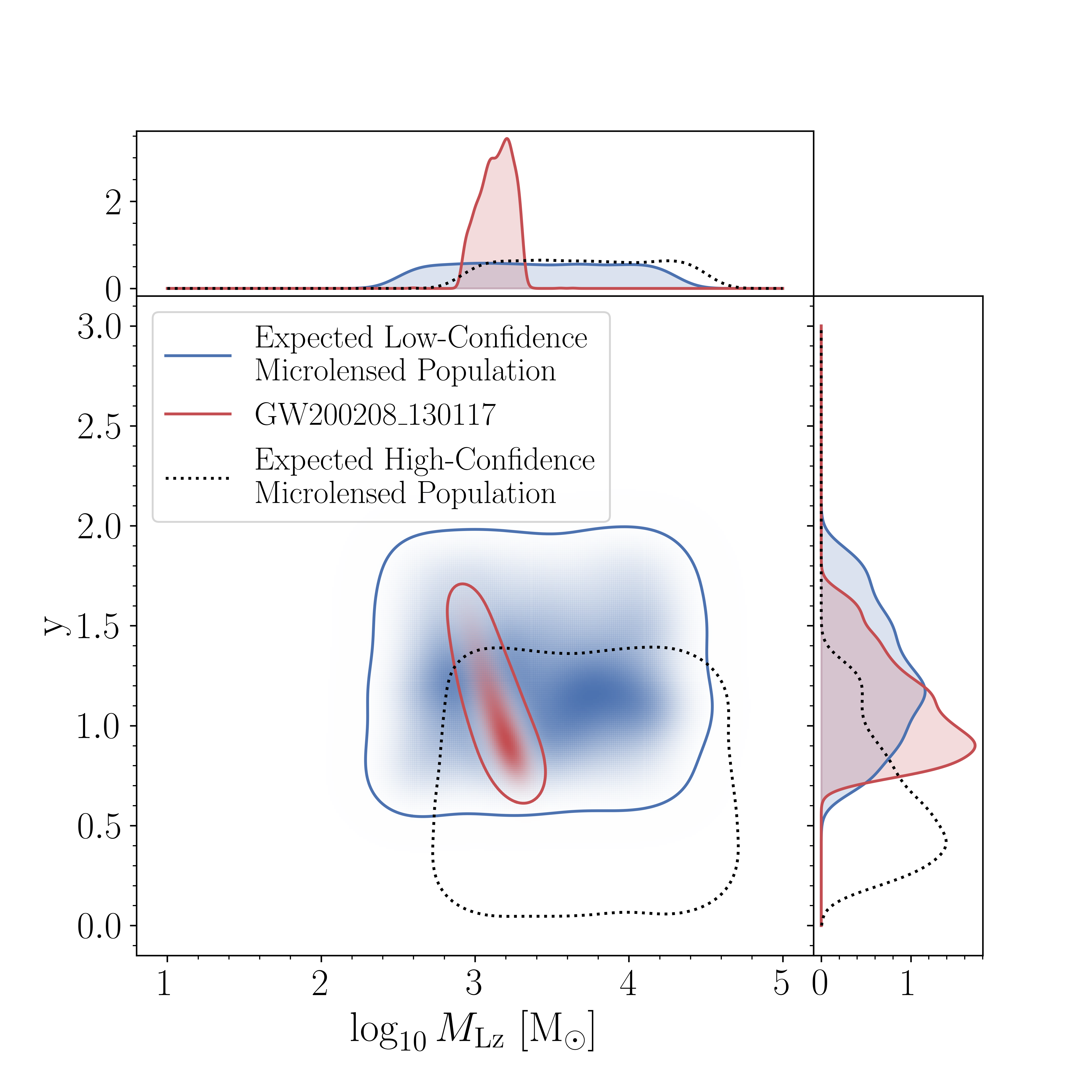}
    \caption{ Comparison between the $1-\sigma$ regions of the posteriors for the microlensed parameters of event GW200208\_130117 (highlighted in red) and the predicted $1-\sigma$ region of the low-confidence microlensed population derived from our population study (highlighted in blue). Additionally, the $1-\sigma$ contour for the predicted high-confidence microlensed population is shown for reference (marked with a black dotted line).}
    \label{fig:gw200208_comparison}
\end{figure}

In \Fref{fig:corner_plots}, we provide corner plots for the microlens parameters, showcasing the distribution of microlensed events that were detected and (potentially) identified as microlensed within our microlensed population. The contour plot in the 2D space and the red lines in the 1D distributions represent credible regions with quantile values of $16\%$, $50\%$, and $84\%$. These correspond to the median with $1\sigma$ uncertainty on either side.
The left panel of the figure depicts cases where the evidence for the microlensing (ML) hypothesis is positive, specifically when $\log_{10}\mathcal{B}^\mathrm{ML}_\mathrm{UL} \in (3\log_{10}(e),~9\log_{10}(e))$. 
In simpler terms, the left panel predicts the distribution of microlensed events that would be interesting candidates in the search for microlensed events but may not be definitively confirmed as such. The conclusions drawn from these events would likely remain \textit{inconclusive} due to various systematics that could mimic similar behaviour. 
Up to an uncertainty of $1-$sigma, \textit{the most probable parameters\footnote{We note that a lot of reasonable assumptions have gone into making such a prediction. The population is generated assuming an O4-like sensitivity. The Bayes factor estimation is not rigorous and we ignore some other factors such as noise systematics, its degeneracy with other physical effects like eccentricity, etc.} that will show only positive evidence for ML are $(\log_{10}M_\mathrm{Lz},~y) = (3.35${\raisebox{0.5ex}{\tiny$^{+1.13}_{-0.93}$}},$~1.21${\raisebox{0.5ex}{\tiny$^{+0.68}_{-0.51}$}}$)$}. 
Meanwhile, the right panel highlights events that would be identified as microlensed with a high degree of confidence, with $\log_{10}\mathcal{B}^\mathrm{ML}_\mathrm{UL} > 9\log_{10}(e)$. \textit{The most probable parameters for such confidently detected microlensed events would be $(\log_{10}M_\mathrm{Lz},~y) = (3.76${\raisebox{0.5ex}{\tiny$^{+0.86}_{-0.88}$}},$~0.58${\raisebox{0.5ex}{\tiny$^{+0.70}_{-0.37}$}}$)$.}

It is interesting to note that among all the super-threshold events detected by LIGO and Virgo detectors so far, the most compelling candidate in the microlensing search was GW$200208\_130117$ during the third observing run \citep{LIGOScientific:2023bwz, Janquart:2023mvf}. This event exhibited the highest Bayes factor value for the evidence of microlensing over the unlensed hypothesis, with a value of $\log_{10}\mathcal{B}^\mathrm{ML}_\mathrm{UL}\sim0.9$\footnote{This value differs from the quoted value of $0.8$ in \cite{LIGOScientific:2023bwz} as it has been recomputed by the authors using the GWMAT framework (Mishra, A., in prep.).}. 
However, the paper concluded the data is inconclusive about the microlensing hypothesis, and it was hinted that the effect could be due to some short-duration noise fluctuations in one of the detectors.
The recovered microlens parameter values for the event, with median values and 1-sigma errors, are $(\log_{10}M_\mathrm{Lz},~y) = (3.15${\raisebox{0.5ex}{\tiny$^{+0.18}_{-0.21}$}},$~1.07${\raisebox{0.5ex}{\tiny$^{+0.61}_{-0.32}$}}$)$.
Interestingly, we note that this recovered value is remarkably close to our predicted value of $(\log_{10}M_\mathrm{Lz},~y) = (3.35${\raisebox{0.5ex}{\tiny$^{+1.13}_{-0.93}$}},$~1.21${\raisebox{0.5ex}{\tiny$^{+0.68}_{-0.51}$}}$)$ (see \Fref{fig:gw200208_comparison} for the comparison) for events that would only positively support the microlensed hypothesis.
\textit{Hence, based on our population study, there is suggestive evidence in favour of the microlensing hypothesis for the event} GW$200208\_130117$.
However, it is important to acknowledge that this study is not rigorous enough to claim lensing with certainty, and therefore, the ultimate nature of this event remains inconclusive.

Furthermore, in Table \ref{table:ML_pop_statistics}, we present the derived statistics from our population, particularly focusing on the fractional loss of signals caused by employing unlensed templates during the search for microlensed signals. The fractional loss, $\epsilon^\mathrm{ML}_\mathrm{UL}$, is defined as:
\begin{equation}
    \epsilon^\mathrm{ML}_\mathrm{UL} \equiv 1 - \frac{n(\mathrm{UL})}{n(\mathrm{ML})},
    \label{eq:fractional_loss}
\end{equation}
where $n(\mathrm{UL})$ and $n(\mathrm{ML})$ represent the number of events quoted in the row labelled as `UL' and `ML', respectively. These labels indicate scenarios where unlensed templates and microlensed templates are used to recover the signals, respectively.
Within the parameter space of $M_\mathrm{Lz} \in (10,~10^5)$ and $y \in (0.01,~3.00)$, we observe an approximate loss of $8\%$ for the microlensed signals (refer to the first column, third row). 
However, this fraction is dependent on our chosen parameter space and does not fully capture the impact on potentially identifiable signals.
Therefore, we further estimate the fractional loss for events that satisfy specific conditions: (i) $\log_{10}\mathcal{B}^\mathrm{ML}_\mathrm{UL}>3\log_{10}(e)$ (referred to as $N_1$; column 2), and (ii) $\log_{10}\mathcal{B}^\mathrm{ML}_\mathrm{UL}>9\log_{10}(e)$ (referred to as $N_2$; column 3). We find that the fractional loss of events in case (i) is $\sim 20\%$, while for case (ii) it increases to about $29\%$. It is expected that the fractional loss would increase with higher threshold values on Bayes factors, as the greater the microlensing effects, the more significant the loss in their SNR during the search process. 
Moreover, considering that real searches utilize template banks that discretely cover the parameter space, typically constructed with a maximum loss threshold of $3\%$, there is an additional loss of such microlensed signals during the search process. Specifically, for case (i) and case (ii), we find that the total fractional losses can reach up to $27\%$ and $35\%$, respectively. 
\textit{This observation suggests that neglecting the loss of microlensed signals during the search process~\citep[e.g.,][]{Basak:2021ten} may impose an over-constraint on the fraction of compact dark matter based on the non-detection of microlensed gravitational wave signals.}

\begin{table*}
\centering
\begin{center}
\caption{
Lens system information, with the lens and source redshifts denoted by $z_\mathrm{L}$ and $z_\mathrm{S}$, respectively, is listed here. The macro ($\mu_\mathrm{macro}$) and smooth ($\mu_\mathrm{smooth}$) magnifications, the microlens density ($\Sigma_\odot$), and the network optimal SNR for both the SL-only case, $\rho^\mathrm{net}_\mathrm{opt}(h_\mathrm{SL})$, and the SL+ML case, $\rho^\mathrm{net}_\mathrm{opt}(h_\mathrm{SL+ML})$, of our chosen GW150914-like source are also tabulated.
}
\label{table:SLML_injs_table}
\begin{tabular}{c c c c c c c c c c}
\hline
system & $z_\mathrm{L}$ & $z_\mathrm{S}$ & image & $\mu_\mathrm{macro}$ & $\mu_\mathrm{smooth}$ & $\Sigma_\odot$ $[$M$_\odot/$pc$^2]$ & $\rho^\mathrm{net}_\mathrm{opt}(h_\mathrm{SL})$ & $\rho^\mathrm{net}_\mathrm{opt}(h_\mathrm{SL+ML})$ & match$(h_\mathrm{SL},~h_\mathrm{SL+ML})$ [\%]\\
\hline
1 & $0.50$ & $0.56$ & I & $4.75$ & $3.27$ &$1030.5$ & $23.7$ & $24.6$ & $99.92$\\
  &  &  & II & $10.85$ & $4.66$ &$1419.5$ & $34.9$ & $38.5$ & $99.21$\\
  &  &  & III & $-11.22$ & $+13.52$ &$2121.6$ & $35.4$ & $34.8$ & $99.96$\\
  &  &  & IV & $-1.56$ & $-2.57$ &$4905.8$ & $11.5$ & $11.4$ & $99.93$\\
\hline
2 & $0.28$ & $0.50$ & I & $6.92$ & $4.41$ &$304.7$ & $32.3$ & $35.2$ & $99.80$\\
  &  &  & II & $11.14$ & $5.54$ &$351.1$ & $50.5$ & $56.5$ & $99.97$\\
  &  &  & III & $-12.64$ & $+25.99$ &$513.31$ & $55.1$ & $50.5$ & $99.89$\\
  &  &  & IV & $-2.59$ & $-4.33$ &$864.94$ & $24.9$ & $23.9$ & $99.97$\\
\hline
\end{tabular}
\end{center}
\end{table*}

\section{Effect of microlens population on the signatures of strong lensing }
\label{sec:ml_effect_on_PO_analysis}
In this section, we study the effect of microlensing from a population of microlenses on the search for strongly lensed gravitational wave signals.
The intervening galaxy or galaxy cluster acting as a macrolens contains substructures in the form of microlens population that can further perturb the signal due to microlensing effects \citep[e.g.,][]{Diego:2019lcd, Mishra:2021xzz, Meena:2022unp}. We simulate such signals and investigate their effects on strong lensing searches, particularly on the interpretation of \textit{posterior overlap} analysis \citep{Haris:2018vmn}. To accomplish this, we perform a set of parameter estimation runs. The strongly lensed signal, which is taken to further undergo microlensing (to produce an ``SL+ML" signal), is generated after computing the amplification factor for such systems using the methodology described in \cite{Mishra:2021xzz} and Mishra, A. (in preparation).

Posterior overlap analysis is a fast and robust method for identifying potential strongly lensed pairs of GW signals. The method relies on two primary observations: (i) strongly lensed images should originate from the same patch of the sky, and (ii) the gravitational lensing does not affect the GW phasing, which means that the parameter estimation for the intrinsic parameters should remain unaffected. Consequently, the sky and the intrinsic parameters should exhibit similarity between the two images of a strongly lensed system. So for any two GW signals, one can compute the overlap between the posteriors for the aforementioned parameters and develop a statistic to assess its significance. Given the posteriors of two events $d_1$ and $d_2$, the Bayes factor for the (strongly-) lensed hypothesis over the unlensed hypothesis can be defined as \citep{Haris:2018vmn}
\begin{equation}
    \mathcal{B}^\text{\scriptsize L}_\text{\scriptsize U} = \int{d\pmb{\theta}} \frac{P(\pmb{\theta}|d_1)P(\pmb{\theta}|d_2)}{P(\pmb{\theta})},
\end{equation}
where $\pmb{\theta}$ is the set of parameters over which we compute the overlap. As mentioned above, $\pmb{\theta}$ is at 
most
a 9D quantity, i.e., $\pmb{\theta}=\{\mathcal{M},~q,~a_1,~a_2,~\theta_1,~\theta_2,~\theta_\mathrm{JN},~\alpha,~\delta\}$, where the symbols have their usual meaning as described in Sec. \ref{subsec:basics_of_GW_data_analysis}.

For the lensing systems, we selected two (quadruply) lensed systems from the catalog described in \cite{More:2021kpb}. We specifically chose systems where the source redshift was relatively lower to ensure high SNR events, and where the brightest image had a (macro-)magnification of $\gtrsim 10$. 
As mentioned in \Sref{sec:ml_effect_on_PE}, we used a GW$150914$-like event as the source, including its spins. The properties of both systems are provided in Table \ref{table:SLML_injs_table}.
As expected, our choice of lens systems with low source redshifts led to smaller Einstein angle~($0.03''$ and $0.54''$ for system-1 and system-2, respectively) compared to typical lens systems in EM observations where Einstein angle is~$\sim1''$. However, as we specifically focus on high SNR systems for our introductory analysis, we proceed with these systems in our current work and leave a more detailed analysis for future work.

\begin{figure*}
    \centering
    \includegraphics[width=0.98\linewidth]{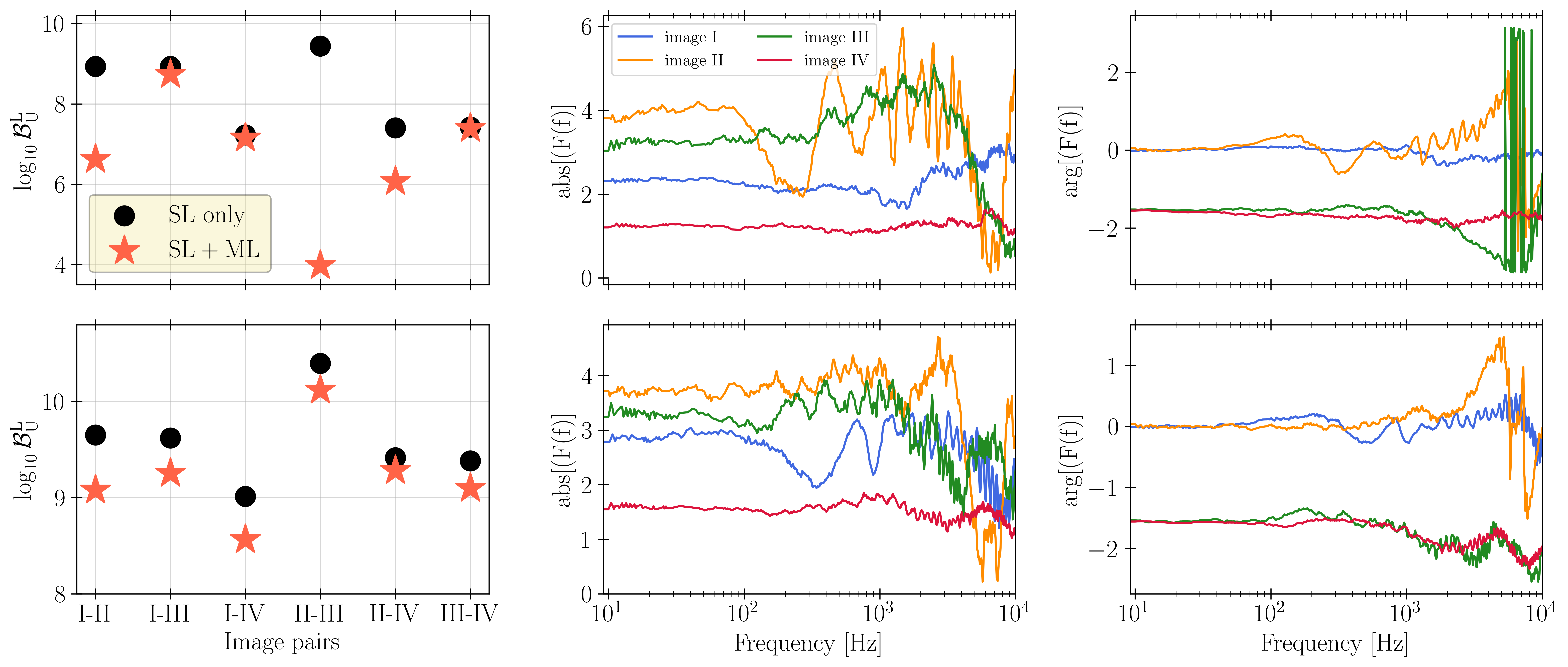}
    \caption{Effect of microlensing on strongly lensed gravitational wave signals due to a population of microlenses. \textit{Left panel:} The comparison between the Bayes factors in favour of lensing using the posterior overlap analysis is shown for two quadruple-lensed systems (\textit{top} and \textit{bottom} rows), indicating both the strongly lensed systems ("SL only``; represented using black circled markers) and the systems further undergoing microlensing ("SL+ML``; represented using star-shaped red markers). \textit{Middle and Right panels:} A realisation of the (micro-)lensing amplification factor $F(f)$ is displayed for the four macroimages resulting due to the presence of microlens population in the vicinity of the macroimages.}
    \label{fig:PO_analysis_1}
\end{figure*}

The results of the posterior overlap analysis are shown in \Fref{fig:PO_analysis_1}. This figure compares the Bayes factors in favour of lensing obtained from the posterior overlap analysis for both the strongly lensed macroimages (black solid circles) and the macroimages that further undergo microlensing (red solid stars). The middle and right panels of the figure also display the corresponding (micro-)lensing amplification factors $F(f)$ for the four macroimages. The small-scale fluctuations observed in the $F(f)$ curves are numerical artifacts that are mitigated by applying a high-pass filter before their utilization.
Firstly, for system 1 (top row), we note that the Bayes factor values can significantly reduce in extreme cases of microlensing in the path of a strongly lensed signal. For example, in the top-left panel, we see orders of magnitude drop in the $\mathcal{B}^\mathrm{L}_\mathrm{U}$ for image pairs I-II, II-III and II-IV, i.e., all image pairs with the second image. This behaviour can be explained if we note that the corresponding amplification factor, $F(f)$, curves in the middle and right panels of the first row. Even though the $F(f)$ curves for the third and fourth images show large modulations, they do so only at high frequencies ($>10^3$ Hz). On the other hand, one can clearly notice from visual inspection that only the $F(f)$ curves for image II (orange-coloured curve) show significant modulations at low frequencies, where most of the power of the gravitational wave is contained. 
For system 2 (bottom row), we still see Bayes factor values to drop for all image pairs but relatively lesser than that for system 1. In this case, $F(f)$ curves for image I and III showed significant modulation at lower frequencies compared to other images. One can see this from the phase plots in the lower-right panel, where blue and green curves start deviating from orange and red curves at around $\sim 100$ Hz. 


One can further ask which of the parameters incorporated for the computation of the posterior overlap is responsible for such a drop in the Bayes factor values for some of the microlensing cases. To that end, in \Fref{fig:PO_analysis_2}, we show 1D marginalised posteriors corresponding to the parameters used for computing the overlap. For ease of representation, we denote spin components using $\{\chi_\mathrm{eff},~\chi_\mathrm{p}\}$ instead of $\{a_1,~a_2,~\theta_1,~\theta_2\}$. The differently coloured curves correspond to posteriors associated with different images, as shown in \Fref{fig:PO_analysis_1}. The dotted curves represent cases with only strong lensing (SL only), while the solid curves depict recoveries for signals that undergo both strong lensing and microlensing (SL+ML). The dashed black vertical lines depict the injected values. 
We notice that the posteriors for most of the parameters are well recovered around the true injected value for both cases (SL only and SL+ML). 
However, in system 1, the posteriors representing the recoveries for the second image of SL+ML case show the maximum deviation from the injected value among different parameters (solid orange-coloured curves). 
In the case of system 2, although the deviation between the recovered parameters for the SL and SL+ML scenarios is largely similar, the posterior distributions for parameters in the SL+ML case exhibit slightly greater variability, which can be attributed to the variability in $F(f)$ values for different images.

Among all the parameters considered here, the sky-position parameters, i.e., RA and Dec ($\alpha$ and $\delta$), are the best-recovered parameters, as their posteriors are sharply peaked around the injected value for both the systems. Therefore, the sky-position parameters contribute the most to the posterior overlap values. The drop in the Bayes factor values is then mostly coming from the biased recoveries of parameters like $\{\mathcal{M}_\mathrm{det}$, $\chi_\mathrm{eff}$, $\chi_\mathrm{p}$, $\theta_\mathrm{JN}\}$ for SL+ML case.

We note that in this section, we studied only a few scenarios of microlensing due to a population of microlenses affecting strongly lensed GWs. Due to our selection of systems with high SNR values, even small deviations (high match values in \Tref{table:SLML_injs_table}; also see \Fref{fig:PO_analysis_2}) led to a significant decrease in the Bayes factor. However, it is important to conduct a more comprehensive statistical study to generalise the effects of microlensing on strong lensing searches. We leave this to future investigations.

\section{Discussion and Conclusion}
\label{sec:conclusion}
In this work, we primarily examine the impact of microlensing caused by isolated microlenses on GW signals. We begin by illustrating how the time delay between microimages divides the microlens parameter space into three distinct regions. Next, we investigate how microlensing can significantly influence the observed SNR, match, and fitting factor values. Subsequently, we analyze the microlensing-induced bias in the observed GW source parameters. Furthermore, we explore the statistical properties of microlensed GW signals and estimate the fraction of missed GW signals if we employ unlensed templates in the search. The distribution of Bayes factors for the population reveals certain regions in the microlensing parameter space that are more likely to be correctly identified as microlensed signals. Finally, we examine more complex and realistic scenarios involving the interaction of strongly lensed GW signals with a population of microlenses residing within lensing galaxies and study their effect on posterior overlap analysis.

Based on our analysis, the results are as follows:
\begin{enumerate}
\item Employing unlensed waveforms to search for microlensed GW signals can significantly decrease the fitting factor (FF), reaching as low as~$\sim70\%$. The FF values decrease as we increase (decrease) the value of $M_\mathrm{Lz}$ ($y$). Consequently, the observed SNR also decreases. However, microlensing itself amplifies the signal and can significantly increase the SNR, with values exceeding 10 times higher in extreme cases. This behaviour overall increases the detector horizon and can even allow us to detect GW signals from high redshifts $z\gtrsim2$, beyond the peak of the star-formation rate.

\item  The correlation study reveals a strong correlation between the microlens parameters and the luminosity distance. Specifically, the parameter $y$ exhibits a significant anti-correlation with the distance, reaching values exceeding $90\%$ in certain cases. Moreover, we observe that the correlations between the microlens parameters and GW signal parameters are generally opposite in nature. For instance, a positive correlation of $M_\mathrm{Lz}$ with a GW signal parameter often implies an anti-correlation of $y$ with the same parameter.
Recovering microlensed GW signals using an unlensed waveform model introduces strong degeneracies among the source parameters, particularly when the microlensing effects are significant and slowly varying, such as in the bottom-right corner of the long-wavelength regime (\Fref{fig:f_ML_cplot_for_Mlz_vs_y}). These degeneracies exhibit a highly nonlinear relationship with variations in the microlens parameters. In other words, our analysis indicates that microlensing can lead to a rotation of the correlation among different pairs of parameters.

\item Recovering microlensed GW signals with unlensed GW signals can lead to significant bias in the estimated parameter values, particularly when the microlenses belong to the wave-dominated zone, where $f\tau_{\rm d}\sim 1$.
Among intrinsic parameters, the in-plane spin components, particularly the precession effective spin $\chi_\mathrm{p}$, are most affected, suggesting a degeneracy between the effects of microlensing and the modulations arising from spin-induced precession. With SNR~$\sim50$, the errors increase for longer signals (lighter binaries) and can even exceed $>90\%$ in cases such as $(M_\mathrm{Lz},~y)=(10^2~{\rm M}_\odot,~1)$, which is a modest representative of microlensing through an intermediate-mass-black-hole (IMBH). 
This suggests that any signal showing signs of precession must also be analysed for the presence of microlensing signatures to avoid any erroneous claims regarding the presence of precession. However, vice-versa may not be true, i.e., it is unlikely that the presence of precession can bias microlensing searches. This is because the parameter space of unlensed signals always falls within the subset of the microlensed parameter space. Hence, unless significant waveform systematics are involved in inferring the precession of a signal, such biasing is not expected.
In addition, other intrinsic parameters related to binary component masses, chirp mass and mass ratio, can also be significantly affected. Although their relative errors is mostly within $10\%$, it can even exceed $50\%$ when microlensing effects are strong. Moreover, KS-statistics show higher sensitivity of the posterior distribution to microlensing effects compared to the recovered best-fit values.    

\item Among extrinsic parameters, the recoveries of luminosity distance are affected the most. In contrast, the trigger time and the sky-position parameters, RA and Dec ($\alpha$ and $\delta$), are the best-recovered source parameters. This is expected since the localization of GW sources is mainly based on the observed time delays between each pair of interferometers and microlensing does not significantly affect them. 

\item A population study of microlensed signals reveals that the fraction of potentially identifiable microlensed signals missed due to the use of usual unlensed templates during the search is around $\epsilon^\mathrm{ML}_\mathrm{UL} \in (20\%,~30\%)$. Hence, neglecting the loss of microlensed signals during the search process~\citep[e.g.,][]{Basak:2021ten} may impose an over-constraint on the fraction of compact dark matter based on the non-detection of microlensed GW signals. Furthermore, investigating the impact of selection bias on the distribution of microlens parameters in the observed signals reveals a significant deviation of the PDF of the impact parameter, $p(y)$, at low values of $y\lesssim 0.1$. Therefore, in contrast to the commonly used lower limit of $0.1$ in microlensing searches of real data \citep[e.g.,][]{LIGOScientific:2021izm, LIGOScientific:2023bwz}, a value of $y=0.01$ is not as insignificant as previously thought. On the other hand, we only observe a mild preference for lower~$M_{\rm Lz}~(<10^3~{\rm M_\odot})$ compared to larger ones, which primarily arises from the use of unlensed waveforms in recovering microlensed signals.

\item A model comparison study highlights the challenges in confidently identifying microlensing by $\lesssim 100~{\rm M_\odot}$ microlenses, especially with average SNR values of $\sim 12$ \citep{Schutz:2011tw}, unless the impact parameter $y$ is very low (i.e., $y < 0.1$). However, for high SNR~($\sim50$) events, even microlenses with masses~$M_\mathrm{Lz} \gtrsim 20~{\rm M}_\odot$ can be detected~(assuming a characteristic value of $y = 1$). On the other hand, microlensing signatures for an event with $(M_\mathrm{Lz},~y)=(10^2~{\rm M}_\odot,~1)$ is not detectable up to an SNR value of around $25$.    

\item The Bayes factor analysis of our population of microlensed signals indicates certain region in $M_\mathrm{Lz}-y$ parameter space have a higher probability of being detected and accurately identified as microlensed. The analysis reveals that events identified as only positively\footnote{We use the terms "positive" and "strong" to characterise the strength of evidence, in accordance with the terminology used in Jeffreys' or Kass-Raftery's scale for interpreting Bayes Factor values \citep{Deutsch:1999gs, Kass:1995loi}.} indicating microlensing would typically fall within the parameter space $(\log_{10}M_\mathrm{Lz},~y) = (3.35${\raisebox{0.5ex}{\tiny$^{+1.13}_{-0.93}$}},$~1.21${\raisebox{0.5ex}{\tiny$^{+0.68}_{-0.51}$}}$)$. On the other hand, events that are expected to favour the microlensing hypothesis strongly would typically lie within the parameter space $(\log_{10}M_\mathrm{Lz},~y) = (3.76${\raisebox{0.5ex}{\tiny$^{+0.86}_{-0.88}$}},$~0.58${\raisebox{0.5ex}{\tiny$^{+0.70}_{-0.37}$}}$)$.

\item In the GWTC-3 catalog \citep{KAGRA:2021vkt}, the most compelling candidate in the microlensing search thus far is the event GW$200208\_130117$, which exhibited the highest Bayes factor of $\log_{10}\mathcal{B}^\mathrm{ML}_\mathrm{UL}\sim0.9$ \citep{LIGOScientific:2023bwz, Janquart:2023mvf}. The recovered values of the microlens parameters for this event, including median values and 1-sigma errors, are $(\log_{10}M_\mathrm{Lz},~y) = (3.15${\raisebox{0.5ex}{\tiny$^{+0.18}_{-0.21}$}},$~1.07${\raisebox{0.5ex}{\tiny$^{+0.61}_{-0.32}$}}$)$. Interestingly, we note that this recovered value is remarkably close to our predicted value of $(\log_{10}M_\mathrm{Lz},~y) = (3.35${\raisebox{0.5ex}{\tiny$^{+1.13}_{-0.93}$}},$~1.21${\raisebox{0.5ex}{\tiny$^{+0.68}_{-0.51}$}}$)$ (see \Fref{fig:gw200208_comparison}) for events that would only positively support the microlensing hypothesis. Hence, based on our population study, there is suggestive evidence in favour of the microlensing hypothesis for the event GW$200208\_130117$. However, it is important to acknowledge that further work is required to confirm lensing with certainty, and the true nature of this event remains inconclusive.

\item Finally, to study the effect of microlensing on the search of strongly lensed gravitational wave signals, specifically the posterior overlap analysis, we focused our attention on a much more complex scenario of microlensing when a strongly lensed GW signal encounters a population of $\mathcal{O}(10^4)$ microlenses present in the lensing galaxy. We find that, in general, the presence of microlens population decreases the measured Bayes factor in favour of strong lensing~(see \Fref{fig:PO_analysis_1}). However, the exact amount of drop is sensitive to the magnitude of microlensing effects in the signal, which in turn depends primarily on the strong lensing magnification and properties of the microlens population. This suggests that, in extreme cases, the presence of microlensing may pose challenges in accurately identifying and characterizing strongly lensed GW signals. However, a more detailed study is required to generalise the above inferences.

\end{enumerate}

In summary, this extensive investigation across various sections sheds light on the diverse effects of microlensing on GW signals. The findings contribute to our understanding of the detectability, parameter estimation biases, and population characteristics associated with microlensed signals.

In future research, it is crucial to distinguish the effects of microlensing from other physical effects, such as eccentricity, precession, tidal heating, etc., as microlensing has the potential to alter the morphology of signals. Furthermore, it is important to investigate whether these effects can lead to false triggers in various tests of general relativity (GR). Additionally, there is a need to delve deeper into the impact of microlensing on strongly lensed GWs and explore their implications for future searches.

\section{Acknowledgements}
A. Mishra would like to thank the University Grants Commission (UGC), India, for financial support as a research fellow.
AKM acknowledges support by grant 2020750 from the United States-Israel Binational Science Foundation (BSF) and grant~2109066 from the United States National Science Foundation (NSF), and by the Ministry of Science \& Technology, Israel.
This material is based upon work supported by NSF's LIGO Laboratory which is a major facility fully funded by the National Science Foundation.
Authors gratefully acknowledge the use of high-performance computing facilities at IUCAA, Pune. The authors are also grateful for the computational resources provided by the LIGO Laboratory supported by National Science Foundation Grants PHY-0757058 and PHY-0823459.  
Additionally, the authors would like to express their sincere gratitude to Apratim Ganguly for their valuable discussions and suggestions; Paolo Cremonese for carefully reading the manuscript; Sumit Kumar and K. Haris for providing the script to compute the posterior overlap; Samanwaya Mukherjee for the useful discussion regarding the Fisher analysis. 
Lastly, the authors would like to thank Marek Basieda for reviewing this manuscript.
A. Mishra would also like to acknowledge the fruitful interactions with the following individuals during the course of this work: P. Ajith, K. Chandra, T. Ghosh, D. Keitel, D. Rana, A. Sharma, K. Soni, and A. Vijaykumar. 

The work utilises the following software packages:
\texttt{Cython} \citep{Behnel:2011ddo},
\texttt{NumPy} \citep{Harris:2020xlr}, 
\texttt{SciPy} \citep{Wu:2009yr}, 
\texttt{dynesty} \citep{Speagle:2019ivv}, 
\texttt{GWPopulation} \citep{Talbot:2019okv}, 
\texttt{corner} \citep{corner}, 
\texttt{seaborn} \citep{Waskom:2021psk}, 
\texttt{Matplotlib} \citep{Madaras:1991ycz}, and
\texttt{Jupyter notebook} \citep{Madaras:1991ycz}.

\section{Data Availability}
The microlensing simulation data can be made available upon reasonable request to the corresponding author.
The simulated strong lens system data is available from the authors of 
\citet{More:2021kpb}.

\bibliographystyle{mnras}
\bibliography{bibliography}

\appendix

\section{Statistical Uncertainties in the measurement of lensing parameters using Fisher Analysis}
In this section, we will estimate the statistical uncertainties in the inference of lensing parameters of a point-lens, namely the redshifted lens mass $M_\mathrm{Lz}$ and the impact parameter $y$, using Fisher analysis (or Fisher information-matrix formalism) \citep{Finn:1992wt,  Vallisneri:2007ev, Borhanian:2020ypi, Antonelli:2021vwg, Mukherjee:2022wws}. Here, we give a brief overview of the formalism. 

Under the assumption that noise, $n(t)=d(t)-h(t)$, is stationary and Gaussian with zero mean, the (log) likelihood of observing a specific data stream realisation can be written as
\begin{equation}
    \log p(D|\pmb{\lambda}) \propto {-\frac{1}{2}\bra{(D - H(\pmb{\lambda})}\ket{D - H(\pmb{\lambda})}},
    \label{eq:likelihood}
\end{equation}
where $D$ and $H$ are Fourier transforms of $d$ and $h$, respectively, and $\pmb{\lambda}$ is the parameter vector that determines a particular waveform.
Next, the formalism exploits the fact that for a sufficiently high SNR, 
the deviation in strain can be approximated as a linear function of parameter errors around the true value at the leading order,
called linear signal approximation (LSA) \citep{Finn:1992wt}. 
Since the best-fit parameter $\pmb{\lambda}_\mathrm{best-fit}$ can be assumed to be a perturbation from the true parameter $\pmb{\lambda}_\mathrm{true}$ in the presence of noise, one can write $\pmb{\lambda}_\mathrm{best-fit}=\pmb{\lambda}_\mathrm{true} + \Delta \pmb{\lambda}$. Thus, using LSA, the waveform model in the vicinity of the best-fit parameters can be written as 
\begin{equation}
    h(t;\pmb{\lambda}_\mathrm{best-fit}) \approx h(t;\pmb{\lambda}_\mathrm{true})  + \partial_i h(t;\pmb{\lambda}_\mathrm{true}) \Delta \lambda^i,
    \label{eq:LSA}
\end{equation}
where we make use of the Einstein-summation convention and $\partial_i\equiv \partial/\partial \lambda_i$. The expression is valid for $|\Delta \lambda^i|\ll 1$. Substituting \Eref{eq:LSA} into \Eref{eq:likelihood}, one obtains
\begin{align}
-2 \log p(D|\pmb{\lambda})  &= (\Delta\lambda^i - \Delta\lambda^i_\mathrm{noise})\Gamma_{ij}(\Delta\lambda^j - \Delta\lambda^j_\mathrm{noise}),\\
\Delta\lambda^i_\mathrm{noise} &= (\Gamma^{-1})^{ij} \bra{\partial_j h}\ket{n}   
\end{align}
where $\Gamma_{ij}$ is the Fisher matrix defined by
\begin{equation}
    \Gamma_{ij} = \bra{\partial_i h}\ket{\partial_j h}.
\end{equation}
Defining the statistic $\widehat{\Delta\lambda^i} = \Delta\lambda^i_\mathrm{noise}$, one finds
\begin{equation}
    \mathbb{E}[\widehat{\Delta\lambda^i}]=0,~~~\mathrm{Cov}(\widehat{\Delta\lambda^i},~\widehat{\Delta\lambda^i}) \equiv \Sigma = (\Gamma^{-1})^{ij} + \mathcal{O}(\rho^{-1}).
\end{equation}
The diagonal and off-diagonal elements of the covariance matrix $\Sigma$ denote the variances and covariances of the parameters, respectively, due to the uncertainty introduced by the detector noise and give 1$\sigma$-uncertainty estimates via $\sigma_{\lambda_i} = \sqrt{\Sigma}_{ii}$.

Note that the validity of Fisher analysis demands a high-SNR where LSA is valid. Furthermore, the Fisher matrix needs to be 'well-conditioned' for invertibility, which could even be compromised due to the limited arithmetic precision. 
See, e.g., the excellent discussion in Ref.~\cite{Vallisneri:2007ev} of these issues related to Fisher analysis.

We compute the statistical uncertainties in
$M_\mathrm{Lz}$ and $y$ in the lensing parameter space spanning $\log_{10}M_\mathrm{Lz}\in(0,~5)$M$_\odot$ and $y\in(0.01,~3)$. 
We introduce microlensing effects to a GW150914-like system with no spins and adjust the luminosity distance to maintain an optimal network SNR of $50$ across the Hanford, Livingston, and Virgo detector network, using projected O4 PSDs \citep{KAGRA:2013rdx}. To compute the covariance matrix, we utilize the publicly available package \texttt{GWBENCH} \citep{Borhanian:2020ypi}. To ensure a well-conditioned Fisher matrix, we only vary the parameters $\Theta = \{\mathcal{M},\eta,~a_1,a_2,\log_{10} M_\mathrm{Lz},\ln y\}$. We employ the \texttt{IMRPhenomXPHM} waveform approximant with lower and upper frequency cutoffs set at $20$Hz and $1024~$Hz, respectively, with a bin size of $2^{-4}~$Hz, which is adequate for the signal's duration in this context. 
\begin{figure}
     \centering
     \begin{subfigure}[b]{0.485\textwidth}
         \centering
         \includegraphics[width=\textwidth]{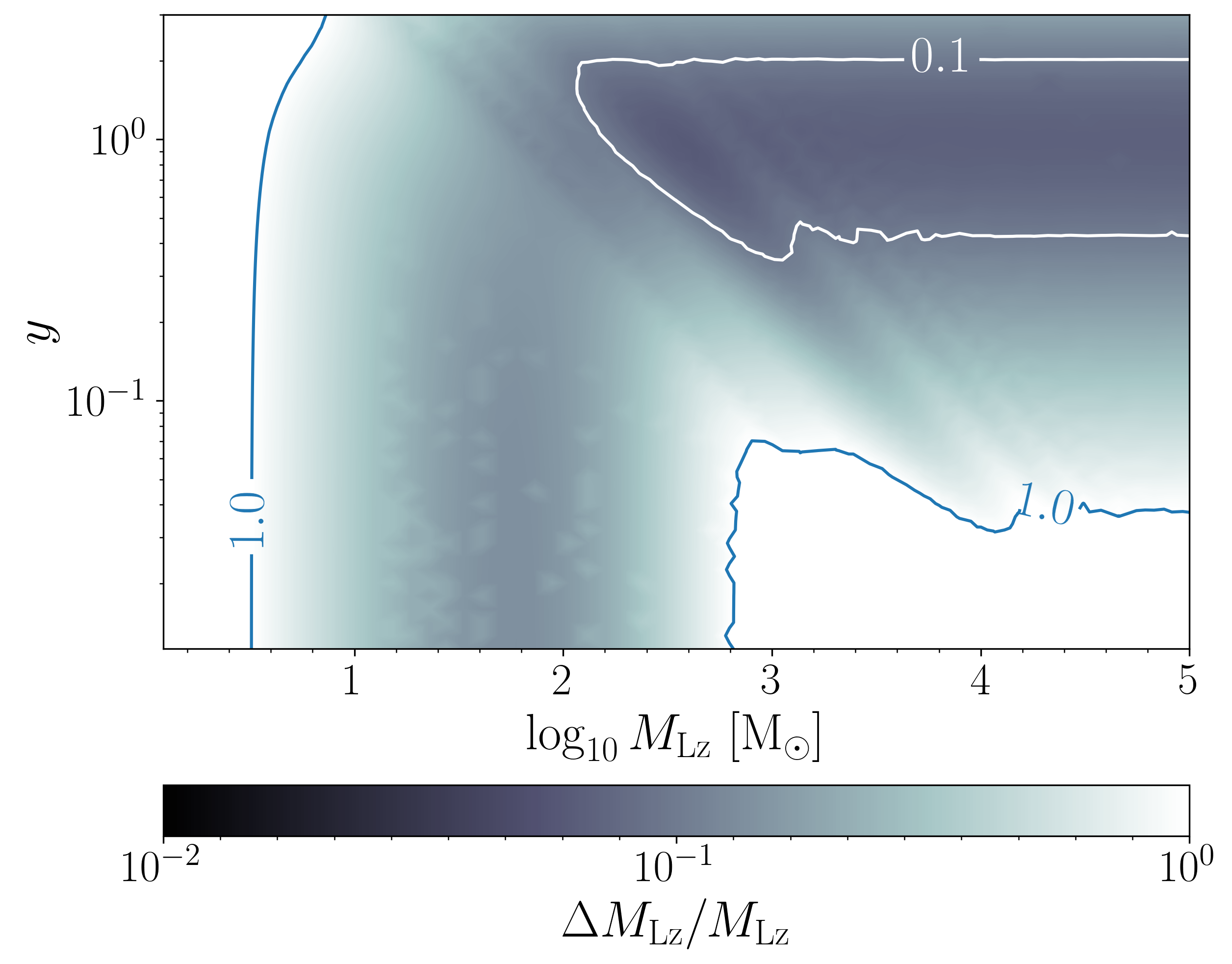}
         \caption*{}
         \label{fig:y rel_err_Mlz}
     \end{subfigure}
     \hfill
     \begin{subfigure}[b]{0.485\textwidth}
         \centering
         \includegraphics[width=\textwidth]{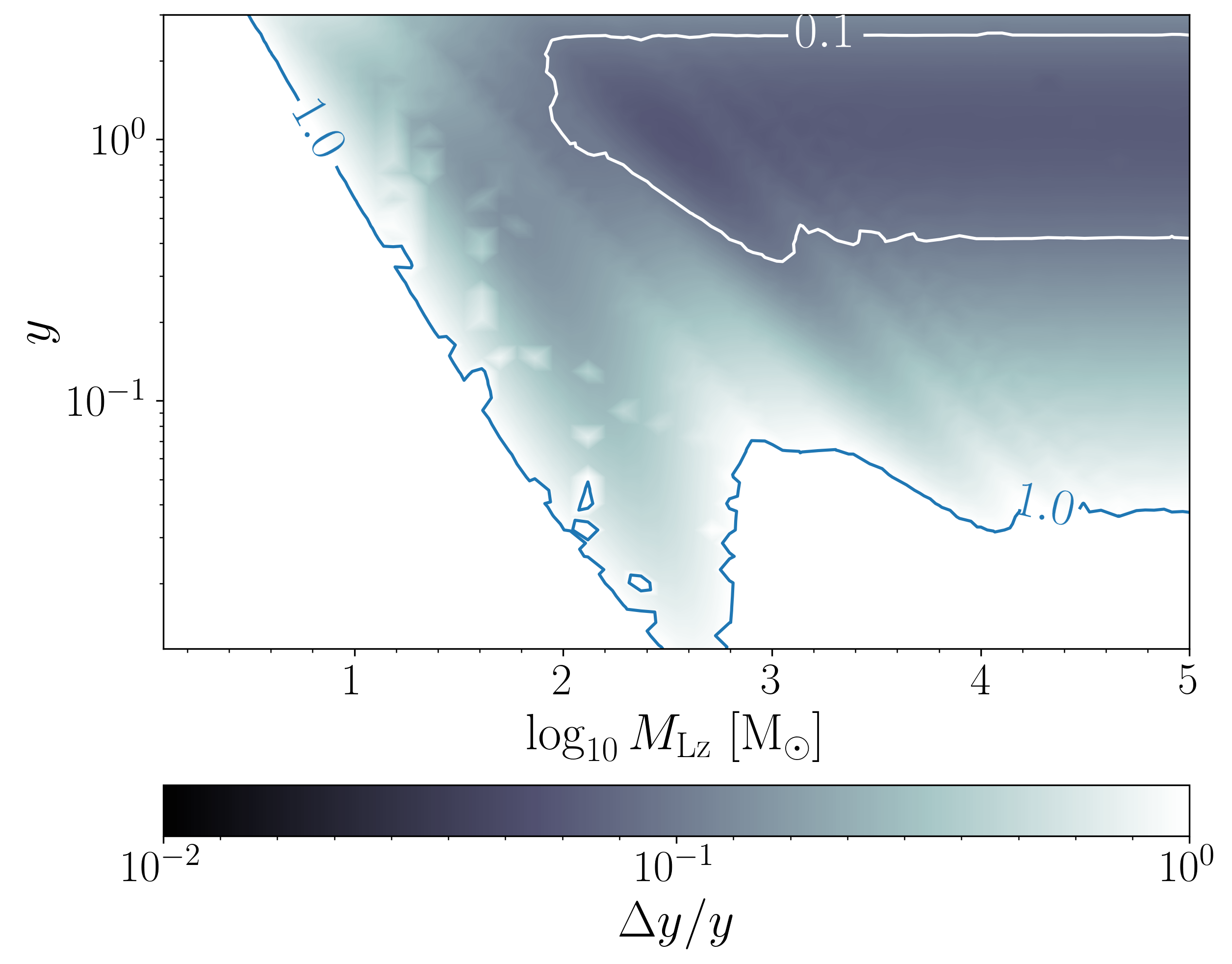}
         \caption*{}
         \label{fig:rel_err_y}
     \end{subfigure}
        \caption{Statistical $1$-$\sigma$ relative uncertainties in the measurement of the redshifted lens mass $\Delta M_\mathrm{Lz}/ M_\mathrm{Lz}$ (top panel) and impact parameter $\Delta y/y$ (bottom panel) in the $(M_\mathrm{Lz},~y)$ plane for a point lens. The white and blue lines lines correspond to $10\%$ and $100\%$ relative errors. In the white regions, the relative errors are larger than $100\%$. The system comprises GW150914-like signals with added microlensing effects. The SNR is kept fixed to $50$ in the detector network of Hanford, Livingston and Virgo using projected O4 sensitivities. }
        \label{fig:rel_errs_fisher}
\end{figure}

\begin{figure}
    \centering
    \includegraphics[width=0.5\textwidth]{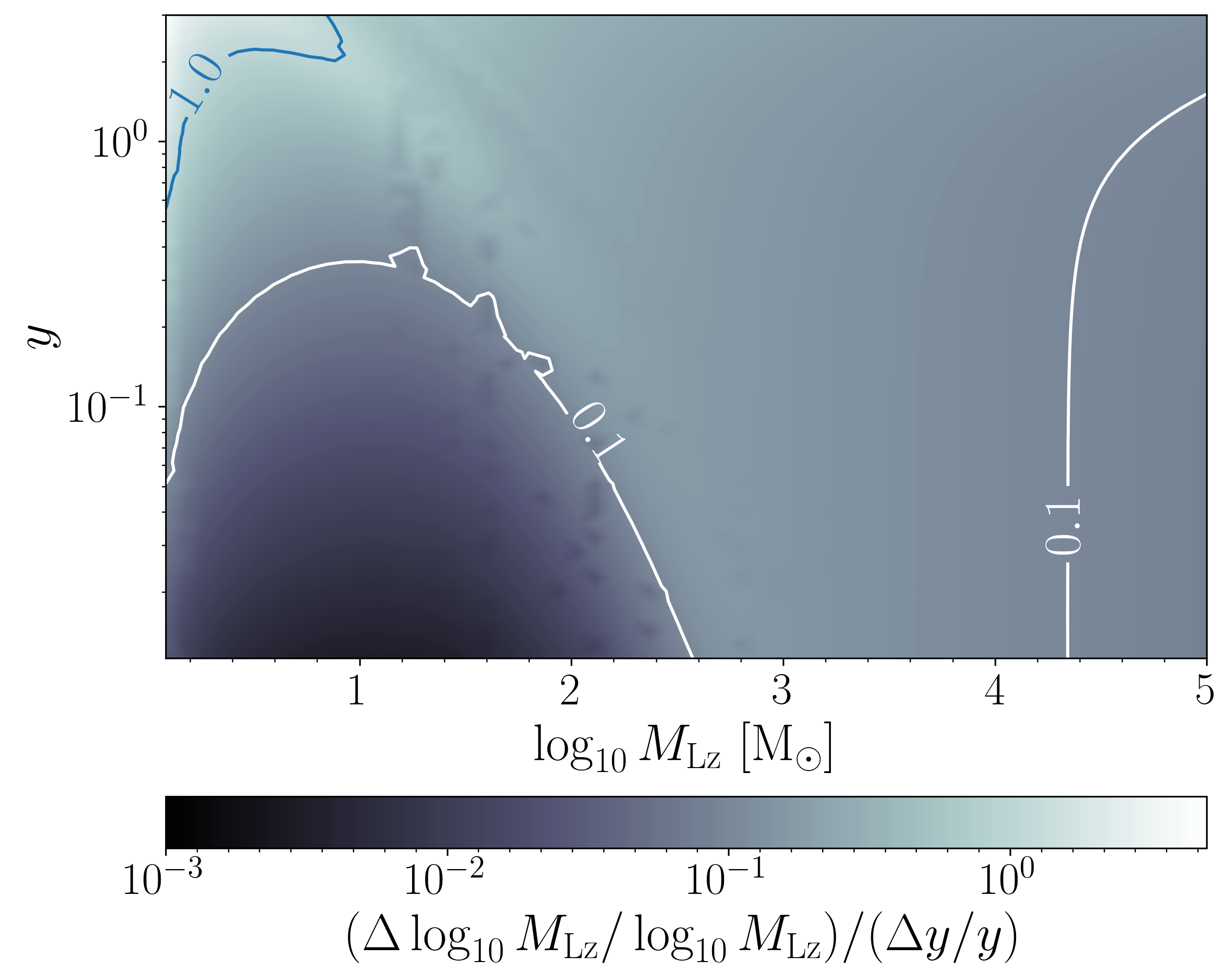}
         \caption{Comparison of the relative $1$-$\sigma$ uncertainties in the (log) redshifted lens mass $\log_{10} M_\mathrm{Lz}$ and $y$. The relative uncertainties in the measurement of $\log_{10} M_\mathrm{Lz}$ are almost always less than that in $y$.}
    \label{fig:rel_errs_comparison}
\end{figure}

The results are shown in Fig. \ref{fig:rel_errs_fisher}, where we plot the statistical relative uncertainties in the measurement of the redshifted lens mass $\Delta M_\mathrm{Lz}/ M_\mathrm{Lz}$ (top panel) and impact parameter $\Delta y/y$ (bottom panel) in the $(M_\mathrm{Lz},$~y$)$ plane for a point lens. The white and blue lines correspond to $10\%$ and $100\%$ relative errors, respectively. The relative errors in the white regions are larger than $100\%$.  

Firstly, if we focus on the geometrical-optics regime in Fig. \ref{fig:rel_errs_fisher} (top-right corner; see \Fref{fig:f_ML_cplot_for_Mlz_vs_y} for reference), we observe mostly similar trends between $\Delta M_\mathrm{Lz}/ M_\mathrm{Lz}$ and $\Delta y/y$. Notably, in the top-right corner of both panels, we observe that the relative uncertainties become independent of variation in the lens mass, i.e., they become constant for a given y value (see, for example, the white contour lines in the top-right corner.). Similarly, we observe that as we decrease $y$ below $\sim0.5$ keeping $\log_{10}M_\mathrm{Lz}$ to be high $\gtrsim 3$, the uncertainties increase drastically and can even exceed $100\%$ for low $y<0.1$. 
This is because in the geometrical-optics regime, the uncertainties in both the parameters depend only on $y$ and the SNR of the signal. As we go away from $y=1$, the uncertainties increase. For $y\gg1$, they are proportional to, roughly, $\sim \sqrt{y}$, while for low $y\ll1$, they increase as, roughly, $\sqrt{1/y}$. Since we have kept the SNR fixed, the uncertainties become roughly constant for a given $y$ value when lens mass is high $\log_{10}M_\mathrm{Lz}\gtrsim 3$. These results are consistent with \cite{2003ApJ...595.1039T}, where a thorough investigation of relative uncertainties in the geometrical-optics regime is illustrated. We note that although the region in the bottom-right corner is not where geometrical optics is a good approximation, the divergence in the uncertainties is still well-captured by the expression obtained for that regime \citep{2003ApJ...595.1039T}.

In the long-wavelength regime (mainly bottom-left region; see \Fref{fig:f_ML_cplot_for_Mlz_vs_y} for reference), where the microlensing effects are weak, we notice that the uncertainties in $y$ are much larger than those in $M_\mathrm{Lz}$. This is explained by the fact that in this regime, the modulations are proportional to the dimensionless frequency $\omega = 8\pi GM_{\rm Lz} f/c^3$ in the leading order \citep{Tambalo:2022plm}. Hence, the estimation of $M_{\rm Lz}$ is better than that of $y$, which leads to the fact that for a given $M_{\rm Lz}$ in this regime, $y$ is only poorly constrained leading to high relative errors.  

Lastly, in Fig. \ref{fig:rel_errs_comparison}, we plot the ratio of relative uncertainties in the measurement of $\log_{10}M_\mathrm{Lz}$ and $y$ for comparison. We notice that in the parameter space of interest, $\log_{10}M_\mathrm{Lz}$ is almost always better measured than $y$, owing to the majority of region having a value less than unity.


\section{Additional Figures}

\begin{figure*}
    \centering
    \includegraphics[width=\linewidth]{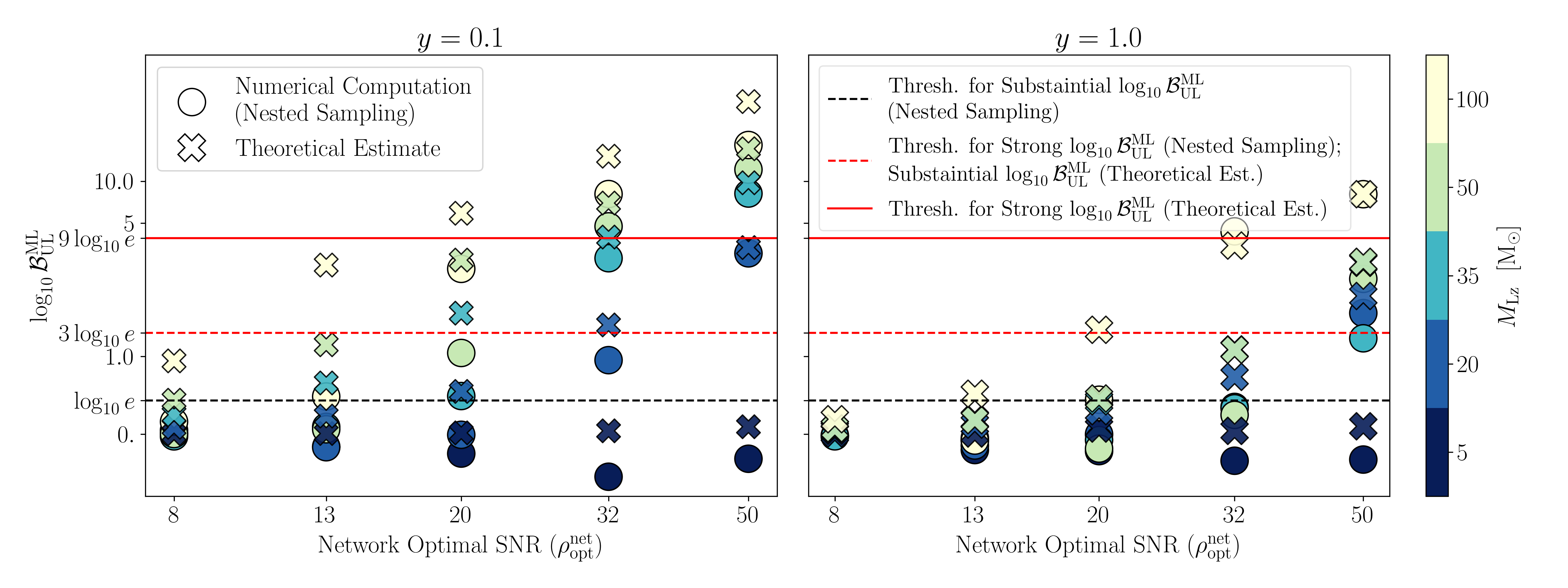}
    \caption{Same as \Fref{fig:ul_vs_ml_hypothesis_comb}, but with added theoretical estimates (cross marks) using \Eref{eq:Bayes_fac_crude_estimate} for comparison. For nested sampling, the dashed black and red lines have a similar meaning as in \Fref{fig:ul_vs_ml_hypothesis_comb}, representing the threshold values for positive and strong evidence for microlensing, respectively. For the theoretical estimate, these values are depicted using dashed and solid red lines, respectively. }
    \label{fig:ul_vs_ml_hypothesis_comb_BF_err_study}
\end{figure*}

\begin{figure*}
    \centering
    \includegraphics[width=0.98\linewidth]{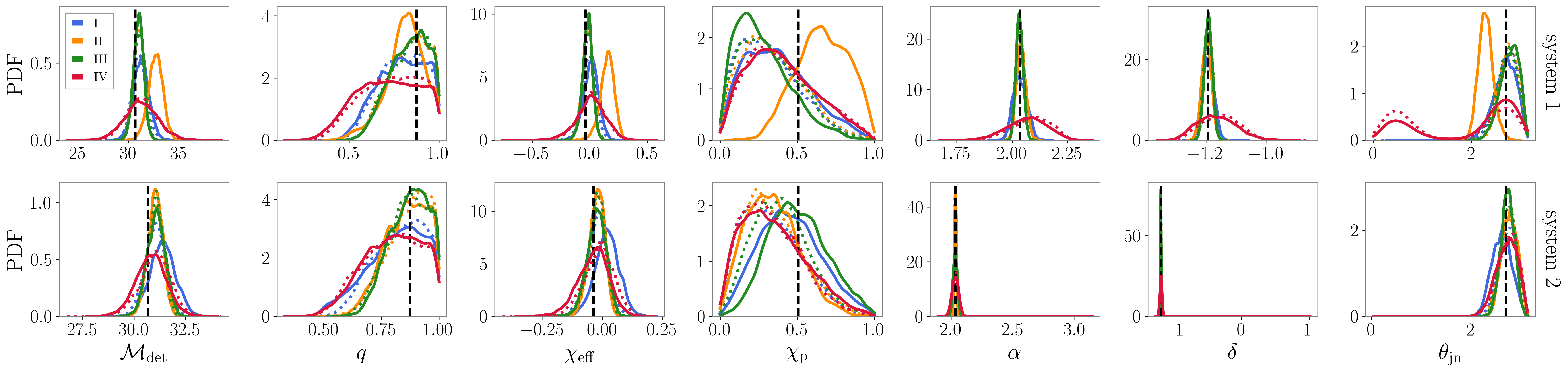}
    \caption{Effect of microlensing due to a population of microlenses on strongly lensed gravitational wave signals. The panels display 1D marginalised posterior distributions for a set of parameters, as labelled on the $x$-axis, for two systems (along the row). These parameters were utilised to compute the posterior overlap, with the exception that we have condensed the spin parameters to $\{\chi_\mathrm{eff},\chi_\mathrm{p}\}$ instead of $\{a_1,a_2,\theta_1,\theta_2\}$ for the ease of representation here. The differently coloured curves correspond to posteriors associated with different images, as shown in Figure \ref{fig:PO_analysis_1}. The dotted curves represent cases with only strong lensing, while the solid curves depict recoveries for signals that undergo both strong lensing and microlensing. The dashed black vertical lines represent the injected values.}
    \label{fig:PO_analysis_2}
\end{figure*}

\bsp	
\label{lastpage}
\end{document}